\documentclass{article}
\usepackage[utf8]{inputenc}
\usepackage{rotating}

\usepackage{enumerate}
\usepackage{mathtools}
\mathtoolsset{showonlyrefs=true}
\usepackage{todonotes}

\usepackage{url}
\usepackage{amsfonts}
\usepackage{amsmath,amssymb,color}
\usepackage{graphicx}
\usepackage{epstopdf}
\usepackage{multirow}
\usepackage{bm}
\usepackage{natbib}

\usepackage{verbatim}
\usepackage{float}
\usepackage[toc,page]{appendix}
\usepackage[doublespacing]{setspace}
\usepackage{setspace}
\usepackage{adjustbox}
\usepackage{rotating}
\usepackage{subcaption}
\usepackage{amsthm}
\usepackage{booktabs,dcolumn,caption}

\usepackage{color}
\usepackage{ulem}

\usepackage{tikz}
\usepackage{multirow}
\usepackage{graphicx}
\usepackage{adjustbox}
\usepackage{float}
\usepackage{footnote}
\usepackage{array}
\usepackage{longtable}
\usepackage{threeparttablex}
\usepackage{booktabs,dcolumn,caption}
\usepackage{amsmath,amssymb,color}
\usepackage{fnbreak}
\usetikzlibrary{arrows.meta,shapes.multipart,positioning}
\usepackage{mathtools}

\usepackage{mathabx}
\usepackage{array}
\usepackage{longtable}
\usepackage{threeparttablex}

\usepackage{chngcntr}
\usepackage{pdflscape}

\usepackage{longtable}
\usepackage{threeparttablex}

\usepackage{placeins}

\usepackage{tikz} 
\usepackage{algpseudocode,algorithm,algorithmicx}

\let\hat\widehat
\let\tilde\widetilde

\usepackage{refcount}

\newcolumntype{L}[1]{>{\raggedright\let\newline\\\arraybackslash\hspace{0pt}}p{#1}}
\newcolumntype{C}[1]{>{\centering\let\newline\\\arraybackslash\hspace{0pt}}p{#1}}
\newcolumntype{R}[1]{>{\raggedleft\let\newline\\\arraybackslash\hspace{0pt}}p{#1}}

\newcommand{\Ch}{\mbox{Ch}}

\newcommand{\argmin}{\operatornamewithlimits{arg\,min}}

\newtheorem{theorem}{Theorem}

\newtheorem{condition}{Condition}

\newtheorem{lemma}{Lemma}

\newtheorem{proposition}{Proposition}

\newtheorem{remark}{Remark}

\title{Exploratory Hierarchical Factor Analysis with an Application to Psychological Measurement}
\author{Jiawei Qiao, Yunxiao Chen and Zhiliang Ying}
\date{}
\begin{document}

\maketitle

\begin{abstract}

\bigskip
\noindent

Hierarchical factor models, which include the bifactor model as a special case, are useful in social and behavioural sciences for measuring hierarchically structured constructs. Specifying a hierarchical factor model involves imposing hierarchically structured zero constraints on a factor loading matrix, which is often challenging. Therefore, an exploratory analysis is needed to learn the hierarchical factor structure from data. Unfortunately, there does not exist an identifiability theory for the learnability of this hierarchical structure, nor a computationally efficient method with provable performance. The method of Schmid-Leiman transformation, which is often regarded as the default method for exploratory hierarchical factor analysis, is flawed and likely to fail. The contribution of this paper is three-fold. First, an identifiability result is established for general hierarchical factor models, which shows that the hierarchical factor structure is learnable under mild regularity conditions. Second, a computationally efficient divide-and-conquer approach is proposed for 
learning the hierarchical factor structure. Finally, asymptotic theory is established for the proposed method, showing that it can consistently recover the true hierarchical factor structure as the sample size grows to infinity.
 The power of the proposed method is shown via simulation studies and a real data application to a personality test. The computation code for the proposed method is publicly available at {https://github.com/EmetSelch97/EHFA/}.

\end{abstract}

\section{Introduction}\label{sec:intro}

Many constructs in social and behavioural sciences are conceptualized to be hierarchically structured, such as psychological traits \citep[e.g.,][]{carroll1993human,deyoung2006higher},  economic factors \citep[e.g.,][]{kose2008understanding,moench2013dynamic}, health outcomes measures \citep[e.g.,][]{chen2006comparison,reise2007role}, and constructs in marketing research \citep[e.g.,][]{sharma2022using}. Hierarchical factor models \citep{brunner2012tutorial,schmid1957development,thomson1939factorial,yung1999relationship}, which include the bi-factor model \citep{holzinger1937bi} as a special case with two factor layers, are commonly used  to measure hierarchically structured constructs. In these models, hierarchically structured zero constraints are imposed on factor loadings to define the hierarchical factors. 
When the hierarchical factor structure is known or hypothesized a priori, the statistical inference of a hierarchical factor model only requires standard confirmatory factor analysis techniques \citep{brunner2012tutorial}. However,  for many real-world scenarios, little prior information about the hierarchical factor structure is available, so we need to learn this structure from data. This analysis is referred to as exploratory hierarchical factor analysis.

{Exploratory hierarchical factor analysis is a structured extension of classical exploratory factor analysis \citep[e.g.,][]{anderson2003introduction,chen2019joint}. In conventional exploratory factor analysis, rotation methods \citep[e.g.,][]{browne2001overview} are typically employed to achieve a sparse loading structure \citep{Thurstone1947FABook} for interpreting the factors. Exploratory hierarchical factor analysis builds on this principle but imposes a hierarchical sparsity pattern on the loading matrix, requiring that zero loadings be placed nonarbitrarily and follow a hierarchical structure.} {Compared with classical exploratory factor analysis,} exploratory hierarchical factor analysis faces theoretical and computational challenges. First, we lack a theoretical understanding of its identifiability, i.e., the conditions under which the hierarchical factor structure is uniquely determined by the distribution of manifest variables. This is an important question, as learning a hierarchical factor structure is only sensible when it is identifiable. 
Although identifiability theory has been established for exploratory bi-factor analysis in \cite{Qiao_Chen_Ying_2025}, to our knowledge, no results are available under the general hierarchical factor model. Second, 
learning the hierarchical factor structure is a model selection problem, which is computationally challenging due to its combinatorial nature. For a moderately large number of manifest variables, it is computationally infeasible to compare all the possible hierarchical factor structures using relative fit measures. However, it is worth noting that a computationally efficient method is available and commonly used for this problem, known as the Schmid–Leiman transformation \citep{schmid1957development}. This method involves constructing a constrained higher-order factor model by iteratively applying an exploratory factor analysis method with oblique rotation and, further, performing orthogonal transformations to turn the higher-order factor model solution into a hierarchical factor model solution. However, as shown in \cite{yung1999relationship}, the Schmid-Leiman transformation imposes unnecessary proportionality constraints on the factor loadings. As a result, it may not work well for more general hierarchical factor models.  
\cite{jennrich2011exploratory} gave an example in which the Schmid–Leiman transformation fails to recover a bi-factor loading structure. Not only theoretically flawed, 
the implementation of the Schmid–Leiman transformation can also be a challenge for practitioners due to several decisions one needs to make,  including the choice of oblique rotation method for the exploratory factor analysis and how the number of factors is determined in each iteration.  

This paper fills these gaps. Specifically, we establish an identifiability result for exploratory hierarchical factor analysis, showing that the hierarchical factor structure is learnable under mild regularity conditions. We also propose a computationally efficient divide-and-conquer approach for learning the hierarchical factor structure.  
This approach divides the learning problem into many subtasks of learning the factors nested within a factor, also known as the child factors of this factor. It conquers these subtasks layer by layer, starting from the one consisting only of the general factor. 
Our method for solving each subtask has two building blocks -- (1) a constraint-based continuous optimization algorithm and (2) a search algorithm based on an information criterion. The former is used to explore the number and 
loading structure of the child factors, and the latter serves as a refinement step that ensures the true structure of the child factors is selected with high probability. Finally, asymptotic theory is established for the proposed method, showing that it can consistently recover the true hierarchical factor structure as the sample size grows to infinity.

The proposed method is closely related to the method proposed in \cite{Qiao_Chen_Ying_2025} for exploratory bi-factor analysis, which can be seen as a special case of the current method when the hierarchical factor structure is known to have only two layers. However, we note that the current problem is substantially more challenging as the complexity of a hierarchical factor structure grows quickly as the number of factor layers increases. Nevertheless, 
the constraint-based continuous optimization algorithm that serves as a building block of the proposed method is similar to the algorithm used for exploratory bi-factor analysis in \cite{Qiao_Chen_Ying_2025}. This algorithm turns a computationally challenging combinatorial model selection problem into a relatively easier-to-solve continuous optimization problem, enabling a more efficient global search of the factor structure. 

The rest of the paper is organized as follows.  In Section~\ref{sec:ehfa}, we establish the identifiability of the general hierarchical factor model and, further, propose 
a divide-and-conquer approach for exploratory hierarchical factor analysis and establish its consistency. In Section~\ref{sec:computation}, the computation of the divide-and-conquer approach is discussed. Simulation studies and a real data example
are presented in Sections~\ref{sec:sim} and \ref{sec:real data}, respectively, to evaluate the performance of the proposed method. We conclude with discussions in Section~\ref{sec:diss}.

\section{Exploratory Hierarchical Factor Analysis}\label{sec:ehfa}

\subsection{Constraints of hierarchical factor model}\label{subsec:constraints}

Consider a factor model for $J$ observed variables, with $K$ orthogonal factors. The population covariance matrix can be decomposed as  
$\Sigma = \Lambda \Lambda^\top + \Psi,$ where $\Lambda = (\lambda_{jk})_{J\times K}$ is the loading matrix and $\Psi$ is a $J\times J$ diagonal matrix, which is typically referred as the unique variance matrix \citep[see, e.g.,][]{fabrigar2012exploratory}, with diagonal entries $\psi_1, \ldots, \psi_J > 0$ that record the unique variances. We say this factor model is a hierarchical factor model if the loading matrix $\Lambda$ satisfies certain zero constraints that encode a factor hierarchy.

Specifically, let $v_k = \{j: \lambda_{jk} \neq 0\}$ be the variables loading on the $k$th factor.  The factor model becomes a hierarchical factor model if $v_1, \ldots, v_K$ satisfy the following constraints:  

\begin{enumerate} 

\item[C1.] $v_1 = \{1, \ldots, J\}$ corresponds to a general factor that is loaded on by all the items.  

\item[C2.] For any
$k< l$, it holds that either $v_l \subsetneq v_k$ or $v_l \subset \{1, \ldots, J\} \setminus {v}_k$. That is, the variables that load on factor $l$ are either a subset of those that load on factor $k$ or do not overlap with them. When $v_l \subsetneq v_k$, we say factor $l$ is a descendant factor of factor $k$. If further that there does not exist $k’$ such that $k < k’ < l$ and $v_l \subsetneq v_{k’} \subsetneq v_{k}$, we say factor $l$ is a child factor of factor $k$, and factor $k$ is a parent factor of factor $l$.

\item[C3.] For a given factor $k$, we denote all its child factors as $\mbox{\Ch}_k$. Then its cardinality $|\mbox{\Ch}_k |$  satisfies that $|\mbox{Ch}_k | = 0$ or $|\mbox{\Ch}_k | \geq 2$. That is, a factor either does not have any child factor or at least two child factors. Moreover, when a factor $k$ has two or more child factors, these child factors satisfy that  
$v_{l} \cap v_{l'} = \emptyset,$
for any $l, l' \in \mbox{Ch}_k$, and   
$\cup_{l \in \mbox{Ch}_k} v_l = v_k.$
That is, the sets of variables that load on the child factors of a factor are a partition of the variables that load on this factor. We note that one child node is not allowed due to identification issues. To avoid ambiguity in the labelling of the factors, we further require that 
\begin{enumerate}
    \item $k < l$ if factors $k$ and $l$ are the child factors of the same factor and $\min\{v_k\} < \min\{v_l\}$. That is, we label the child factors of the same factor based on the labels of the variables that load on each factor. 
     \item $k < l$ if factors $k$ and $l$ do not have the same parent factor, and the parent factor of $k$ has a smaller label than the parent factor of $l$. 
\end{enumerate}

\end{enumerate} 

The requirement $|\mbox{Ch}_k | = 0$ or $|\mbox{\Ch}_k | \geq 2$ in constraint C3 is necessary for the hierarchical factor model to be identifiable. 
When a factor $k$ has a unique child factor (i.e. $|\mbox{Ch}_{k}|=1$), it is easy to show that the two columns of the loading matrix that correspond to factor $k$ and its single child factor are not determined up to an orthogonal rotation. 
 
We note that when the above constraints hold, the hierarchical factor structure can be visualized as a tree, where each internal node represents a factor, and each leaf node represents an observed variable. In this tree, factor $l$ being a child factor of factor $k$, is represented by node $l$ being a child node of node $k$. The variables that load on each factor are indicated by its descendant leaf nodes. 

When the factors follow a hierarchical structure, we can classify the factors into layers. The first factor layer only includes the general factor, denoted by $L_1 = \{1\}$. The rest of the layers can be defined recursively. That is, if a factor $k$ is in the $t$th layer, then its child factors are in the $(t+1)$th layer. Let $T$ be the total number of layers and $L_1, \ldots, L_T$ be the sets of factors for the $T$ layers.  It is worth noting that the way the layers are labelled here is opposite to how they are labelled in the literature. That is, we label the layers from the top to the bottom of the hierarchy of the factors. In contrast, they are labelled from the bottom to the top in the literature \citep[see, e.g.,][]{yung1999relationship}. We adopt the current labelling system because it is more convenient for the proposed method in Section~\ref{subsec:overview method} that learns the factor hierarchy from top to bottom.

\begin{figure}[tb!]
    \centering

    \begin{subfigure}[b]{0.9\linewidth}
        \centering

\tikzset{every picture/.style={line width=0.75pt}} 

\begin{tikzpicture}[x=0.75pt,y=0.75pt,yscale=-1,xscale=1]

\draw   (270.17,31) -- (291.17,31) -- (291.17,51) -- (270.17,51) -- cycle ;
\draw   (311.17,71) -- (332.17,71) -- (332.17,91) -- (311.17,91) -- cycle ;
\draw   (435,71) -- (456,71) -- (456,91) -- (435,91) -- cycle ;
\draw   (66,111) -- (87,111) -- (87,131) -- (66,131) -- cycle ;
\draw   (128.5,71) -- (149.5,71) -- (149.5,91) -- (128.5,91) -- cycle ;
\draw   (190,111) -- (211,111) -- (211,131) -- (190,131) -- cycle ;
\draw   (19.67,151) -- (40.67,151) -- (40.67,172.5) -- (19.67,172.5) -- cycle ;
\draw   (49.67,151) -- (70.67,151) -- (70.67,172.5) -- (49.67,172.5) -- cycle ;
\draw   (80.67,151) -- (101.67,151) -- (101.67,172.5) -- (80.67,172.5) -- cycle ;
\draw   (110.67,151) -- (131.67,151) -- (131.67,172.5) -- (110.67,172.5) -- cycle ;
\draw   (144.67,151) -- (165.67,151) -- (165.67,172.5) -- (144.67,172.5) -- cycle ;
\draw   (174.67,151) -- (195.67,151) -- (195.67,172.5) -- (174.67,172.5) -- cycle ;
\draw   (205.67,151) -- (226.67,151) -- (226.67,172.5) -- (205.67,172.5) -- cycle ;
\draw   (235.67,151) -- (256.67,151) -- (256.67,172.5) -- (235.67,172.5) -- cycle ;
\draw   (265.67,151) -- (286.67,151) -- (286.67,172.5) -- (265.67,172.5) -- cycle ;
\draw   (295.67,151) -- (316.67,151) -- (316.67,172.5) -- (295.67,172.5) -- cycle ;
\draw   (326.67,151) -- (347.67,151) -- (347.67,172.5) -- (326.67,172.5) -- cycle ;
\draw   (357.67,151) -- (378.67,151) -- (378.67,172.5) -- (357.67,172.5) -- cycle ;
\draw   (387.67,151) -- (408.67,151) -- (408.67,172.5) -- (387.67,172.5) -- cycle ;
\draw   (418.67,151) -- (439.67,151) -- (439.67,172.5) -- (418.67,172.5) -- cycle ;
\draw   (449.67,151) -- (470.67,151) -- (470.67,172.5) -- (449.67,172.5) -- cycle ;
\draw   (481.67,151) -- (502.67,151) -- (502.67,172.5) -- (481.67,172.5) -- cycle ;
\draw    (280.67,51) -- (280.67,61) ;
\draw    (139,61) -- (445.5,61) ;
\draw    (139,61) -- (139,71) ;
\draw    (321.67,61) -- (321.67,71) ;
\draw    (445.5,61) -- (445.5,71) ;
\draw    (76.5,100) -- (200.5,100) ;
\draw    (139,91) -- (138.5,100) ;
\draw    (76.5,100) -- (76.5,111) ;
\draw    (200.5,100) -- (200.5,111) ;
\draw    (30.17,141) -- (121.17,141) ;
\draw    (76.5,131) -- (76.5,141) ;
\draw    (30.17,141) -- (30.17,151) ;
\draw    (60.17,141) -- (60.17,151) ;
\draw    (91.17,141) -- (91.17,151) ;
\draw    (121.17,141) -- (121.17,151) ;
\draw    (155.17,141) -- (246.17,141) ;
\draw    (200.5,131) -- (200.67,141) ;
\draw    (155.17,141) -- (155.17,151) ;
\draw    (185.17,141) -- (185.17,151) ;
\draw    (216.17,141) -- (216.17,151) ;
\draw    (246.17,141) -- (246.17,151) ;
\draw    (276.17,100) -- (368.17,100) ;
\draw    (276.17,100) -- (276.17,151) ;
\draw    (306.17,100) -- (306.17,151) ;
\draw    (321.67,91) -- (322.17,100) ;
\draw    (337.17,100) -- (337.17,151) ;
\draw    (368.17,100) -- (368.17,151) ;
\draw    (398.17,100) -- (492.17,100) ;
\draw    (445.5,91) -- (445.17,100) ;
\draw    (398.17,100) -- (398.17,151) ;
\draw    (429.17,100) -- (429.17,151) ;
\draw    (460.17,100) -- (460.17,151) ;
\draw    (492.17,100) -- (492.17,151) ;

\draw (272.17,34) node [anchor=north west][inner sep=0.75pt]   [align=left] {$\displaystyle v_{1}$};
\draw (130.5,75) node [anchor=north west][inner sep=0.75pt]   [align=left] {$\displaystyle v_{2}$};
\draw (313.17,75) node [anchor=north west][inner sep=0.75pt]   [align=left] {$\displaystyle v_{3}$};
\draw (437,75) node [anchor=north west][inner sep=0.75pt]   [align=left] {$\displaystyle v_{4}$};
\draw (68,115) node [anchor=north west][inner sep=0.75pt]   [align=left] {$\displaystyle v_{5}$};
\draw (192,115) node [anchor=north west][inner sep=0.75pt]   [align=left] {$\displaystyle v_{6}$};
\draw (25.17,155) node [anchor=north west][inner sep=0.75pt]   [align=left] {1};
\draw (55.17,155) node [anchor=north west][inner sep=0.75pt]   [align=left] {2};
\draw (86.17,155) node [anchor=north west][inner sep=0.75pt]   [align=left] {3};
\draw (116.17,155) node [anchor=north west][inner sep=0.75pt]   [align=left] {4};
\draw (150.17,155) node [anchor=north west][inner sep=0.75pt]   [align=left] {5};
\draw (180.17,155) node [anchor=north west][inner sep=0.75pt]   [align=left] {6};
\draw (211.17,155) node [anchor=north west][inner sep=0.75pt]   [align=left] {7};
\draw (241.17,155) node [anchor=north west][inner sep=0.75pt]   [align=left] {8};
\draw (271.17,155) node [anchor=north west][inner sep=0.75pt]   [align=left] {9};
\draw (297.17,155) node [anchor=north west][inner sep=0.75pt]   [align=left] {10};
\draw (328.67,155) node [anchor=north west][inner sep=0.75pt]   [align=left] {11};
\draw (359.17,155) node [anchor=north west][inner sep=0.75pt]   [align=left] {12};
\draw (389.17,155) node [anchor=north west][inner sep=0.75pt]   [align=left] {13};
\draw (420.17,155) node [anchor=north west][inner sep=0.75pt]   [align=left] {14};
\draw (451.17,155) node [anchor=north west][inner sep=0.75pt]   [align=left] {15};
\draw (483.17,155) node [anchor=north west][inner sep=0.75pt]   [align=left] {16};
\draw (533,33) node [anchor=north west][inner sep=0.75pt]   [align=left] {$\displaystyle L_{1}$};
\draw (533,75) node [anchor=north west][inner sep=0.75pt]   [align=left] {$\displaystyle L_{2}$};
\draw (533,115) node [anchor=north west][inner sep=0.75pt]   [align=left] {$\displaystyle L_{3}$};

\end{tikzpicture}

        \caption{The hierarchical factor structure of a three-layer hierarchical factor model.}
        \label{fig: example hierarchy}
    \end{subfigure}

\hfill
\hfill

    \begin{subfigure}[b]{0.9\linewidth}
        \centering

\tikzset{every picture/.style={line width=0.75pt}} 

\begin{tikzpicture}[x=0.75pt,y=0.75pt,yscale=-1,xscale=1]

\draw   (39.67,171) -- (60.67,171) -- (60.67,192.5) -- (39.67,192.5) -- cycle ;
\draw   (100.67,171) -- (121.67,171) -- (121.67,192.5) -- (100.67,192.5) -- cycle ;
\draw   (69.67,171) -- (90.67,171) -- (90.67,192.5) -- (69.67,192.5) -- cycle ;
\draw   (130.67,171) -- (151.67,171) -- (151.67,192.5) -- (130.67,192.5) -- cycle ;
\draw   (164.67,171) -- (185.67,171) -- (185.67,192.5) -- (164.67,192.5) -- cycle ;
\draw   (225.67,171) -- (246.67,171) -- (246.67,192.5) -- (225.67,192.5) -- cycle ;
\draw   (194.67,171) -- (215.67,171) -- (215.67,192.5) -- (194.67,192.5) -- cycle ;
\draw   (255.67,171) -- (276.67,171) -- (276.67,192.5) -- (255.67,192.5) -- cycle ;
\draw   (285.67,171) -- (306.67,171) -- (306.67,192.5) -- (285.67,192.5) -- cycle ;
\draw   (315.67,171) -- (336.67,171) -- (336.67,192.5) -- (315.67,192.5) -- cycle ;
\draw   (346.67,171) -- (367.67,171) -- (367.67,192.5) -- (346.67,192.5) -- cycle ;
\draw   (377.67,171) -- (398.67,171) -- (398.67,192.5) -- (377.67,192.5) -- cycle ;
\draw   (407.67,171) -- (428.67,171) -- (428.67,192.5) -- (407.67,192.5) -- cycle ;
\draw   (438.67,171) -- (459.67,171) -- (459.67,192.5) -- (438.67,192.5) -- cycle ;
\draw   (469.67,171) -- (490.67,171) -- (490.67,192.5) -- (469.67,192.5) -- cycle ;
\draw   (501.67,171) -- (522.67,171) -- (522.67,192.5) -- (501.67,192.5) -- cycle ;
\draw   (271.92,65.75) .. controls (271.92,58.71) and (277.63,53) .. (284.67,53) .. controls (291.71,53) and (297.42,58.71) .. (297.42,65.75) .. controls (297.42,72.79) and (291.71,78.5) .. (284.67,78.5) .. controls (277.63,78.5) and (271.92,72.79) .. (271.92,65.75) -- cycle ;
\draw   (208.42,247.75) .. controls (208.42,240.71) and (214.13,235) .. (221.17,235) .. controls (228.21,235) and (233.92,240.71) .. (233.92,247.75) .. controls (233.92,254.79) and (228.21,260.5) .. (221.17,260.5) .. controls (214.13,260.5) and (208.42,254.79) .. (208.42,247.75) -- cycle ;
\draw   (325.42,292.75) .. controls (325.42,285.71) and (331.13,280) .. (338.17,280) .. controls (345.21,280) and (350.92,285.71) .. (350.92,292.75) .. controls (350.92,299.79) and (345.21,305.5) .. (338.17,305.5) .. controls (331.13,305.5) and (325.42,299.79) .. (325.42,292.75) -- cycle ;
\draw   (145.25,292.75) .. controls (145.25,285.71) and (150.96,280) .. (158,280) .. controls (165.04,280) and (170.75,285.71) .. (170.75,292.75) .. controls (170.75,299.79) and (165.04,305.5) .. (158,305.5) .. controls (150.96,305.5) and (145.25,299.79) .. (145.25,292.75) -- cycle ;
\draw   (451.42,292.75) .. controls (451.42,285.71) and (457.13,280) .. (464.17,280) .. controls (471.21,280) and (476.92,285.71) .. (476.92,292.75) .. controls (476.92,299.79) and (471.21,305.5) .. (464.17,305.5) .. controls (457.13,305.5) and (451.42,299.79) .. (451.42,292.75) -- cycle ;
\draw    (284.67,78.5) -- (52.03,170.27) ;
\draw [shift={(50.17,171)}, rotate = 338.47] [color={rgb, 255:red, 0; green, 0; blue, 0 }  ][line width=0.75]    (10.93,-3.29) .. controls (6.95,-1.4) and (3.31,-0.3) .. (0,0) .. controls (3.31,0.3) and (6.95,1.4) .. (10.93,3.29)   ;
\draw    (284.67,78.5) -- (81.99,170.18) ;
\draw [shift={(80.17,171)}, rotate = 335.66] [color={rgb, 255:red, 0; green, 0; blue, 0 }  ][line width=0.75]    (10.93,-3.29) .. controls (6.95,-1.4) and (3.31,-0.3) .. (0,0) .. controls (3.31,0.3) and (6.95,1.4) .. (10.93,3.29)   ;
\draw    (284.67,78.5) -- (112.93,170.06) ;
\draw [shift={(111.17,171)}, rotate = 331.94] [color={rgb, 255:red, 0; green, 0; blue, 0 }  ][line width=0.75]    (10.93,-3.29) .. controls (6.95,-1.4) and (3.31,-0.3) .. (0,0) .. controls (3.31,0.3) and (6.95,1.4) .. (10.93,3.29)   ;
\draw    (284.67,78.5) -- (176.69,169.71) ;
\draw [shift={(175.17,171)}, rotate = 319.81] [color={rgb, 255:red, 0; green, 0; blue, 0 }  ][line width=0.75]    (10.93,-3.29) .. controls (6.95,-1.4) and (3.31,-0.3) .. (0,0) .. controls (3.31,0.3) and (6.95,1.4) .. (10.93,3.29)   ;
\draw    (284.67,78.5) -- (206.47,169.48) ;
\draw [shift={(205.17,171)}, rotate = 310.68] [color={rgb, 255:red, 0; green, 0; blue, 0 }  ][line width=0.75]    (10.93,-3.29) .. controls (6.95,-1.4) and (3.31,-0.3) .. (0,0) .. controls (3.31,0.3) and (6.95,1.4) .. (10.93,3.29)   ;
\draw    (284.67,78.5) -- (142.85,169.92) ;
\draw [shift={(141.17,171)}, rotate = 327.19] [color={rgb, 255:red, 0; green, 0; blue, 0 }  ][line width=0.75]    (10.93,-3.29) .. controls (6.95,-1.4) and (3.31,-0.3) .. (0,0) .. controls (3.31,0.3) and (6.95,1.4) .. (10.93,3.29)   ;
\draw    (284.67,78.5) -- (237.1,169.23) ;
\draw [shift={(236.17,171)}, rotate = 297.67] [color={rgb, 255:red, 0; green, 0; blue, 0 }  ][line width=0.75]    (10.93,-3.29) .. controls (6.95,-1.4) and (3.31,-0.3) .. (0,0) .. controls (3.31,0.3) and (6.95,1.4) .. (10.93,3.29)   ;
\draw    (284.67,78.5) -- (266.56,169.04) ;
\draw [shift={(266.17,171)}, rotate = 281.31] [color={rgb, 255:red, 0; green, 0; blue, 0 }  ][line width=0.75]    (10.93,-3.29) .. controls (6.95,-1.4) and (3.31,-0.3) .. (0,0) .. controls (3.31,0.3) and (6.95,1.4) .. (10.93,3.29)   ;
\draw    (284.67,78.5) -- (295.92,169.02) ;
\draw [shift={(296.17,171)}, rotate = 262.91] [color={rgb, 255:red, 0; green, 0; blue, 0 }  ][line width=0.75]    (10.93,-3.29) .. controls (6.95,-1.4) and (3.31,-0.3) .. (0,0) .. controls (3.31,0.3) and (6.95,1.4) .. (10.93,3.29)   ;
\draw    (284.67,78.5) -- (325.35,169.18) ;
\draw [shift={(326.17,171)}, rotate = 245.84] [color={rgb, 255:red, 0; green, 0; blue, 0 }  ][line width=0.75]    (10.93,-3.29) .. controls (6.95,-1.4) and (3.31,-0.3) .. (0,0) .. controls (3.31,0.3) and (6.95,1.4) .. (10.93,3.29)   ;
\draw    (284.67,78.5) -- (355.93,169.43) ;
\draw [shift={(357.17,171)}, rotate = 231.91] [color={rgb, 255:red, 0; green, 0; blue, 0 }  ][line width=0.75]    (10.93,-3.29) .. controls (6.95,-1.4) and (3.31,-0.3) .. (0,0) .. controls (3.31,0.3) and (6.95,1.4) .. (10.93,3.29)   ;
\draw    (284.67,78.5) -- (386.68,169.67) ;
\draw [shift={(388.17,171)}, rotate = 221.79] [color={rgb, 255:red, 0; green, 0; blue, 0 }  ][line width=0.75]    (10.93,-3.29) .. controls (6.95,-1.4) and (3.31,-0.3) .. (0,0) .. controls (3.31,0.3) and (6.95,1.4) .. (10.93,3.29)   ;
\draw    (284.67,78.5) -- (416.52,169.86) ;
\draw [shift={(418.17,171)}, rotate = 214.72] [color={rgb, 255:red, 0; green, 0; blue, 0 }  ][line width=0.75]    (10.93,-3.29) .. controls (6.95,-1.4) and (3.31,-0.3) .. (0,0) .. controls (3.31,0.3) and (6.95,1.4) .. (10.93,3.29)   ;
\draw    (284.67,78.5) -- (447.42,170.02) ;
\draw [shift={(449.17,171)}, rotate = 209.35] [color={rgb, 255:red, 0; green, 0; blue, 0 }  ][line width=0.75]    (10.93,-3.29) .. controls (6.95,-1.4) and (3.31,-0.3) .. (0,0) .. controls (3.31,0.3) and (6.95,1.4) .. (10.93,3.29)   ;
\draw    (284.67,78.5) -- (478.36,170.14) ;
\draw [shift={(480.17,171)}, rotate = 205.32] [color={rgb, 255:red, 0; green, 0; blue, 0 }  ][line width=0.75]    (10.93,-3.29) .. controls (6.95,-1.4) and (3.31,-0.3) .. (0,0) .. controls (3.31,0.3) and (6.95,1.4) .. (10.93,3.29)   ;
\draw    (284.67,78.5) -- (510.31,170.25) ;
\draw [shift={(512.17,171)}, rotate = 202.13] [color={rgb, 255:red, 0; green, 0; blue, 0 }  ][line width=0.75]    (10.93,-3.29) .. controls (6.95,-1.4) and (3.31,-0.3) .. (0,0) .. controls (3.31,0.3) and (6.95,1.4) .. (10.93,3.29)   ;
\draw    (338.17,280) -- (297.03,194.3) ;
\draw [shift={(296.17,192.5)}, rotate = 64.36] [color={rgb, 255:red, 0; green, 0; blue, 0 }  ][line width=0.75]    (10.93,-3.29) .. controls (6.95,-1.4) and (3.31,-0.3) .. (0,0) .. controls (3.31,0.3) and (6.95,1.4) .. (10.93,3.29)   ;
\draw    (338.17,280) -- (326.44,194.48) ;
\draw [shift={(326.17,192.5)}, rotate = 82.19] [color={rgb, 255:red, 0; green, 0; blue, 0 }  ][line width=0.75]    (10.93,-3.29) .. controls (6.95,-1.4) and (3.31,-0.3) .. (0,0) .. controls (3.31,0.3) and (6.95,1.4) .. (10.93,3.29)   ;
\draw    (338.17,280) -- (356.74,194.45) ;
\draw [shift={(357.17,192.5)}, rotate = 102.25] [color={rgb, 255:red, 0; green, 0; blue, 0 }  ][line width=0.75]    (10.93,-3.29) .. controls (6.95,-1.4) and (3.31,-0.3) .. (0,0) .. controls (3.31,0.3) and (6.95,1.4) .. (10.93,3.29)   ;
\draw    (338.17,280) -- (387.17,194.24) ;
\draw [shift={(388.17,192.5)}, rotate = 119.74] [color={rgb, 255:red, 0; green, 0; blue, 0 }  ][line width=0.75]    (10.93,-3.29) .. controls (6.95,-1.4) and (3.31,-0.3) .. (0,0) .. controls (3.31,0.3) and (6.95,1.4) .. (10.93,3.29)   ;
\draw    (464.17,280) -- (419.1,194.27) ;
\draw [shift={(418.17,192.5)}, rotate = 62.27] [color={rgb, 255:red, 0; green, 0; blue, 0 }  ][line width=0.75]    (10.93,-3.29) .. controls (6.95,-1.4) and (3.31,-0.3) .. (0,0) .. controls (3.31,0.3) and (6.95,1.4) .. (10.93,3.29)   ;
\draw    (464.17,280) -- (449.5,194.47) ;
\draw [shift={(449.17,192.5)}, rotate = 80.27] [color={rgb, 255:red, 0; green, 0; blue, 0 }  ][line width=0.75]    (10.93,-3.29) .. controls (6.95,-1.4) and (3.31,-0.3) .. (0,0) .. controls (3.31,0.3) and (6.95,1.4) .. (10.93,3.29)   ;
\draw    (464.17,280) -- (479.81,194.47) ;
\draw [shift={(480.17,192.5)}, rotate = 100.36] [color={rgb, 255:red, 0; green, 0; blue, 0 }  ][line width=0.75]    (10.93,-3.29) .. controls (6.95,-1.4) and (3.31,-0.3) .. (0,0) .. controls (3.31,0.3) and (6.95,1.4) .. (10.93,3.29)   ;
\draw    (464.17,280) -- (511.2,194.25) ;
\draw [shift={(512.17,192.5)}, rotate = 118.75] [color={rgb, 255:red, 0; green, 0; blue, 0 }  ][line width=0.75]    (10.93,-3.29) .. controls (6.95,-1.4) and (3.31,-0.3) .. (0,0) .. controls (3.31,0.3) and (6.95,1.4) .. (10.93,3.29)   ;
\draw    (158,280) -- (51.72,193.76) ;
\draw [shift={(50.17,192.5)}, rotate = 39.06] [color={rgb, 255:red, 0; green, 0; blue, 0 }  ][line width=0.75]    (10.93,-3.29) .. controls (6.95,-1.4) and (3.31,-0.3) .. (0,0) .. controls (3.31,0.3) and (6.95,1.4) .. (10.93,3.29)   ;
\draw    (158,280) -- (81.5,193.99) ;
\draw [shift={(80.17,192.5)}, rotate = 48.35] [color={rgb, 255:red, 0; green, 0; blue, 0 }  ][line width=0.75]    (10.93,-3.29) .. controls (6.95,-1.4) and (3.31,-0.3) .. (0,0) .. controls (3.31,0.3) and (6.95,1.4) .. (10.93,3.29)   ;
\draw    (158,280) -- (112.11,194.26) ;
\draw [shift={(111.17,192.5)}, rotate = 61.84] [color={rgb, 255:red, 0; green, 0; blue, 0 }  ][line width=0.75]    (10.93,-3.29) .. controls (6.95,-1.4) and (3.31,-0.3) .. (0,0) .. controls (3.31,0.3) and (6.95,1.4) .. (10.93,3.29)   ;
\draw    (158,280) -- (141.54,194.46) ;
\draw [shift={(141.17,192.5)}, rotate = 79.11] [color={rgb, 255:red, 0; green, 0; blue, 0 }  ][line width=0.75]    (10.93,-3.29) .. controls (6.95,-1.4) and (3.31,-0.3) .. (0,0) .. controls (3.31,0.3) and (6.95,1.4) .. (10.93,3.29)   ;
\draw    (158,280) -- (174.78,194.46) ;
\draw [shift={(175.17,192.5)}, rotate = 101.1] [color={rgb, 255:red, 0; green, 0; blue, 0 }  ][line width=0.75]    (10.93,-3.29) .. controls (6.95,-1.4) and (3.31,-0.3) .. (0,0) .. controls (3.31,0.3) and (6.95,1.4) .. (10.93,3.29)   ;
\draw    (158,280) -- (204.22,194.26) ;
\draw [shift={(205.17,192.5)}, rotate = 118.33] [color={rgb, 255:red, 0; green, 0; blue, 0 }  ][line width=0.75]    (10.93,-3.29) .. controls (6.95,-1.4) and (3.31,-0.3) .. (0,0) .. controls (3.31,0.3) and (6.95,1.4) .. (10.93,3.29)   ;
\draw    (158,280) -- (234.83,193.99) ;
\draw [shift={(236.17,192.5)}, rotate = 131.78] [color={rgb, 255:red, 0; green, 0; blue, 0 }  ][line width=0.75]    (10.93,-3.29) .. controls (6.95,-1.4) and (3.31,-0.3) .. (0,0) .. controls (3.31,0.3) and (6.95,1.4) .. (10.93,3.29)   ;
\draw    (158,280) -- (264.61,193.76) ;
\draw [shift={(266.17,192.5)}, rotate = 141.03] [color={rgb, 255:red, 0; green, 0; blue, 0 }  ][line width=0.75]    (10.93,-3.29) .. controls (6.95,-1.4) and (3.31,-0.3) .. (0,0) .. controls (3.31,0.3) and (6.95,1.4) .. (10.93,3.29)   ;
\draw   (82.92,247.75) .. controls (82.92,240.71) and (88.63,235) .. (95.67,235) .. controls (102.71,235) and (108.42,240.71) .. (108.42,247.75) .. controls (108.42,254.79) and (102.71,260.5) .. (95.67,260.5) .. controls (88.63,260.5) and (82.92,254.79) .. (82.92,247.75) -- cycle ;
\draw    (95.67,235) -- (51.63,193.87) ;
\draw [shift={(50.17,192.5)}, rotate = 43.05] [color={rgb, 255:red, 0; green, 0; blue, 0 }  ][line width=0.75]    (10.93,-3.29) .. controls (6.95,-1.4) and (3.31,-0.3) .. (0,0) .. controls (3.31,0.3) and (6.95,1.4) .. (10.93,3.29)   ;
\draw    (95.67,235) -- (80.85,194.38) ;
\draw [shift={(80.17,192.5)}, rotate = 69.96] [color={rgb, 255:red, 0; green, 0; blue, 0 }  ][line width=0.75]    (10.93,-3.29) .. controls (6.95,-1.4) and (3.31,-0.3) .. (0,0) .. controls (3.31,0.3) and (6.95,1.4) .. (10.93,3.29)   ;
\draw    (95.67,235) -- (110.48,194.38) ;
\draw [shift={(111.17,192.5)}, rotate = 110.04] [color={rgb, 255:red, 0; green, 0; blue, 0 }  ][line width=0.75]    (10.93,-3.29) .. controls (6.95,-1.4) and (3.31,-0.3) .. (0,0) .. controls (3.31,0.3) and (6.95,1.4) .. (10.93,3.29)   ;
\draw    (95.67,235) -- (139.71,193.87) ;
\draw [shift={(141.17,192.5)}, rotate = 136.95] [color={rgb, 255:red, 0; green, 0; blue, 0 }  ][line width=0.75]    (10.93,-3.29) .. controls (6.95,-1.4) and (3.31,-0.3) .. (0,0) .. controls (3.31,0.3) and (6.95,1.4) .. (10.93,3.29)   ;
\draw    (221.17,235) -- (176.64,193.86) ;
\draw [shift={(175.17,192.5)}, rotate = 42.74] [color={rgb, 255:red, 0; green, 0; blue, 0 }  ][line width=0.75]    (10.93,-3.29) .. controls (6.95,-1.4) and (3.31,-0.3) .. (0,0) .. controls (3.31,0.3) and (6.95,1.4) .. (10.93,3.29)   ;
\draw    (221.17,235) -- (205.87,194.37) ;
\draw [shift={(205.17,192.5)}, rotate = 69.37] [color={rgb, 255:red, 0; green, 0; blue, 0 }  ][line width=0.75]    (10.93,-3.29) .. controls (6.95,-1.4) and (3.31,-0.3) .. (0,0) .. controls (3.31,0.3) and (6.95,1.4) .. (10.93,3.29)   ;
\draw    (221.17,235) -- (235.5,194.39) ;
\draw [shift={(236.17,192.5)}, rotate = 109.44] [color={rgb, 255:red, 0; green, 0; blue, 0 }  ][line width=0.75]    (10.93,-3.29) .. controls (6.95,-1.4) and (3.31,-0.3) .. (0,0) .. controls (3.31,0.3) and (6.95,1.4) .. (10.93,3.29)   ;
\draw    (221.17,235) -- (264.71,193.87) ;
\draw [shift={(266.17,192.5)}, rotate = 136.64] [color={rgb, 255:red, 0; green, 0; blue, 0 }  ][line width=0.75]    (10.93,-3.29) .. controls (6.95,-1.4) and (3.31,-0.3) .. (0,0) .. controls (3.31,0.3) and (6.95,1.4) .. (10.93,3.29)   ;

\draw (45.17,175) node [anchor=north west][inner sep=0.75pt]   [align=left] {1};
\draw (75.17,175) node [anchor=north west][inner sep=0.75pt]   [align=left] {2};
\draw (106.17,175) node [anchor=north west][inner sep=0.75pt]   [align=left] {3};
\draw (136.17,175) node [anchor=north west][inner sep=0.75pt]   [align=left] {4};
\draw (170.17,175) node [anchor=north west][inner sep=0.75pt]   [align=left] {5};
\draw (200.17,175) node [anchor=north west][inner sep=0.75pt]   [align=left] {6};
\draw (231.17,175) node [anchor=north west][inner sep=0.75pt]   [align=left] {7};
\draw (261.17,175) node [anchor=north west][inner sep=0.75pt]   [align=left] {8};
\draw (291.17,175) node [anchor=north west][inner sep=0.75pt]   [align=left] {9};
\draw (317.17,175) node [anchor=north west][inner sep=0.75pt]   [align=left] {10};
\draw (348.67,175) node [anchor=north west][inner sep=0.75pt]   [align=left] {11};
\draw (379.17,175) node [anchor=north west][inner sep=0.75pt]   [align=left] {12};
\draw (409.17,175) node [anchor=north west][inner sep=0.75pt]   [align=left] {13};
\draw (440.17,175) node [anchor=north west][inner sep=0.75pt]   [align=left] {14};
\draw (471.17,175) node [anchor=north west][inner sep=0.75pt]   [align=left] {15};
\draw (503.17,175) node [anchor=north west][inner sep=0.75pt]   [align=left] {16};
\draw (275.67,59) node [anchor=north west][inner sep=0.75pt]   [align=left] {$\displaystyle F_{1}$};
\draw (455.17,286) node [anchor=north west][inner sep=0.75pt]   [align=left] {$\displaystyle F_{4}$};
\draw (86.67,242) node [anchor=north west][inner sep=0.75pt]   [align=left] {$\displaystyle F_{5}$};
\draw (329.17,286) node [anchor=north west][inner sep=0.75pt]   [align=left] {$\displaystyle F_{3}$};
\draw (149,286) node [anchor=north west][inner sep=0.75pt]   [align=left] {$\displaystyle F_{2}$};
\draw (212.17,242) node [anchor=north west][inner sep=0.75pt]   [align=left] {$\displaystyle F_{6}$};

\end{tikzpicture}
        \caption{The path diagram corresponding to the hierarchical factor model in Panel (a).}
        \label{fig: example path}
    \end{subfigure}
    
    \caption{The illustrative example of a three-layer hierarchical factor model.}
    \label{fig:hier example}
\end{figure}

An illustrative example of a three-layer hierarchical factor model is given in Figure~\ref{fig:hier example}, 
where Panel (a) shows the variables that load on each factor from the top layer to the bottom layer, and Panel (b) shows the corresponding path diagram. In this example, $J = 16$, $K=6$, $v_1 = \{1,2,\ldots,16\}$, $v_2=\{1,\ldots,8\}$, $v_3 = \{9,\ldots,12\}$, $v_4 = \{13,\ldots,16\}$, $v_5 = \{1,\ldots,4\}$ and $v_6 = \{5,\ldots,8\}$. The factors are labeled following the constraints C3(a) and C3(b). Based on this hierarchical structure, we have $T=3$, $L_1 = \{1\}$, $L_2 = \{2,3,4\}$ and $L_{3} = \{5,6\}$.
The loading matrix $\Lambda$ under the hierarchical structure takes the form 
\begin{equation}\label{eq:hier example}
\tiny
\fontsize{9}{9}
\Lambda = \left(
    \begin{array}{cccccccccccccccc}
        \lambda_{11} & \lambda_{21} & \lambda_{31} & \lambda_{41} & \lambda_{51} & \lambda_{61} & \lambda_{71} & \lambda_{81} &\lambda_{91} & \lambda_{10,1} & \lambda_{11,1} & \lambda_{12,1} & \lambda_{13,1} & \lambda_{14,1} & \lambda_{15,1}&\lambda_{16,1}\\
        \lambda_{12} & \lambda_{22} & \lambda_{32} & \lambda_{42} & \lambda_{52}&\lambda_{62} & \lambda_{72} & \lambda_{82} &0&0 & 0 & 0 & 0 & 0 & 0 & 0 \\
        0 & 0 & 0 & 0 & 0 & 0 & 0 & 0 & \lambda_{93}&\lambda_{10,3} & \lambda_{11,3} &\lambda_{12,3} & 0 & 0 & 0 & 0\\
        0 & 0 & 0 & 0 & 0 & 0 & 0 & 0 & 0 & 0 & 0 & 0 & \lambda_{13,4} & \lambda_{14,4} & \lambda_{15,4}& \lambda_{16,4} \\    
        \lambda_{15} & \lambda_{25} & \lambda_{35} & \lambda_{45} & 0& 0 & 0 & 0 & 0 & 0 & 0 & 0 & 0 & 0 & 0 & 0  \\
        0 & 0 & 0 & 0 & \lambda_{56}& \lambda_{66} & \lambda_{76}&\lambda_{86}&0& 0 & 0 & 0 & 0 & 0 & 0 & 0 \\
    \end{array}
\right)^\top. 
\end{equation}

Under a confirmatory setting, the number of factors $K$ and the variables associated with each factors, $v_1, v_2, \ldots, v_K$, are known. In that case, estimating the hierarchical factor model is a relatively simple problem, which involves solving an optimization problem with suitable zero constraints on the loading parameters. However, in many real-world applications, we do not have prior knowledge about the hierarchical structure of the loading matrix. In these cases, we are interested in exploratory hierarchical factor analysis, i.e., simultaneously learning the hierarchical structure from data and estimating the corresponding parameters.

Before presenting a method for exploratory hierarchical factor analysis, we first show that the true factor hierarchy is unique under mild conditions, which is essential for the true structure to be learnable. Suppose that we are given a true covariance matrix $\Sigma^* = \Lambda^* (\Lambda^*)^\top + \Psi^* $, where the true loading matrix $\Lambda^* $ satisfies the constraints of a hierarchical factor model. Theorem~\ref{thm:identifiability} below shows that the true loading matrix $\Lambda^*$ is unique up to column sign-flips and thus yields the same hierarchical structure. 

The following notation is needed in the rest of the paper. Given a hierarchical factor structure with loading sets 
$v_{i}$, let $D_i = \{j: v_{j}\subsetneq v_i\}$ be the set of all descendent factors of factor $i$. For example, in the hierarchical structure shown in Figure~\ref{fig:hier example}, $D_{2} = \{5,6\}$. 
 For any matrix $A=(a_{i,j})_{m\times n}$ and sets $\mathcal{S}_1\subset\{1,\ldots,m\}$ and $\mathcal{S}_{2}\subset \{1,\ldots,n\}$, let $A_{[\mathcal{S}_1,\mathcal{S}_2]} = (a_{i,j})_{i\in\mathcal{S}_1, j\in\mathcal{S}_2}$ be the submatrix of $A$ consisting of elements that lie in rows belonging to set $\mathcal{S}_1$ and columns belonging to set $\mathcal{S}_2$, where the rows and columns are arranged in ascending order based on their labels in $\mathcal{S}_1$ and $\mathcal{S}_2$, respectively. 
 For example, consider the loading matrix in~\eqref{eq:hier example}, 
 where $v_2 = \{1, 2, \ldots, 8\}$. Then, $\Lambda_{[v_{2},\{1,2\}]}$ takes the form
 $$\Lambda_{[v_{2},\{1,2\}]} = \left(\begin{array}{cccccccc}
    \lambda_{11} & \lambda_{21} & \lambda_{31} & \lambda_{41} & \lambda_{51} & \lambda_{61}& \lambda_{71}& \lambda_{81}  \\
   \lambda_{12} & \lambda_{22} & \lambda_{32} & \lambda_{42} & \lambda_{52} & \lambda_{62} & \lambda_{72}& \lambda_{82} 
\end{array}\right)^{\top}.$$
For any vector $\boldsymbol{a} = (a_1,\ldots,a_n)^{\top}$ and set $\mathcal S\subset\{1,\ldots,n\}$, we similarly define $\boldsymbol{a}_{[\mathcal S]} = (a_{i})^{\top}_{i\in \mathcal S}$  be the subvector of $\boldsymbol{a}$ consisting of the elements belonging to $\mathcal S$,  where the elements in $\boldsymbol{a}_{[\mathcal S]}$ are arranged in ascending order based on their labels in $\mathcal S$. For any set $\mathcal{S}_1 \subset \{1, 2, \ldots, n\}$, let $\mbox{vec}(\mathcal{S}_1)$  
be a mapping that maps the set $\mathcal{S}_1$ 
to a vector whose elements are the same as $\mathcal{S}_1$ and arranged in ascending order. 
For two sets $\mathcal{S}_1 \subset \{1, 2, \ldots, n\}$ and $\mathcal{S}_2 \subset \{1, 2, \ldots, |\mathcal{S}_1|\}$, we denote $\mathcal{S}_1[\mathcal{S}_2]$ as the subset of $\mathcal{S}_1$, consisting of elements in 
$\mbox{vec}(\mathcal{S}_1)[\mathcal{S}_2]$.

\begin{condition}\label{cond:true para}
The population covariance matrix can be expressed as the form $\Sigma^* = \Lambda^* (\Lambda^*)^\top + \Psi^*$, where the true loading matrix $\Lambda^* $ is of rank $K$ and the loading sets $v^{*}_{k}$ and child factors $\Ch_k^*$ defined by $\Lambda^* $ 
satisfy the constraints C1--C3 of a hierarchical factor model. 
\end{condition}

\begin{condition}\label{cond:seperate}
Given another $J\times K$ matrix $\Lambda$ and $J\times J$ diagonal matrix $\Psi$ such that $\Sigma^* = \Lambda^* (\Lambda^*)^\top + \Psi^*  = \Lambda\Lambda^{\top} + \Psi$, we have $\Lambda\Lambda^{\top} = \Lambda^* (\Lambda^*)^\top$ and $\Psi = \Psi^*$.
\end{condition}

\begin{condition}\label{cond:rank}
Let $D_k^*$ be the corresponding true set of descendant factors of factor $k$. 
For any factor $i$ with $\Ch^{*}_{i}\neq \emptyset$ and any $j\in \text{Ch}^{*}_{i}$, it satisfies that (1) any two rows of $\Lambda^{*}_{[v^{*}_j,\{i,j\}]}$ are linearly independent,  (2) for any $k\in v^{*}_{j}$,  $\Lambda^{*}_{[v^{*}_{j}\setminus\{k\},\{i,j\}\cup D^{*}_{j}]}$ has full column rank, and (3) if $|\Ch_j^*| \geq 2$, then, for any $s_1, s_2 \in \text{Ch}^{*}_{j}$, $k_1, k_2 \in v^{*}_{s_1}$, and $k_3, k_4 \in v^{*}_{s_2}$, $\Lambda^{*}_{[\{k_1, \ldots, k_4\}, \{i,j,s_1,s_2\}]}$ is of full rank.
\end{condition}

\begin{theorem}\label{thm:identifiability}
Suppose that Conditions~\ref{cond:true para}--\ref{cond:rank} hold. If there exists some hierarchical factor structure with $K$ factors such that its loading matrix $\Lambda$ and unique variance matrix $\Psi$ satisfy $\Sigma^* = \Lambda\Lambda^{\top} + \Psi$,  there exists some sign flip matrix $Q\in \mathcal{Q}$ such that $\Lambda = \Lambda^{*}Q$, where $\mathcal{Q}$ consists of all $K\times K$ diagonal matrices $Q$ whose diagonal entries take values $1$ or $-1$. 
\end{theorem}

\begin{remark}
As far as we know, Theorem~\ref{thm:identifiability} is the first identifiability result for exploratory hierarchical factor analysis. This theorem establishes mild regularity conditions under which the true hierarchical factor model is identifiable when we do not know the true hierarchical factor structure. 
Condition 1 assumes that the true model is a hierarchical factor model. Under this model assumption, the identifiability result of Theorem~\ref{thm:identifiability} has two parts. The first part
involves identifying the column space of the loading matrix based on the population covariance
matrix, and the second part entails identifying the factors based on this column space. 
The identifiability result in the first part, which is assumed in Condition 2, has already been well studied in the literature.  For example, Condition~\ref{cond:suff 2} below is a result in Theorem 5.1, \cite{anderson1956statistical}, which 
gives a sufficient condition for Condition~\ref{cond:seperate} to hold. 
On the other hand, the second part is more challenging and relies more on the hierarchical
factor structure.   Theorem~\ref{thm:identifiability} focuses on proving the second part. 

\end{remark}

\begin{condition}\label{cond:suff 2}
For each $j\in\{1,\ldots,J\}$, there exist two disjoint set $E_1, E_2 \subset \{1,\ldots,J\}\setminus\{j\}$ with $|E_{1}| = |E_{2}| = K$ such that $\Lambda^{*}_{[E_1, :]}$ and $\Lambda^{*}_{[E_2, :]}$ are of full rank, where  $\Lambda^{*}_{[E_1, :]}$ and $\Lambda^{*}_{[E_2, :]}$ are the submatrices of $\Lambda^{*}$ consisting of the rows belonging to $E_1$ and $E_2$.
\end{condition}

\begin{remark}
Condition 2 implicitly imposes some minimum requirements on the parameter space for identifiable hierarchical factor models. In fact, Proposition~\ref{prop:seperate necessary} below implies a necessary condition for Condition 2. This necessary condition leads to the following constraint: 
\begin{enumerate}
    \item[C4.]  For all $k = 1, \ldots, K$, $|v_{k}|\geq 3$, and $|v_{k}|\geq 7$ if factor $k$ has two or more child factors.
\end{enumerate}
\end{remark}

\begin{proposition}\label{prop:seperate necessary}
There exists another $J\times K$ matrix $\Lambda$ following the same hierarchical factor structure as the true model and a $J\times J$ diagonal matrix $\Psi$ such that $\Sigma^* = \Lambda^* (\Lambda^*)^\top + \Psi^*  = \Lambda\Lambda^{\top} + \Psi$, if there exists a factor $k$ such that (1) $|v^{*}_k| \leq 2$ or (2)  $|\mbox{\Ch}^{*}_k | \geq 2$ and $|v^{*}_k| \leq 6$. 
\end{proposition}
{Proposition \ref{prop:seperate necessary} follows directly from Theorem 1 in \cite{fang2021identifiability}.}

\begin{remark}

Condition \ref{cond:rank} imposes three requirements. First, it requires that there do not exist two variables loading on factor $j$
such that their loadings on any factor $i$ and its child node $j$ are linearly dependent. This is a mild assumption satisfied by almost all the models in the full parameter space of hierarchical factor models. Second, it requires that the submatrix $\Lambda^{*}_{[v_{j}^*,\{i,j\}\cup D^{*}_{j}]}$, which corresponds to variables in $v_{j}^*$ and factors $i$, $j$,  $j$'s descendants, are still of full column rank after deleting any row. This condition mainly imposes a restriction on the number of descendant factors each factor can have. That is,  the full-column-rank requirement implies that  $|v_{j}^*| \geq 3 + |D^{*}_j|$. As shown via Proposition \ref{prop:structure C4} below, 
this requirement automatically holds for all the identifiable hierarchical factor models that satisfy constraints C1--C4. Other than that, the full-column-rank requirement is easily satisfied by most hierarchical factor models. 
These two requirements can be seen as an extension of Condition 2 of \cite{Qiao_Chen_Ying_2025} to hierarchical factor models, where \cite{Qiao_Chen_Ying_2025} consider a bi-factor model with possibly correlated bi-factors. Third, we require that when factor $j$ has child factors $s_1$ and $s_2$, for any two variables $k_1$, $k_2$ loading on factor $s_1$ and any variables $k_3$, $k_4$ loading on factor $s_2$, the sub-loading matrix corresponding to variables $k_1,\ldots,k_4$ and factors $i,j,s_1,s_2$ is of full rank. Although the requirements in Condition 3 are quite mild, we acknowledge that they may be further weakened. For example, instead of requiring any two roles of $\Lambda^{*}_{[v_{j}^{*},\{i,j\}]}$ to be linearly independent, we may only need to require a sufficient number pair of the rows of $\Lambda^{*}_{[v_{j}^{*},\{i,j\}]}$ to be linearly independent; see Appendix~\ref{append: discussion on condition 3} for further discussions. We leave the refinement of the condition for future investigation.
\end{remark}

\begin{proposition}\label{prop:structure C4}
Suppose that the hierarchical factor structure satisfies constraints C1--C4. Then $|v_{j}^*| \geq 3 + |D^{*}_j|$ holds for each factor j.
\end{proposition}

\subsection{An Overview of Proposed Method}\label{subsec:overview method}
As the proposed method is quite sophisticated, we start with an overview of the proposed method to help readers understand it. 
Consider a dataset with $N$ observation units from a certain population and $J$ observed variables.  Let $S$ be the sample covariance matrix of observed data. The proposed method takes $S$ as the input and outputs estimators: 

\begin{enumerate}
    \item $\hat T$ and $\hat K$ for the number of layers $T$ and the number of factors $K$. 
    
    \item $\hat{L}_1, \ldots,\hat{L}_{\hat{T}}$ for the factor layers $L_1, \ldots, L_T$ and 
    $\hat v_1, \hat v_2, \ldots, \hat v_{\hat K}$ for the sets of variables loading on the $K$ factors, $v_1, \ldots, v_K$. 
    \item $\hat \Lambda$  and $\hat \Psi$ for the loading matrix $\Lambda$ and 
    unique variance matrix $\Psi$. 
     
\end{enumerate}
As shown in Theorem~\ref{cor:consistency} below, with the sample size $N$ going to infinity, these estimates will converge to their true values. 

The proposed method learns the hierarchical factor structure from the top to the bottom of the factor hierarchy. It divides the learning problem into many subproblems and conquers them layer by layer, starting from the first layer  
$\hat L_1 = \{1\}$ with $\hat{v}_1 = \{1, \ldots, J\}$. 
For each step $t$, $t = 2,3, \ldots, $ suppose the first to the $(t-1)$th layers have been learned. These layers are 
denoted by $\hat L_i = \{k_{i-1}+1, \ldots, k_i\}$, $i = 1, \ldots, t-1$, where $k_0 = 0$ and $k_1 = 1$, and the associated sets of variables are denoted by $\hat v_1, \ldots, \hat v_{k_{t-1}}$. 
We make the following decisions in the $t$th step: 

\begin{enumerate}
    \item For each factor $k \in \hat L_{t-1}$, learn 
    its child factors under the constraints C3 and C4. This is achieved by an Information-Criterion-Based (ICB) method described in Section~\ref{subsubsec:ICB} below. The labels of the child factors are denoted by $\hat \Ch_k$. When $\hat \Ch_k \neq \emptyset$, we denote the associated sets of variables as $\hat v_l, l \in \hat \Ch_k$.

     \item If $\hat\Ch_k = \emptyset$ for all $k \in  \hat L_{t-1}$, stop the learning algorithm and conclude that the factor hierarchy has $\hat T = t-1$ layers.

     \item Otherwise, let $\hat L_t = \{{k_{t-1}}+1, \ldots, k_t\} = \cup_{k \in \hat L_{t-1}} \hat\Ch_k$ and proceed to the $(t+1)$th step. 
     
\end{enumerate}

We iteratively learn the structure of each layer until the preceding stopping criterion is met. Then we obtain the estimates $\hat \Lambda$ and $\hat \Psi$ by maximum likelihood estimation given $\hat K = k_{\hat T}$, $\hat v_1, \ldots, \hat v_{\hat K}$: 
\begin{equation}\label{eq:refit}
\begin{aligned}
(\hat \Lambda, \hat{\Psi}) =& \argmin_{\Lambda,\Psi}l(\Lambda\Lambda^{\top} + \Psi ;S),\\
\mbox{s.t.~} & \lambda_{ij}=0,i\notin \hat{v}_j,i=1,\ldots,J, j=1,\ldots,\hat{K},\\
& \Psi_{[\{i\},\{i\}]}\geq0, \Psi_{[\{i\},\{j\}]}=0, i=1,\ldots,J, j\neq i,
\end{aligned}
\end{equation}
where
$
l(\Lambda\Lambda^{\top} + \Psi ;S) = N\big(\log(\text{det}(\Lambda\Lambda^\top + \Psi)) + \textnormal{tr}(S (\Lambda\Lambda^\top + \Psi)^{-1}) -\log(\text{det}(S)) - J\big)
$
equals twice the negative log-likelihood of the observed data up to a constant. 
We output $\hat T$, $\hat K$, $\hat L_1, \ldots, \hat L_{\hat T}$, $\hat v_1, \ldots, \hat v_{\hat K}$, $\hat \Lambda$ and $\hat \Psi$ as our final estimate of the hierarchical factor model.

To illustrate, consider the example in Figure~\ref{fig:hier example}. In the first step, we start with $\hat{L}_{1} = \{1\}$ and $\hat{v}_1 = \{1,\ldots,16\}$. In the second step, we learn the child factors of Factor 1. If they are correctly learned, then we obtain $\hat{\text{Ch}}_1 = \{2,3,4\}$ with $\hat{v}_2 = \{1,\ldots,8\}$, $\hat{v}_3 = \{9,\ldots,12\}$ and $\hat{v}_4 = \{13,\ldots,16\}$. This leads to $\hat{L}_{2} = \{2,3,4\}$. In the third step, we learn the child factors of Factors 2, 3 and 4, one by one. 
If correctly learned, we have $\hat{\text{Ch}}_2 = \{5,6\}$, $\hat{\text{Ch}}_3 = \emptyset$, $\hat{\text{Ch}}_4 = \emptyset$, $\hat{L}_{3} = \{5,6\}$, $\hat{v}_5 = \{1,\ldots,4\}$ and 
 $\hat{v}_6 = \{5,\ldots,8\}$. In the fourth step, if correctly learned, we have $\hat{\text{Ch}}_5= \hat{\text{Ch}}_6 = \emptyset$, and the learning algorithm stops. We then have $\hat{T} = 3$, $\hat{K} = 6$, $\hat{L}_1,\ldots,\hat{L}_{3}$, $\hat{v}_1,\ldots,\hat{v}_6$ and further obtain $\hat{\Lambda}$ and $\hat{\Psi}$ using \eqref{eq:refit} given $\hat K$ and $\hat{v}_{1},\ldots,\hat{v}_{6}$.

We summarise the steps of the proposed method in Algorithm \ref{algo:divide and conquer} below. 

\begin{algorithm}
\caption{A Divide-and-Conquer method for learning the hierarchical factor structure}
\label{algo:divide and conquer}
\begin{algorithmic}[1]
\Require Sample covariance matrix $S\in\mathbb{R}^{J\times J}$.
\State Set $\hat{L}_1 = \{1\}$ with $\hat{v}_{1} = \{1,\ldots,J\}$.
\State Determine $\hat{\mbox{Ch}}_1$, the child factors of Factor 1, and $\hat{v}_i$ for all $i\in\hat{\mbox{Ch}}_1$, the sets of variables loading on these child factors, by the ICB method {in Algorithm \ref{algo:ICB}}.
\State Set $\hat{L}_2 = \hat{\mbox{Ch}}_1$ and $t=2$,
\While{$\hat{L}_t \neq \emptyset$}
\For{$k\in\hat{L}_t$}
\State Determine $\hat{\mbox{Ch}}_k$ and $\hat{v}_i$ for all $i\in\hat{\mbox{Ch}}_k$ by the ICB method {in Algorithm \ref{algo:ICB}}.
\EndFor
\State Set $\hat{L}_{t+1} = \cup_{k\in\hat{L}_t}\hat{\mbox{Ch}}_k$.
\State $t=t+1$.
\EndWhile
\State Set $\hat{T} = t-1$, $\hat{K} = \Sigma_{l=1}^{\hat{T}}|\hat{L}_{l}|$.
\State Obatin $\hat{\Lambda}$ and $\hat{\Psi}$ using \eqref{eq:refit} given $\hat{K}$ and $\hat{v}_1,\ldots,\hat{v}_{\hat{K}}$.
\Ensure $\hat{T}$, $\hat{K}$, $\hat{L}_{1},\ldots,\hat{L}_{\hat{T}}$, $\hat{v}_{1},\ldots,\hat{v}_{\hat{K}}$, $\hat{\Lambda}$ and $\hat{\Psi}$.
\end{algorithmic}
\end{algorithm}

\subsection{ICB Method for Learning Child Factors}\label{subsubsec:ICB}
From the overview of the proposed method described above, we see that the proposed method solves the learning problem by iteratively applying an ICB method to learn the child factors of each given factor. We now give the details of this method. We start with the ICB method for learning the child factors of Factor~1, i.e., the general factor. In this case, the main questions the ICB method answers are: (1) how many child factors does Factor 1 have?  and (2) what variables load on each child factor? It is worth noting that when learning these from data, we need to account for the fact that each child factor can have an unknown number of descendant factors. However, with a divide-and-conquer spirit, we do not learn the structure of the descendant factors (i.e., the hierarchical structure of these descendant factors and the variables loading on them) of each child factor in this step because this structure is too complex to learn at once. 
 
The ICB method answers the two questions above by learning a loading matrix $\tilde \Lambda_1$ with zero patterns that encode the 
number and loading structure of the child factors of Factor 1. More specifically, $\tilde \Lambda_1$ is searched among the space of loading matrices that satisfy certain zero constraints that encode a hierarchical factor model. This space is defined as 
$$\mathcal A_1 = \cup_{c \in \{0, 2, \ldots, c_{\max}\}, d_1, \ldots, d_c \in \{1, \ldots, d_{\max}\}}\mathcal A^1(c, d_1, \ldots, d_c),$$
where, if $c \geq 2$, for a pre-specified constant $\tau > 0$, 
\begin{equation}
\small
\begin{aligned}
\mathcal A^1(c, d_1, \ldots, d_c) = &\{A=(a_{ij})_{{J}\times (1+d_1 + \cdots +d_c)}: \mbox{ there exists a partition of } \{1, \ldots, J\}, \mbox{~denoted}\\
&~~\mbox{by~} v_{1}^{1},\ldots,v_{c}^{1}, \mbox{satisfying } 
\min\{v_1^1\} < \min\{v_2^1\} < \cdots < \min\{v_c^1\},\mbox{~such that } \\
&~~A_{[v_{s}^1, \{j\}]} = \mathbf 0, \mbox{ for all }s = 1, \ldots, c, \mbox{ and } j \notin \{1,  2+\sum_{s'<s}d_{s'}, 3+\sum_{s'<s}d_{s'}, \ldots,\\
&~~1+\sum_{s'\leq s}d_{s'} \} \mbox{~and }  |a_{ij}| \leq \tau, \mbox{~~for all~} i=1,\ldots,J \mbox{~and~} j =1,\ldots,1+\sum_{s=1}^{c}d_c.\},
 \end{aligned}
 \end{equation}
and, if $c = 0$,
$\mathcal A^1(0) = \{A = (a_{ij})_{J \times 1}: |a_{ij}|\leq \tau\}.$ Here, $c_{\max}$ and $d_{\max}$ are pre-specified constants typically decided by domain knowledge. $\tau$ is a universal upper bound for the loading parameters, which is needed for technical reasons for our theory. The space $\mathcal A^1(c, d_1, \ldots, d_c)$ includes all possible loading matrices for a hierarchical factor structure, where Factor 1 has $c$ child factors, and each child factor has $d_s -1$ descendant factors. 
The space  $\mathcal A_1$  is the union of all the possible $\mathcal A^1(c, d_1, \ldots, d_c)$ for different combinations of the numbers of child factors and their descendant factors.

For example, consider the hierarchical factor model example in Figure~\ref{fig:hier example}, for which $\hat{v}_1 = \{1,\ldots,16\}$. Then, the matrix
\begin{equation}\label{eq:block example}
\fontsize{8}{8}
\begin{aligned}
\Lambda_1 =\left(
    \begin{array}{cccccccccccccccc}
        \lambda_{11} & \lambda_{21} & \lambda_{31} & \lambda_{41} & \lambda_{51} & \lambda_{61} & \lambda_{71} & \lambda_{81} & \lambda_{91} & \lambda_{10,1} & \lambda_{11,1} & \lambda_{12,1} & \lambda_{13,1} & \lambda_{14,1} & \lambda_{15,1}& \lambda_{16,1} \\
        \lambda_{12} & \lambda_{22} & \lambda_{32} & \lambda_{42} & \lambda_{52} & \lambda_{62} & \lambda_{72} & \lambda_{82} &0&0 & 0 & 0 & 0 & 0 & 0 & 0 \\
         \lambda_{13} & \lambda_{23} & \lambda_{33} & \lambda_{43} & \lambda_{53} & \lambda_{63} & \lambda_{73} & \lambda_{83} & 0 & 0 & 0 & 0 & 0 & 0 & 0 & 0  \\
        \lambda_{14} & \lambda_{24} & \lambda_{34} & \lambda_{44} & \lambda_{54}& \lambda_{64} & \lambda_{74}&\lambda_{84}&0& 0 & 0 & 0 & 0 & 0 & 0  & 0\\
        0 & 0 & 0 & 0 & 0 & 0 & 0 & 0 & \lambda_{95}& \lambda_{10,5} & \lambda_{11,5} & \lambda_{12,5} & 0 & 0 & 0  & 0\\
        0 & 0 & 0 & 0 & 0 & 0 & 0 & 0 & 0&0 & 0 & 0 & \lambda_{13,6} & \lambda_{14,6} & \lambda_{15,6}& \lambda_{16,6} \\
    \end{array}
\right)^\top
\end{aligned}
\end{equation}
lies in space $\mathcal{A}^{1}(3,3,1,1)$. 
This loading matrix is what the ICB method aims to find,
as it has the same blockwise zero pattern (ignoring the zero constraints implied by the lower-layer factors) as the true loading pattern in \eqref{eq:hier example} after reordering the columns of $\Lambda$ in \eqref{eq:hier example}.

We search for the best possible loading matrix in $\mathcal A_1$ using the information criterion defined as: 
\begin{equation}\label{eq:optIC}
\begin{aligned}
\mbox{IC}_1(c, d_1, \ldots, d_c) = \min_{\Lambda_1,\Psi_1}&l\left( \Lambda_1\Lambda_1^\top + \Psi_1, S\right) + {p_1(\Lambda_1)} \log N,\\
\mbox{s.t.~} & \Lambda_1\in\mathcal{A}^{1}(c,d_1,\ldots,d_c), \kappa_1\leq(\Psi_1)_{[\{i\},\{i\}]}\leq \kappa_2,\\
&(\Psi_1)_{[\{i\},\{j\}]}=0,i=1,\ldots,|\hat{v}_1|,j\neq i,
\end{aligned}
\end{equation}
where {$\kappa_1$ and $\kappa_2$ are pre-specified lower and upper bounds for the unique variance, and}
\begin{equation}
\begin{aligned}
p_1(\Lambda_1) = \left\{
    \begin{array}{l}
    \Sigma_{s=1}^{c}(|v_{s}^{1}|d_s - d_s(d_s-1)/2) \mbox{~if~} d_s \leq |v_{s}^{1}| ~\mbox{~for all~} s = 1,\ldots c,\\
    \infty, \mbox{~otherwise},
    \end{array}
\right.
\end{aligned}
\end{equation}
is a penalty on the number of free parameters
for a matrix $\Lambda_1$ in $\mathcal{A}^{1}(c,d_1,\ldots,d_c)$. The penalty ensures that in the selected factor loadings, one plus the number of descendant factors of each child factor of Factor 1 will not exceed the number of items loading on the corresponding child factor.

Ideally, we hope to find the loading matrix in $\mathcal A_1$ that minimises $\mbox{IC}_1(c, d_1, \ldots, d_c)$ among all  $c \in \{0, 2, \ldots, c_{\max}\}$ and $d_1, \ldots, d_c \in \{1, \ldots, d_{\max}\}$. 
More specifically, we define 
\begin{equation}\label{eq:naive IC opt}
(\bar{c}_1, \bar{d}_1^1, \ldots, \bar{d}_{\bar{c}_1}^1) = \argmin_{c\in\{0,2,\ldots,c_{\max}\}, 1\leq d_{s} \leq d_{\max}, s = 1, \ldots, c} \mbox{IC}_1(c, d_1, \ldots, d_c)
\end{equation}
and further
\begin{equation}\label{eq:naive est}
\begin{aligned}
(\bar \Lambda_1, \bar \Psi_1) = \argmin_{\Lambda_1, \Psi_1} & ~l\left( \Lambda_1\Lambda_1^\top + \Psi_1, S\right) \\
\mbox{s.t.~}& \Lambda_1\in\mathcal{A}^{1}(\bar{c}_1,\bar{d}_{1}^{1},\ldots,\bar{d}_{\bar{c}_{1}}^{1}), \kappa_1\leq(\Psi_1)_{[\{i\},\{i\}]}\leq \kappa_2\\
&(\Psi_1)_{[\{i\},\{j\}]}=0,i=1,\ldots,|\hat{v}_1|,j\neq i.
\end{aligned}
\end{equation}
We determine the variables loading on each child factor of Factor 1 based on the zero pattern of $\bar \Lambda_1$.

However, we note that  $\mathcal A_1$ is highly complex, and thus,  enumerating all the possible loading matrices in $\mathcal A_1$ is computationally infeasible. In other words, while the quantities in \eqref{eq:naive IC opt} and \eqref{eq:naive est} are well-defined mathematically, they cannot be computed within a reasonable time. In this regard, we develop a greedy search method, presented in Algorithm~\ref{algo:ICB}, for searching over the space $\mathcal A_1$.  
This greedy search method will output $\hat c_1$ and $\hat v_{1}^{1},\ldots,\hat v_{\hat c_1}^{1}$. 
As shown in Theorem~\ref{cor:consistency}, with probability tending to 1, they are consistent estimates of the corresponding true quantities for the factors in this layer. In other words, this greedy search is theoretically guaranteed to learn the correct hierarchical factor structure.
Moreover, Algorithm \ref{algo:ICB} also solves a similar optimization as \eqref{eq:naive est}
for loading matrices in $\mathcal{A}^{1}(\hat{c}_1,\hat{d}_1,\ldots,\hat{d}_{\hat c_1})$, from which we obtain a consistent estimate of the first column of the loading matrix, denoted by $\tilde{\boldsymbol{\lambda}}_1$. So far, we have learned the factors in the second layer of the factor hierarchy. 

For $t \geq 3$, suppose that the first to the $(t-1)$th layers have been successfully learned, and we now need to learn the factors in the $t$th layer. This problem can be decomposed into learning the child factors of each factor $k$ in $\hat{L}_{t-1} = \{{k_{t-2}}+1, \ldots, k_{t-1}\}$. At this moment, we have the estimated variables loading on Factor $k$, denoted by $\hat v_k$, and a consistent estimate of the loading parameters for the factors in the first to the $(t-2)$th layer, denoted by $\tilde{\boldsymbol\lambda}_i$, $i=1, \ldots, k_{t-2}$, which are 
obtained as a by-product of the ICB method in the previous steps. 
We define $\tilde \Sigma_{k,0} := \sum_{i=1}^{k_{t-2}} (\tilde{\boldsymbol{\lambda}}_i)_{[\hat{v}_{k}]} (\tilde{\boldsymbol{\lambda}}_i)_{[\hat{v}_{k}]}^\top$ and  $S_{k} := S_{[\hat{v}_{k},\hat{v}_{k}]}$. 
Similar to the learning of child factors of Factor 1, we define the possible space for the loading submatrix associated with the descendant factors of Factor $k$ as
$$\mathcal A_k = \cup_{c \in \{0, 2, \ldots, c_{\max}\}, d_1, \ldots, d_c \in \{1, \ldots, d_{\max}\}}\mathcal A^k(c, d_1, \ldots, d_c),$$
where, if $c \geq 2$, for the same constant $\tau > 0$ as in $\mathcal A_1$
\begin{equation}
\begin{aligned}
\mathcal A^k(c, d_1, \ldots, d_c)= &\{A=(a_{ij})_{{|\hat{v}_k|}\times (1+d_1 + \cdots +d_c)}: \mbox{ there exists a partition of } \{1, \ldots, \vert\hat v_k\vert\}, \mbox{denoted}\\ 
&~\mbox{~by }  
 v_{1}^{k},\ldots,v_{c}^{k}, \mbox{~satisfying~} 
\min\{v_1^k\} < \min\{v_2^k\} < \cdots < \min\{v_c^k\}, \mbox{ such that } \\
&~~A_{[v_{s}^k, \{j\}]} = \mathbf 0, \mbox{ for all~}s = 1, \ldots, c, \mbox{ and } j \notin \{1,  2+\sum_{s'<s}d_{s'}, 3+\sum_{s'<s}d_{s'}, \ldots,\\
&~~1+\sum_{s'\leq s}d_{s'} \} \mbox{~and~}
~|a_{ij}|\leq\tau \mbox{~for all~} i=1,\ldots,\vert\hat v_k\vert \mbox{~and~} j =1,\ldots,1+\sum_{s=1}^{c}d_s\},
 \end{aligned}
 \end{equation}
and, if $c = 0$,
$\mathcal A^k(0) = \{A = (a_{ij})_{|\hat{v}_k| \times 1}: |a_{ij}| \leq \tau\}.$ Here, $c$ and $d_1, \ldots, d_c$ have similar meanings as in $\mathcal A^1(c, d_1, \ldots, d_c)$. That is, $\mathcal A^k(c, d_1, \ldots, d_c)$ includes the corresponding loading submatrices when Factor $k$ has $c$ child factors, and each child factor has $d_s -1$ descendant factors. It should be noted that, however, each matrix in $\mathcal A^k(c, d_1, \ldots, d_c)$ has only $|\hat v_k|$ rows, while those in 
$\mathcal A^1(c, d_1, \ldots, d_c)$ have $J$ rows. This is because, given the results from the previous steps, we have already estimated that 
factor $k$ and its descendant factors are only loaded by the variables in $\hat v_k$. Therefore, we only focus on learning the rows of the loading matrix that correspond to the variables in $\hat v_k$
in the current task. Similar to $\mbox{IC}_1(c, d_1, \ldots, d_c)$, we define 
\begin{equation}\label{eq:opt IC kth}
\begin{aligned}
\mbox{IC}_k(c, d_1, \ldots, d_c) = \min_{\Lambda_k,\Psi_k}&l\left( \tilde{\Sigma}_{k,0}+ \Lambda_k\Lambda_k^\top + \Psi_k, S_k\right) + {p_k(\Lambda_k)} \log N,\\
\mbox{s.t.~} & \Lambda_k\in\mathcal{A}^{k}(c,d_1,\ldots,d_c), \kappa_1\leq(\Psi_k)_{[\{i\},\{i\}]}\leq \kappa_2,\\
&(\Psi_k)_{[\{i\},\{j\}]}=0,i=1,\ldots,|\hat{v}_k|,j\neq i,
\end{aligned}
\end{equation}
where 
\begin{equation}
\begin{aligned}
p_k(\Lambda_k) = \left\{
    \begin{array}{l}
    \Sigma_{s=1}^{c}(|v_{s}^{k}|d_s - d_s(d_s-1)/2) \mbox{~if~} d_s \leq |v_{s}^{k}| ~\mbox{~for all~} s = 1,\ldots c,\\
    \infty, \mbox{~otherwise}\\
    \end{array}
\right.
\end{aligned}
\end{equation}
is a penalty term.

Again, we use the greedy search algorithm, Algorithm~\ref{algo:ICB}, to search for the best possible $\Lambda_k$ in $\mathcal A_k$. It outputs $\hat c_k$ and $\hat v_{1}^{k},\ldots,\hat v_{\hat c_k}^{k}$, and an estimate of the $k$th column of the loading matrix, $\tilde{\boldsymbol{\lambda}}_k$. Under some regularity conditions, 
Theorem~\ref{cor:consistency} shows that $\hat c_k$, $\hat v_{1}^{k},\ldots,\hat v_{\hat c_k}^{k}$, and $\tilde{\boldsymbol{\lambda}}_k$ are consistent estimates of the corresponding true quantities. 

\begin{remark}
The penalty term in the proposed information criterion is essential for learning the correct hierarchical factor structure that satisfies the constraints in C1-C4. It avoids asymptotically rank-degenerated solutions for the loading matrix and, thus, avoids selecting an over-specified hierarchical factor model with redundant parameters in the loading matrix and redundant factors, which affects the interpretation of the estimated factors. Consider the example in Figure~\ref{fig:hier example}. Without the penalty in the proposed information criterion, we may select the structure in Figure~\ref{fig:counter}, which is still a correctly specified model but has a redundant factor (corresponding to {$v_2$}) that is not very interpretable.

\end{remark}

\begin{figure}[tb!]
    \centering

\tikzset{every picture/.style={line width=0.75pt}} 

\begin{tikzpicture}[x=0.75pt,y=0.75pt,yscale=-1,xscale=1]

\draw   (290.17,51) -- (311.17,51) -- (311.17,71) -- (290.17,71) -- cycle ;
\draw   (331.17,131) -- (352.17,131) -- (352.17,151) -- (331.17,151) -- cycle ;
\draw   (455,91) -- (476,91) -- (476,111) -- (455,111) -- cycle ;
\draw   (86,131) -- (107,131) -- (107,151) -- (86,151) -- cycle ;
\draw   (210.17,91) -- (231.17,91) -- (231.17,111) -- (210.17,111) -- cycle ;
\draw   (210,131) -- (231,131) -- (231,151) -- (210,151) -- cycle ;
\draw   (39.67,171) -- (60.67,171) -- (60.67,192.5) -- (39.67,192.5) -- cycle ;
\draw   (69.67,171) -- (90.67,171) -- (90.67,192.5) -- (69.67,192.5) -- cycle ;
\draw   (100.67,171) -- (121.67,171) -- (121.67,192.5) -- (100.67,192.5) -- cycle ;
\draw   (130.67,171) -- (151.67,171) -- (151.67,192.5) -- (130.67,192.5) -- cycle ;
\draw   (164.67,171) -- (185.67,171) -- (185.67,192.5) -- (164.67,192.5) -- cycle ;
\draw   (194.67,171) -- (215.67,171) -- (215.67,192.5) -- (194.67,192.5) -- cycle ;
\draw   (225.67,171) -- (246.67,171) -- (246.67,192.5) -- (225.67,192.5) -- cycle ;
\draw   (255.67,171) -- (276.67,171) -- (276.67,192.5) -- (255.67,192.5) -- cycle ;
\draw   (285.67,171) -- (306.67,171) -- (306.67,192.5) -- (285.67,192.5) -- cycle ;
\draw   (315.67,171) -- (336.67,171) -- (336.67,192.5) -- (315.67,192.5) -- cycle ;
\draw   (346.67,171) -- (367.67,171) -- (367.67,192.5) -- (346.67,192.5) -- cycle ;
\draw   (377.67,171) -- (398.67,171) -- (398.67,192.5) -- (377.67,192.5) -- cycle ;
\draw   (407.67,171) -- (428.67,171) -- (428.67,192.5) -- (407.67,192.5) -- cycle ;
\draw   (438.67,171) -- (459.67,171) -- (459.67,192.5) -- (438.67,192.5) -- cycle ;
\draw   (469.67,171) -- (490.67,171) -- (490.67,192.5) -- (469.67,192.5) -- cycle ;
\draw   (501.67,171) -- (522.67,171) -- (522.67,192.5) -- (501.67,192.5) -- cycle ;
\draw    (300.67,71) -- (300.67,81) ;
\draw    (220.67,81) -- (465.5,81) ;
\draw    (465.5,81) -- (465.5,91) ;
\draw    (96.5,120) -- (342.17,120) ;
\draw    (96.5,120) -- (96.5,131) ;
\draw    (220.67,111) -- (220.5,131) ;
\draw    (50.17,161) -- (141.17,161) ;
\draw    (96.5,151) -- (96.5,161) ;
\draw    (50.17,161) -- (50.17,171) ;
\draw    (80.17,161) -- (80.17,171) ;
\draw    (111.17,161) -- (111.17,171) ;
\draw    (141.17,161) -- (141.17,171) ;
\draw    (175.17,161) -- (266.17,161) ;
\draw    (220.5,151) -- (220.67,161) ;
\draw    (175.17,161) -- (175.17,171) ;
\draw    (205.17,161) -- (205.17,171) ;
\draw    (236.17,161) -- (236.17,171) ;
\draw    (266.17,161) -- (266.17,171) ;
\draw    (296.17,161) -- (388.17,161) ;
\draw    (418.17,120) -- (512.17,120) ;
\draw    (465.5,111) -- (465.17,120) ;
\draw    (418.17,120) -- (418.17,171) ;
\draw    (449.17,120) -- (449.17,171) ;
\draw    (480.17,120) -- (480.17,171) ;
\draw    (512.17,120) -- (512.17,171) ;
\draw    (341.67,151) -- (342.17,161) ;
\draw    (296.17,161) -- (296.17,171) ;
\draw    (326.17,161) -- (326.17,171) ;
\draw    (357.17,161) -- (357.17,171) ;
\draw    (388.17,161) -- (388.17,171) ;
\draw    (342.17,120) -- (342.17,131) ;
\draw    (220.67,81) -- (220.67,91) ;

\draw (292.17,54) node [anchor=north west][inner sep=0.75pt]   [align=left] {$\displaystyle v_{1}$};
\draw (212.17,95) node [anchor=north west][inner sep=0.75pt]   [align=left] {$\displaystyle v_{2}$};
\draw (457,94) node [anchor=north west][inner sep=0.75pt]   [align=left] {$\displaystyle v_{3}$};
\draw (88,135) node [anchor=north west][inner sep=0.75pt]   [align=left] {$\displaystyle v_{4}$};
\draw (212.17,135) node [anchor=north west][inner sep=0.75pt]   [align=left] {$\displaystyle v_{5}$};
\draw (333.67,135) node [anchor=north west][inner sep=0.75pt]   [align=left] {$\displaystyle v_{6}$};
\draw (45.17,175) node [anchor=north west][inner sep=0.75pt]   [align=left] {1};
\draw (75.17,175) node [anchor=north west][inner sep=0.75pt]   [align=left] {2};
\draw (106.17,175) node [anchor=north west][inner sep=0.75pt]   [align=left] {3};
\draw (136.17,175) node [anchor=north west][inner sep=0.75pt]   [align=left] {4};
\draw (170.17,175) node [anchor=north west][inner sep=0.75pt]   [align=left] {5};
\draw (200.17,175) node [anchor=north west][inner sep=0.75pt]   [align=left] {6};
\draw (231.17,175) node [anchor=north west][inner sep=0.75pt]   [align=left] {7};
\draw (261.17,175) node [anchor=north west][inner sep=0.75pt]   [align=left] {8};
\draw (291.17,175) node [anchor=north west][inner sep=0.75pt]   [align=left] {9};
\draw (317.17,175) node [anchor=north west][inner sep=0.75pt]   [align=left] {10};
\draw (348.67,175) node [anchor=north west][inner sep=0.75pt]   [align=left] {11};
\draw (379.17,175) node [anchor=north west][inner sep=0.75pt]   [align=left] {12};
\draw (409.17,175) node [anchor=north west][inner sep=0.75pt]   [align=left] {13};
\draw (440.17,175) node [anchor=north west][inner sep=0.75pt]   [align=left] {14};
\draw (471.17,175) node [anchor=north west][inner sep=0.75pt]   [align=left] {15};
\draw (503.17,175) node [anchor=north west][inner sep=0.75pt]   [align=left] {16};
\draw (553,53) node [anchor=north west][inner sep=0.75pt]   [align=left] {$\displaystyle L_{1}$};
\draw (553,95) node [anchor=north west][inner sep=0.75pt]   [align=left] {$\displaystyle L_{2}$};
\draw (553,135) node [anchor=north west][inner sep=0.75pt]   [align=left] {$\displaystyle L_{3}$};

\end{tikzpicture}
    \caption{{A correctly specified model with a redundant factor corresponding to $v_2$.}}
    \label{fig:counter}
\end{figure}

We present the proposed greedy search algorithm for efficiently searching over the space $\mathcal A_k$ for each $k$. 
Recall that $\tilde \Sigma_{k,0} := \sum_{i=1}^{k_{t-2}} (\tilde{\boldsymbol{\lambda}}_i)_{[\hat{v}_{k}]} (\tilde{\boldsymbol{\lambda}}_i)_{[\hat{v}_{k}]}^\top$ when $k \in \hat{L}_{t-1}$ for $t \geq 3$. We further define 
$\tilde \Sigma_{k,0}$ as a $J\times J$ zero matrix to cover the case when $t = 2$ and $k=1$. We divide the search into two cases. 

\noindent 1. For $c=0$, we simply compute 
\begin{equation}\label{eq:opt zero child}
\begin{aligned}
\tilde{\mbox{IC}}_{k,0} = \min_{\Lambda_k,\Psi_k}&l\left( \tilde{\Sigma}_{k,0}+ \Lambda_k\Lambda_k^\top + \Psi_k, S_k\right),\\
\mbox{s.t.~} & \Lambda_k\in\mathcal{A}^{k}(0), \kappa_1\leq (\Psi_k)_{[\{i\},\{i\}]}\leq \kappa_2,\\
&(\Psi_k)_{[\{i\},\{j\}]}=0,i=1,\ldots,|\hat{v}_k|,j\neq i
\end{aligned}
\end{equation}
and use $(\tilde{\Lambda}_{k,0},\tilde{\Psi}_{k,0})$ to denote the solution to \eqref{eq:opt zero child}. This is a relatively simple continuous optimization problem that a standard numerical solver can solve. 

\noindent 2. Set $d = d_{\max}+2-t$. For each $c \in \{2, \ldots, c_{\max}\}$, we perform the following steps: 

\begin{enumerate}
    \item[(a)] Solve the optimization in $\mbox{IC}_{k}(c,d,\ldots,d)$. It is easy to check that the penalty term in $\mbox{IC}_{k}(c,d,\ldots,d)$ equals to $|\hat v_k|d - c d(d-1)/2 $, which does not depend on the loading matrix $\Lambda_k$ as long as the number of item within each of the corresponding partition is no less than $d$. Therefore, the optimization problem becomes 
\begin{equation}\label{eq:optmain}
\begin{aligned}
\min_{\Lambda_k,\Psi_k}&l\left( \tilde{\Sigma}_{k,0}+ \Lambda_k\Lambda_{k}^{\top} + \Psi_k, S_k\right) ,\\
\mbox{s.t.~} & {\Lambda_k\in\mathcal{A}^{k}(c,d,\ldots,d)},\kappa_1\leq(\Psi_k)_{[\{i\},\{i\}]}\leq \kappa_2,\\
&(\Psi_k)_{[\{i\},\{j\}]}=0,i=1,\ldots,|\hat{v}_k|,j\neq i.
\end{aligned}
\end{equation}
Let $v_1^{k,c}, \ldots, v_c^{k,c}$ be the partition of ${1, \ldots, |\hat v_k|}$ given by the solution to \eqref{eq:optmain}.
We note that \eqref{eq:optmain} is a discrete optimization problem, due to the combinatorial nature of the space $\mathcal{A}^{k}(c,d,\ldots,d)$. The theoretical properties in Theorem~\ref{cor:consistency} are established under the ideal scenario that this optimization is solved exactly for all $k$. In reality, however, exactly solving \eqref{eq:optmain} is computationally infeasible when $J$ and $c$ are large. To search for the solution to \eqref{eq:optmain}, we cast it into a continuous optimization problem with nonlinear zero constraints and solved by an augmented Lagrangian method; see Section~\ref{sec:computation} for the relevant details. 

\item[(b)] Given the partition $v_1^{k,c}, \ldots, v_c^{k,c}$ from the previous step, we define the space for all $d_1, \ldots, d_c \in \{1, \ldots, d_{\max}\}$  
\begin{equation}\label{eq: para space after estimated location}
    \begin{aligned}
    \tilde{\mathcal{A}}^k(c, d_1, \ldots, d_c)= &\{A=(a_{ij})_{|\hat{v}_{k}|\times (1+d_1 + \cdots +d_c)}: A_{[v_s^{k,c}, \{j\}]} = \mathbf 0, \mbox{ for all } s = 1, \ldots, c,
    \mbox{~and } \\
    &~~j \notin \{1,  2+\sum_{s'<s}d_{s'}, 3+\sum_{s'<s}d_{s'}, \ldots, 1+\sum_{s'\leq s}d_{s'} \}, \mbox{~and~} |a_{ij}|\leq\tau \mbox{~for~} \\
    &~\mbox{~all~} i=1,\ldots,|\hat{v}_{k}|, j =1,\ldots,1+\sum_{s=1}^{c}d_s\}
    \end{aligned}
\end{equation}
for the same constant $\tau$ as in $\mathcal{A}^k(c, d_1, \ldots, d_c)$. We note that the space of $\tilde{\mathcal{A}}^k(c, d_1, \ldots,d_c)$ is substantially smaller than ${\mathcal{A}}^k(c, d_1, \ldots, d_c)$ as the partition of the variables is fixed. Based on $\tilde{\mathcal{A}}^k(c, d_1, \ldots, d_c)$, we define 
 information criterion 
\begin{equation}\label{eq:tilde IC}
    \begin{aligned}
    \tilde{\mbox{IC}}_k(c, d_1, \ldots, d_c) = \min_{\Lambda_k,\Psi_k}&l\left( \tilde{\Sigma}_{k,0}+ \Lambda_k\Lambda_k^\top + \Psi_k, S_k\right) + {p_k(\Lambda_k)} \log N,\\
    \mbox{s.t.~} & \Lambda_k\in\tilde{\mathcal{A}}^{k}(c,d_1,\ldots,d_c),\kappa_1\leq (\Psi_k)_{[\{i\},\{i\}]}\leq \kappa_2,\\
    &(\Psi_k)_{[\{i\},\{j\}]}=0,i=1,\ldots,|\hat{v}_k|,j\neq i.
    \end{aligned}
    \end{equation}
As the space $\tilde{\mathcal{A}}^k(c, d_1, \ldots, d_c)$ is relatively simple, the optimization in \eqref{eq:tilde IC} is a relatively simple continuous optimization problem that a standard numerical solver can solve. 
\item[(c)]  We then search for the best values for $d_1, \ldots, d_c$ for the given $c$. They are determined sequentially, one after another. More specifically, we
first determine $d_1$ 
by

\begin{equation}\label{eq:d1}
  \tilde{d}_{1}^{c} = \argmin_{1\leq d_1\leq \min(|v_{1}^{k,c}|,d)}\tilde{\mbox{IC}}_{k}(c,d_1,\min(|v_{2}^{k,c}|,d),\ldots,\min(|v_{c}^{k,c}|,d)),  
\end{equation} 
where we fix the value of $d_{2},\ldots,d_c$ at $\min(|v_{2}^{k,c}|,d),\ldots,\min(|v_{c}^{k,c}|,d)$ and only vary the value of $d_1$. Solving \eqref{eq:d1} involves solving $\min(|v_{1}^{k,c}|,d)$ relatively simple continuous optimization problems. 
Then we proceed to $d_2$ and so on. For $s \geq 2$, suppose that 
we have learned $\tilde{d}_{1}^{c},\ldots,\tilde{d}_{s-1}^{c}$, then $d_s$ is determined by 
\begin{equation*}
\begin{aligned}
\tilde{d}_{s}^{c} = &\argmin_{1\leq d_s\leq \min(|v_{s}^{k,c}|,d)}\tilde{\mbox{IC}}_{k}(c,\tilde{d}_{1}^{c},\ldots,\tilde{d}_{s-1}^{c},d_s,\min(|v_{s+1}^{k,c}|,d),\ldots,\min(|v_{c}^{k,c}|,d)),
\end{aligned}
\end{equation*}
where we fix $d_1,\ldots,d_{s-1}$ at their learned values and further fix $d_{s+1},\ldots, d_c$ at the value of \\ $\min(|v_{s+1}^{k,c}|,d)$,$\ldots,\min(|v_{c}^{k,c}|,d)$.  
\item[(d)] Given $\tilde{d}_{1}^{c},\ldots,\tilde{d}_{c}^{c}$, we define 
\begin{equation}\label{eq:greedy search opt}
        \tilde{\mbox{IC}}_{k,c} = \tilde{\mbox{IC}}_{k}(c,\tilde{d}_{1}^{c},\ldots,\tilde{d}_{c}^{c})
    \end{equation}
and $\tilde{\Lambda}_{k,c}$, $\tilde{\Psi}_{k,c}$ as the solution to \eqref{eq:greedy search opt}.
\end{enumerate}

The above steps yield $\tilde{\mbox{IC}}_{k,c}$, $c \in\{0,2,\ldots,c_{\max}\}$. Then, we estimate the number of child factors of Factor $k$ by the value of $c$ that minimises the modified information criterion $\tilde{\mbox{IC}}_{k,c}$. That is, we let 
$$\hat{c}_{k} = \argmin_{c\in\{0,2,\ldots,c_{\max}\}}\tilde{\mbox{IC}}_{k,c}. $$
Moreover, we define 
$$\hat{v}_{s}^{k} =  \hat{v}_{k}[v_s^{k,\hat c_k}], s = 1, \ldots, \hat c_k,$$
where $v_s^{k,\hat c_k}$, $s = 1, \ldots, \hat c_k$, is the partition of $\{1, \ldots, |\hat v_k|\}$ learned above for $c= \hat c_k$. Then 
$\hat{v}_{s}^{k}$, $s = 1, \ldots, \hat c_k$, give a partition of $\hat v_k$, and we estimate that the $s$th child factor of Factor $k$ is loaded by the variables in $\hat{v}_{s}^{k}$. {As a by-product, we obtain an estimate of the $k$th column of the loading matrix, denoted by $\tilde{\boldsymbol \lambda}_k$, satisfying that $(\tilde{\boldsymbol \lambda}_{k})_{[\hat{v}_k]}$ equals to the first column of $\tilde{\Lambda}_{k,\hat{c}_k}$ and $(\tilde{\boldsymbol \lambda}_{k})_{[\{1, \ldots, J\}\setminus \hat{v}_k]}$ is a zero vector.}

 We summarise the steps described previously in Algorithm \ref{algo:ICB}. 
\begin{algorithm}
\caption{Information-Criterion-based method}
\label{algo:ICB}
\begin{algorithmic}[1]
\Require {$\hat{v}_{k}$, $c_{\max},d_{\max}\in\mathbb{N}^{+}$, $\tilde{\Sigma}_{k,0}$, $S_{k}$ and layer $t$}.
\State {Set $d = \min(|\hat{v}_k|,d_{\max}+2-t)$.}
\State Solve $\tilde{\mbox{IC}}_{k,0}$ defined in \eqref{eq:opt zero child}. Let $(\tilde{\Lambda}_{k,0},\tilde{\Psi}_{k,0})$ as the solution to $\tilde{\mbox{IC}}_{k,0}$.
\For{$c = 2,3,\ldots,c_{\max}$}
\State {Solve the optimization problem~\eqref{eq:optmain}.
Set $v_{1}^{k,c},\ldots,v_{c}^{k,c}$ as the partition of $\{1,\ldots,|\hat{v}_k|\}$ by the solution to~\eqref{eq:optmain}}.
\For{$s=1,\ldots,c$}
\State {Compute $$\tilde{d}_{s}^{c} = \argmin_{1\leq d_s\leq \min(|v_{s}^{k,c}|,d)}\tilde{\mbox{IC}}_{k}(c,\tilde{d}_{1}^{c},\ldots,\tilde{d}_{s-1}^{c},d_s,\min(|v_{s+1}^{k,c}|,d),\ldots,\min(|v_{c}^{k,c}|,d)),$$
where $\tilde{\mbox{IC}}_k$ is defined in \eqref{eq:tilde IC}.
}
\EndFor
\State Define $\tilde{\mbox{IC}}_{k,c} = \tilde{\mbox{IC}}_{k}(c,\tilde{d}_{1}^{c},\ldots,\tilde{d}_{c}^{c})$ and $(\tilde{\Lambda}_{k,c},\tilde{\Psi}_{k,c})$ as the solution to $\tilde{\mbox{IC}}_{k,c}$.
\EndFor
\State Define $\hat{c}_{k} = \argmin_{c\in\{0,2,3,\ldots,c_{\max}\}} \tilde{\mbox{IC}}_{k,c}$.
\State Set $\tilde{v}_{1}^{k},\ldots,\tilde{v}_{\hat{c}_{k}}^{k}$ be the partition of $\{1,\ldots,|\hat{v}_{k}|\}$ associated with $\tilde{\Lambda}_{k,\hat{c}_k}$. Define the partition of $\hat{v}_{k}$ by $\hat{v}_{1}^{k} = \hat{v}_{k}[\tilde{v}_{1}^{k}],\ldots,\hat{v}_{\hat{c}_k}^{k} = \hat{v}_{k}[\tilde{v}_{\hat{c}_k}^{k}]$.
\State Define $\tilde{\boldsymbol{\lambda}}_{k}$ following that $(\tilde{\boldsymbol{\lambda}}_{k})_{[\hat{v}_{k}]}$ equals to the first column of $\tilde{\Lambda}_{k,\hat{c}_k}$ and $(\tilde{\boldsymbol{\lambda}}_{k})_{[\{1,\ldots,J\}\setminus\hat{v}_{k}]} $ is a zero vector.
\Ensure $\hat{c}_{k}$, $\hat{v}_{1}^{k},\ldots,\hat{v}_{\hat{c}_{k}}^{k}$ and  $\tilde{\boldsymbol{\lambda}}_{k}$.
\end{algorithmic}
\end{algorithm}

\begin{remark}
$c\in\{0,2,\ldots,c_{\max}\}$ represents the number of child factors of Factor $k$. 
In other words, $c_{\max}$ is an upper bound on the possible number of child factors of Factor $k$. 
On the one hand, we need to ensure that $c_{\max}$ is not too small so that Condition~\ref{cond:hyper} is satisfied. On the other hand, we want to avoid $c_{\max}$ being too large to reduce the computational cost.  
Since the true value of $c$ should satisfy 
constraints C3 and C4 in Section \ref{subsec:constraints}, 
$c_{\max}$ should be no more than $\lfloor|\hat{v}_{k}|/3\rfloor$ when $|\hat{v}_{k}| \geq 7$ and $c_{\max} = 0$, when $|\hat{v}_{k}| \leq 6$,
where $\lfloor\cdot\rfloor$ is the floor function that returns the greatest integer less than or equal to the input. In the simulation study in Section~\ref{sec:sim}, we set $c_{\max} =  \min(4,\lfloor|\hat{v}_{k}|/3\rfloor)$ when $|\hat{v}_{k}| \geq 7$ and $c_{\max} =  0$ when $|\hat{v}_{k}| \leq 6$, 
which, according to the data generation model, is an upper bound for the value of $c$. For the real data analysis in Section~\ref{sec:real data}, since the true structure is unknown, we set $c_{\max} =  \min(6,\lfloor|\hat{v}_{k}|/3\rfloor)$ when $|\hat{v}_{k}| \geq 7$ and $c_{\max} =  0$ when $|\hat{v}_{k}| \leq 6$ 
as a more conservative choice than that of $c_{\max}$ for the simulation study. In practice, we may adjust our choice based on prior knowledge about the hierarchical factor structure. 
\end{remark}

\begin{remark}
The input hyperparameter $d_{\max}$ is an upper bound of one plus the number of descendant factors of the factors in the second layer. When learning the factors on the $t$th layer for $t\geq 3$, we use $d_{\max}+2-t$ as an upper bound of one plus the number of descendant factors of the factors in the $(t+1)$th layer, as the number of descendant factors each factor has tends to decrease as $t$ increases. 
Similar to the choice of $c_{\max}$, we want to choose a $d_{\max}$ that is neither too large nor too small. In the simulation study in Section~\ref{sec:sim}, we start with $d_{\max} = 6$ when learning the factors in the second layer. In the real data analysis in Section~\ref{sec:real data}, we start with $d_{\max} = 10$. In practice, we may adjust this choice based on the problem size (e.g., the number of variables) and prior knowledge of the hierarchical factor structure.
\end{remark}

\begin{remark}
Efficiently learning the hierarchical structure from data is challenging due to the super-exponential growth of the search space with the number of items $J$, which creates a significant computational bottleneck. To overcome the computational issue, we convert the combinatorial optimization problems in \eqref{eq:optIC} and \eqref{eq:opt IC kth} into the continuous optimization problems in \eqref{eq:optmain}. A similar constraint-based continuous optimization method is proposed for learning directed acyclic graphs (DAGs) in \cite{zheng2018dags} and the bi-factor model in \cite{Qiao_Chen_Ying_2025}. By integrating continuous optimization techniques with a breadth-first search strategy, our approach (presented in Algorithms \ref{algo:divide and conquer} and \ref{algo:ICB}) requires solving only $\mathcal{O}(Jc^{2}_{\max}d_{\max})$ continuous optimization problems, thus significantly improving the computational efficiency.
\end{remark}

\subsection{Theoretical Results}\label{subsub:selection}
We now provide theoretical guarantees for the proposed method based on Algorithms~\ref{algo:divide and conquer} and \ref{algo:ICB}. We start with introducing some notation. We use $\|\cdot\|_{F}$ to denote the Frobenius norm of any matrix and $\|\cdot\|$ as the Euclidean norm of any vector. We also use the notation $a_{N} = O_{\mathbb{P}}(b_{N})$ to denote that $a_{N}/b_{N}$ is bounded in probability. In addition to the conditions required for the identifiability of the true hierarchical factor model, we additionally require Conditions~\ref{cond:bic correct}--\ref{cond:hyper} to ensure the proposed method is consistent.

\begin{condition}\label{cond:bic correct}
For any factor $i$ with $\text{Ch}^{*}_{i}\neq \emptyset$ and any $j\in v_{i}^{*}$, there exist
$E_1, E_2 \subset v^{*}_{i}\setminus\{j\}$ with $|E_1| = |E_2| = 1+|D_{i}^{*}|$ and $E_1\cap E_2 = \emptyset$, such that $\Lambda^{*}_{[E_1,\{i\}\cup D^{*}_{i}]}$ and $\Lambda^{*}_{[E_2,\{i\}\cup D^{*}_{i}]}$ are of full rank.
\end{condition}  

\begin{condition}\label{cond:hier always select correct number}
For any factor $i$ with $\text{Ch}^{*}_{i}\neq \emptyset$ and any $k\in \text{Ch}_{i}^{*}$, there exist $E_1, E_2 \subset v^{*}_{k}$ with $|E_1| = 2+|D_{k}^{*}|$, $|E_2| = 1+|D_{k}^{*}|$ and $E_1\cap E_2 = \emptyset$ such that $\Lambda^{*}_{[E_1,\{i,k\}\cup D^{*}_{k}]}$ and $\Lambda^{*}_{[E_2,\{k\}\cup D^{*}_{k}]}$ are of full rank.
\end{condition}

\begin{condition}\label{cond:correct pilot}
$\|S - \Sigma^*\|_{F}  =  O_{\mathbb{P}}(1/\sqrt{N})$.
\end{condition} 

\begin{condition}\label{cond:hier compact set}
The true loading parameters and unique variance parameters satisfy $ |\lambda_{ij}^*|\leq \tau$ and $\kappa_1\leq\psi_{i}^*\leq \kappa_2$ for all $i,j$, where $\tau$, $\kappa_1$ and $\kappa_2$ are constraints used in the ICB method. 
\end{condition}

\begin{condition}\label{cond:hyper}
When learning the child factors of each true factor $k$, the constants $c_{\max}$ and $d_{\max}$ are chosen such that $c_{\max}\geq |\text{Ch}^{*}_{k}|$ and $d_{\max}\geq\max_{s\in \text{Ch}^{*}_{k}}|D^{*}_{s}|+1$.
\end{condition}

\begin{theorem}\label{cor:consistency}
Suppose that Conditions \ref{cond:true para},\ref{cond:rank}, and \ref{cond:bic correct}--\ref{cond:hyper} hold.  Then, the outputs from Algorithm~\ref{algo:divide and conquer} are consistent. That is,  as $N$ goes to infinity, the probability of $\hat T = T$, $\hat K  = K$, $\hat L_t = L_t$, $t=1, \ldots, T$, and $\hat v_i = v_i^*$, $i=1, \ldots, K$ goes to 1, and   
$\Vert \hat \Lambda - \Lambda^*\hat{Q}\Vert_F = O_{\mathbb{P}}(1/\sqrt{N})$  and $\Vert \hat \Psi - \Psi^*\Vert_F =  O_{\mathbb{P}}(1/\sqrt{N})$, where {$\hat{Q} \in \mathcal Q$} is the diagonal matrix with diagonal entries consisting of the signs of the corresponding entries of $\hat{\Lambda}^{\top}\Lambda^*$.

\end{theorem}

Theorem~\ref{cor:consistency} guarantees that the true hierarchical factor structure can be consistently learned from data and its parameters can be consistently estimated {after adjusting the sign for each column of the loading matrix by $\hat{Q}$}.
  
\begin{remark}

It should be noted that in  Theorem~\ref{cor:consistency}, Algorithm~\ref{algo:divide and conquer} applies Algorithm~\ref{algo:ICB}, which involves some nontrivial optimization problems, including a discrete optimization problem~\eqref{eq:optmain}. 
The theorem is established under the oracle scenario that these optimizations are always solved successfully. However, we should note that this cannot be achieved by polynomial-time algorithms due to the complexity of these optimizations. 
\end{remark}

\begin{remark}
    Theorem~\ref{cor:consistency} does not explicitly require Condition \ref{cond:seperate}, because 
Condition~\ref{cond:bic correct} is a stronger condition that implies Condition~\ref{cond:seperate}, as shown in Lemma \ref{lem:apprxi anderson} in the Appendix~\ref{appen:proof consistency}.
In fact, Condition~\ref{cond:bic correct} is sufficient for Condition~\ref{cond:suff 2}, which further implies Condition~\ref{cond:seperate}. We need a stronger condition (i.e., Condition~\ref{cond:bic correct}) here, for distinguishing between the loading structure and the unique variance at each stage of recursion.
Similar to Condition~\ref{cond:rank}, this condition imposes further requirements on the number of child factors and the number of descendant factors a factor can have. More specifically, for such a partition to exist, we need $|v_i^*| \geq 2|D_i^*| + 3$.  Other than that, the full-rank requirement is easily satisfied by most hierarchical factor models. {Similar to Condition~\ref{cond:bic correct}, Condition~\ref{cond:hier always select correct number} also requires $|v_i^*| \geq 2|D_i^*| + 3$. This condition plays a central role in ensuring that Step 6 in Algorithm~\ref{algo:ICB} is valid.} Condition \ref{cond:correct pilot} is very mild. It is automatically satisfied when the sample covariance matrix is constructed using independent and identically distributed observations from the true model, {and all the fourth-order moments of the i.i.d. data are finite.} {Condition~\ref{cond:hier compact set} requires the true loading and unique variance parameters to satisfy the same boundedness constraints as in the ICB method in Algorithm~\ref{algo:ICB}. Theoretically, these constraints ensure that the parameter space is compact, which is needed for bounding the differences in the loss function of different models. Empirically, we notice that the ICB method works well even without these constraints, and thus omit these constraints in the computation.} Condition \ref{cond:hyper} requires that $c_{\max}$ and $d_{\max}$ are chosen sufficiently large so that the search space covers the true model. 
\end{remark}

\section{Computation}\label{sec:computation}

As mentioned previously, the optimization problem in $\mbox{IC}_{k}(c,d,\ldots,d)$ in Algorithm~\ref{algo:ICB} can be cast into a continuous optimization problem and solved by an augmented Lagrangian method(ALM). In what follows, we provide the details.

With slight abuse of notation, we use the reparameterization of the unique variance matrix such that $\Psi_k = \mbox{diag}(\boldsymbol{\psi}_k^{2})$, where $\mbox{diag}(\cdot)$ is a function that converts a vector to a diagonal matrix with the diagonal entries filled by the vector. Here,  $\boldsymbol{\psi}_{k}^{2} = \{\psi_{k1}^{2},\ldots,\psi_{k,|\hat{v}_{k}|}^{2}|\}$ is a $|\hat{v}_{k}|$-dimensional vector for $\psi_{k1},\ldots,\psi_{k,|\hat{v}_{k}|}\in\mathbb{R}$. We further let 
$\mathcal{B}_{s} = \{2+(s-1)d,\ldots,1+sd\}$ for $s=1,\ldots,c$.  We note that, up to a relabelling of the partition sets or, equivalently, dropping the label ordering constraint $\min\{v_1^k\} < \min\{v_2^k\} < \cdots < \min\{v_c^k\}$, 
the set  $\mathcal{A}^{k}(c,d,\ldots,d)$ can be rewritten as 
$${\{A=(a_{ij})_{{|\hat{v}_k|}\times (1+cd)}: a_{ij}a_{ij'} = 0 \mbox{~for~} i=1,\ldots,|\hat{v}_{k}|, j\in\mathcal{B}_{s}, j'\in\mathcal{B}_{s'},s\neq s', |a_{ij}| \leq \tau\}.}$$

Therefore, we can solve $\mbox{IC}_{k}(c,d,\ldots,d)$ by solving the following continuous optimization problem with nonlinear zero constraints:

\begin{equation}\label{opt:alm aug bbf}
    \begin{aligned}
    \bar{\Lambda}_{k,c},\bar{\boldsymbol{\psi}}_{k,c} =&  \argmin_{\Lambda_k,\boldsymbol{\psi}_k}l\left( \tilde{\Sigma}_{k,0} + \Lambda_{k}(\Lambda_k)^\top + \mbox{diag}(\boldsymbol{\psi}_k^{2}), S_{k}\right)\\
     \mbox{~s.t.~}& \lambda_{k,ij}\lambda_{k,ij'} = 0 \mbox{~for~} i=1,\ldots,|\hat{v}_{k}|, j\in\mathcal{B}_{s}^{k}, j'\in\mathcal{B}_{s'}^{k},s\neq s'.
    \end{aligned}
\end{equation}
{Here, the constraints on the loading and unique variance parameters are omitted for simplicity, as these constraints are always satisfied when we set $\tau$ and $\kappa_2$ to be sufficiently large and $\kappa_1$ to be sufficiently small.} Once this optimization is solved, then for each $i$, there is one and only one $\mathcal B_s$ such that 
$(\bar{\Lambda}_{k,c})_{[\{i\},\mathcal B_s]} \neq  \mathbf 0$. Therefore, we obtain a partition of $1, \ldots, |\hat v_k|$ by the sets
$$\{i:(\bar{\Lambda}_{k,c})_{[\{i\},\mathcal B_s]} \neq \mathbf 0 \}, s = 1, \ldots, c.$$
We obtain $v_{1}^{k,c},\ldots,v_{c}^{k,c}$ by reordering 
$\{i:(\bar{\Lambda}_{k,c})_{[\{i\},\mathcal B_s]} \neq \mathbf 0 \}, s = 1, \ldots, c$  to satisfy the constraint on the labels of these sets. 

We solve \eqref{opt:alm aug bbf} by the ALM algorithm \citep[see, e.g.,][]{bertsekas2014constrained}, which is a standard approach to such problems. This method finds a solution to \eqref{opt:alm aug bbf} by solving a sequence of unconstrained optimization problems. More specifically, in the $t$th iteration, $t=1, 2, \ldots $, the ALM minimizes an augmented Lagrangian function that is constructed based on the result of the previous iteration. 
Details of the ALM are given in Algorithm~\ref{alg:main} below, where the function $h(\cdot)$ returns the second largest values of a vector. The updating rule of $\beta^{(t)}_{jii'}$ and $c^{(t)}$ follows equations (1) and (47) in Chapter 2.2 of \cite{bertsekas2014constrained}, and the convergence of Algorithm \ref{alg:main} {to a stationary point of \eqref{opt:alm aug bbf}} is guaranteed by Proposition 2.7 of \cite{bertsekas2014constrained}. We follow the recommended choices of $c_{\theta} = 0.25$ and $c_{\sigma} = 10$ in \cite{bertsekas2014constrained} for the tuning parameters in Algorithm~\ref{alg:main}.

\begin{algorithm}
\caption{An augmented Lagrangian method for solving $\mbox{IC}_{k}(c,d,\ldots,d)$\label{alg:main}}
\begin{algorithmic}[1]
\Require Initial value $\Lambda^{(0)}$ and $\boldsymbol{\psi}^{(0)}$, initial Lagrangian parameters $\beta_{ijj'}^{(0)}$ for $i=1,\ldots,|\hat{v}_k|$, $j\in\mathcal{B}_{s}$, $j'\in\mathcal{B}_{s'}$ and $s\neq s'$, initial penalty coefficient  $c^{(0)}>0$, constants $c_{\theta} \in (0,1)$ and $c_{\sigma} > 1$, tolerances $\delta_1, \delta_2>0$, {maximal iteration number $M_{\max}$}.

\For{$t = 1,2,\ldots,M_{\max}$}
\State Solve the following problem:
\begin{equation*}
\begin{aligned}
\Lambda_{k}^{(t)}, \boldsymbol{\psi}_{k}^{(t)}
=&\argmin_{\Lambda_{k}, \boldsymbol{\psi}_k}~~~  l\left( \tilde{\Sigma}_{k,0} + \Lambda_{k}(\Lambda_k)^\top + \mbox{diag}(\boldsymbol{\psi}_k), S_{k}\right) \\
&~~~~~~~~~~~+ \left(\sum_{i=1}^{|\hat{v}_{k}|}\sum_{j\in\mathcal{B}_{s},j'\in \mathcal{B}_{s'},s\neq s'} \beta_{ijj'}^{(t)} \lambda_{k,ij}\lambda_{k,ij'} \right)\\
&~~~~~~~~~~~+  \frac{1}{2}c^{(t)} \left(\sum_{i=1}^{|\hat{v}_{k}|} \sum_{j\in\mathcal{B}_{s},j'\in \mathcal{B}_{s'},s\neq s'} (\lambda_{k,ij}\lambda_{k,ij'})^2 \right).
\end{aligned}
\end{equation*}
\State Update $\beta_{ijj'}^{(t)}$ and $c^{(t)}$ according to equations \eqref{eq:beta rule} and \eqref{eq:c rule} 

\begin{equation}\label{eq:beta rule}
\begin{aligned}
\beta_{ijj'}^{(t)} = \beta_{ijj'}^{(t-1)} + c^{(t-1)}\lambda^{(t)}_{k,ij}\lambda^{(t)}_{k,ij'},
\end{aligned}
\end{equation}

and

\begin{equation}\label{eq:c rule}
\begin{aligned}
c^{(t)} = \left\{
    \begin{array}{l}
        c_{\sigma}c^{(t-1)} \mbox{~~if~~}  \left(\sum_{i=1}^{|\hat{v}_{k}|} \sum_{j\in\mathcal{B}_{s},j'\in \mathcal{B}_{s'},s\neq s'} (\lambda_{k,ij}^{(t)}\lambda_{k,ij'}^{(t)})^2 \right)^{1/2} \\
        ~~~~~~~~~~~~~~~~~~~~~> c_{\theta} \left(\sum_{i=1}^{|\hat{v}_{k}|} \sum_{j\in\mathcal{B}_{s},j'\in \mathcal{B}_{s'},s\neq s'} (\lambda_{k,ij}^{(t-1)}\lambda_{k,ij'}^{(t-1)})^2 \right)^{1/2}, \\
        c^{(t-1)} \mbox{~otherwise}.
    \end{array}
\right.
\end{aligned}
\end{equation}
\If{ $$\left(\Vert \Lambda_{k}^{(t)} - \Lambda_{k}^{(t-1)} \Vert_{F}^{2} + \Vert \boldsymbol{\psi}_{k}^{(t)} -  \boldsymbol{\psi}_k^{(t-1)}  \Vert^{2}\right)^{1/2}/ \left(|\hat{v}_k|(2+d)\right)^{1/2}<\delta_1$$ and $$\max_{i \in \{1, \ldots, |\hat{v}_{k}|\}} h(\max_{j\in\mathcal{B}_1}|\lambda_{k,ij}^{(t)}|, \max_{j\in\mathcal{B}_2}|\lambda_{k,ij}^{(t)}| \ldots, \max_{j\in\mathcal{B}_c}|\lambda_{k,ij}^{(t)}|)<\delta_2,$$} 
\State \Return $\Lambda^{(t)}_{k}, \boldsymbol{\psi}^{(t)}_{k}$.
\State \textbf{Break}
\EndIf
\EndFor
\Ensure $\Lambda^{(t)}_{k}, \boldsymbol{\psi}^{(t)}_{k}$.
\end{algorithmic}
\end{algorithm}

We remark on the stopping criterion in the implementation of Algorithm~\ref{alg:main}. We monitor the convergence of the algorithm based on two criteria: (1) the change in parameter values in two consecutive steps, measured by $$\left(\Vert \Lambda_{k}^{(t)} - \Lambda_{k}^{(t-1)} \Vert_{F}^{2} + \Vert \boldsymbol{\psi}_{k}^{(t)} -  \boldsymbol{\psi}_k^{(t-1)}  \Vert^{2}\right)^{1/2}/\left(|\hat{v}_k|(2+d)\right)^{1/2},$$
and (2) the distance between the estimate and the space $\mathcal{A}^{k}(c,d,\ldots,d)$ measured by 
$$\max_{i \in \{1, \ldots, |\hat{v}_{k}|\}} h(\max_{j\in\mathcal{B}_1}|\lambda_{k,ij}^{(t)}|, \max_{j\in\mathcal{B}_2}|\lambda_{k,ij}^{(t)}| \ldots, \max_{j\in\mathcal{B}_c}|\lambda_{k,ij}^{(t)}|).$$ 
When both criteria are below their pre-specified thresholds, $\delta_1$ and $\delta_2$, respectively, we stop the algorithm.  Let $M$ be the last iteration number. 
Then the selected partition of $\{1,\ldots,\hat{v}_{k}\}$, denoted by $v_{1}^{k,c},\ldots,v_{c}^{k,c}$, is given by 
$v_{s}^{k,c} = \{j: |\lambda_{k,ij}^{(M)}| < \delta_2 \mbox{~for all~} j\notin\mathcal{B}_{s}\}.$
For the analyses in Sections~\ref{sec:sim} and  \ref{sec:real data}, we choose $\delta_1 = \delta_2 = 0.01$.

Algorithm~\ref{alg:main} can suffer from slow convergence when the penalty terms become large, resulting in an ill-conditioned optimization problem. When the algorithm does not converge within $M_{\max}$ iterations, we suggest restarting the algorithm, 
using the current parameter value as a warm start. We set $M_{\max} = 100$ in the simulation study in Section~\ref{sec:sim} and the real data analysis in Section~\ref{sec:real data} and keep the maximum number of restarting times to be five. In addition, 
since the optimization problem~\eqref{opt:alm aug bbf} is non-convex,  Algorithm~\ref{alg:main}
may only converge to a local optimum and this local solution may not satisfy condition C4. Therefore,  we recommend running it with multiple random starting points and then 
finding the best solution that satisfies condition C4. In our implementation, each time to solve \eqref{opt:alm aug bbf}, we start by running Algorithm~\ref{alg:main} 100 times, each with a random starting point. {If more than 50 of the solutions satisfy C4, then we stop and proceed to Steps 5--8 in Algorithm~\ref{algo:ICB}. Otherwise, we rerun Algorithm~\ref{alg:main} 100 times with random starting points, until either 50 solutions satisfy C4 or the algorithm has been restarted five times.} 

\section{Simulation Study}\label{sec:sim}
In this section, we examine the recovery of the hierarchical structure as well as the accuracy in estimating the loading matrix and the unique variance matrix of the proposed method. Suppose that $\hat{v}_1,\ldots,\hat{v}_{\hat{K}}$ are the estimated sets of variables loading on each factor, where $\hat{K}$ is the estimated number of factors, $\hat{\Lambda}$ is the estimated loading matrix and $\hat{\Psi}$ is the estimated unique variance matrix. To examine the recovery of the hierarchical factor structure, we measure the matching between the true sets of variables loading on each factor and the estimated sets of variables. More specifically, the following evaluation criteria are considered:
\begin{enumerate}
    \item Exact Match Criterion (EMC): $\mathbf{1}(\hat{K}=K)\prod_{k=1}^{\min(\hat{K},K)}\mathbf{1}(\hat{v}_{k} = v^{*}_{k})$, which equals to 1 when the hierarchical structure is fully recovered and 0 otherwise. 
    \item Layer Match Criterion (LMC): $\mathbf{1}(\{\hat{v}_k\}_{k\in\hat{L}_t} = \{v_{k}^{*}\}_{k\in L_t})$, which is defined for each layer $t$. It equals 1 if the sets of variables loading on the factors in the $t$th layer are correctly learned and 0 otherwise for $t=1,\ldots,T$. 
\end{enumerate}

We then examine the accuracy in estimating the loading matrix and the unique variance matrix. We calculate the mean square error(MSE) for $\hat{\Lambda}$ and $\hat{\Psi}$, after adjusting for the sign indeterminacy shown in Theorem~\ref{thm:identifiability}. More specifically, recall that $\mathcal{Q}$ is the set of sign flip matrices defined in Theorem~\ref{thm:identifiability}. When the proposed method outputs a correct estimate of the hierarchical structure (i.e. $\mbox{EMC}=1$), we define the MSEs for $\hat{\Lambda}$ and $\hat{\Psi}$ as 
$\text{MSE}_{\Lambda} = \min_{ Q\in \mathcal Q}\Vert \hat{\Lambda} - \Lambda^{*}Q\Vert_{F}^{2}/(JK), \mbox{~and~} \text{MSE}_{\Psi} = \Vert \hat{\Psi} - \Psi^{*}\Vert_{F}^{2}/J.$  

We consider the following hierarchical factor structure shown in Figure \ref{fig:sim example} with the number of variables {$J\in\{36,54\}$}, the number of layers $T=4$, the number of factors $K=10$, $L_1=\{1\}$, $L_2=\{2,3\}$, $L_3=\{4,\ldots,8\}$, $L_4=\{9,10\}$ and $v^{*}_{1} = \{1,\ldots,J\}$, $v^{*}_{2} = \{1,\ldots,J/3\}$, $v^{*}_{3} = \{1+J/3,\ldots, J\}$, $v^{*}_4 =  \{1,\ldots,J/6\}$, $v^{*}_5 = \{1+J/6,\ldots,J/3\}$, $v^{*}_6 = \{1+J/3,\ldots,5J/9\}$, $v^{*}_7 = \{1+5J/9,\ldots,7J/9\}$, $v^{*}_8 = \{1+7J/9,\ldots,J\}$, $v^{*}_9 = \{1+J/3,\ldots,4J/9\}$, $v^{*}_{10} = \{1+4J/9,\ldots,5J/9\}$. In the data generation model, $\Psi^{*}$
is either a $J\times J$ identity matrix or $\Psi^{*} = \mbox{diag}(\psi_{1}^{*2},\ldots,\psi_{J}^{*2})$ with $\psi_{j}^{*}, j=1,\ldots,J$, i.i.d., following a Uniform$(0.5,1.5)$ distribution,
and $\Lambda^*$ is generated by
\begin{equation}
\begin{aligned}
\lambda_{jk}^{*} = \left\{
    \begin{array}{l}
        u_{jk} \mbox{~~if~~} k=1; \\
        0 \mbox{~~if~~} k>1 ,j \notin v_{k}^{*}; \\
        (1-2x_{jk})u_{jk} \mbox{~~if~~} k>1, j\in v_{k}^{*},
    \end{array}
\right.
\end{aligned}
\end{equation}
for $j = 1,\ldots,J$, and $k = 1,\ldots, K$. Here, $u_{jk}$s are i.i.d., following a Uniform$(0.5,2)$ distribution and $x_{jk}$s are i.i.d., following 
a Bernoulli$(0.5)$ distribution. 
For each value of $J$, we generate the true loading matrix $\Lambda^*$ and the true unique variance matrix $\Psi^{*}$ once and use it for all its simulations. 

\begin{figure}[tb!]
    \centering

\tikzset{every picture/.style={line width=0.75pt}} 
\resizebox{0.9\linewidth}{!}{
\begin{tikzpicture}[x=0.95pt,y=0.9pt,yscale=-1,xscale=1]

\draw   (310.5,122) -- (331.5,122) -- (331.5,142) -- (310.5,142) -- cycle ;
\draw   (155.5,122) -- (176.5,122) -- (176.5,142) -- (155.5,142) -- cycle ;
\draw   (105.5,70) -- (126.5,70) -- (126.5,90) -- (105.5,90) -- cycle ;
\draw   (421.5,70) -- (442.5,70) -- (442.5,90) -- (421.5,90) -- cycle ;
\draw   (273.5,20.5) -- (294.5,20.5) -- (294.5,40.5) -- (273.5,40.5) -- cycle ;
\draw   (364,171) -- (385,171) -- (385,191) -- (364,191) -- cycle ;
\draw   (467.5,122) -- (488.5,122) -- (488.5,142) -- (467.5,142) -- cycle ;
\draw   (257.5,171) -- (278.5,171) -- (278.5,191) -- (257.5,191) -- cycle ;
\draw   (53.5,122) -- (74.5,122) -- (74.5,142) -- (53.5,142) -- cycle ;
\draw   (572.5,122) -- (593.5,122) -- (593.5,142) -- (572.5,142) -- cycle ;
\draw   (18,222) -- (50,222) -- (50,243.5) -- (18,243.5) -- cycle ;
\draw   (80,222) -- (112,222) -- (112,243.5) -- (80,243.5) -- cycle ;
\draw   (121,222) -- (153,222) -- (153,243.5) -- (121,243.5) -- cycle ;
\draw   (180,222) -- (212,222) -- (212,243.5) -- (180,243.5) -- cycle ;
\draw   (222,222) -- (254,222) -- (254,243.5) -- (222,243.5) -- cycle ;
\draw   (284,222) -- (316,222) -- (316,243.5) -- (284,243.5) -- cycle ;
\draw   (432,222) -- (464,222) -- (464,243.5) -- (432,243.5) -- cycle ;
\draw   (327,222) -- (359,222) -- (359,243.5) -- (327,243.5) -- cycle ;
\draw   (390,222) -- (422,222) -- (422,243.5) -- (390,243.5) -- cycle ;
\draw   (494,222) -- (526,222) -- (526,243.5) -- (494,243.5) -- cycle ;
\draw   (537,222) -- (569,222) -- (569,243.5) -- (537,243.5) -- cycle ;
\draw   (599,222) -- (631,222) -- (631,243.5) -- (599,243.5) -- cycle ;
\draw    (268,191) -- (268,201) ;
\draw    (238,201) -- (300,201) ;
\draw    (238,201) -- (238,222) ;
\draw    (300,201) -- (300,222) ;
\draw    (343,201) -- (406,201) ;
\draw    (374.5,191) -- (374.5,201) ;
\draw    (343,201) -- (343,222) ;
\draw    (406,201) -- (406,222) ;
\draw    (321,142) -- (321,152) ;
\draw    (268,152) -- (374.5,152) ;
\draw    (268,152) -- (268,171) ;
\draw    (374.5,152) -- (374.5,171) ;
\draw    (448,152) -- (448,222) ;
\draw    (448,152) -- (510,152) ;
\draw    (510,152) -- (510,222) ;
\draw    (478,142) -- (478,152) ;
\draw    (553,152) -- (615,152) ;
\draw    (583,142) -- (583,152) ;
\draw    (553,152) -- (553,222) ;
\draw    (615,151) -- (615,222) ;
\draw    (64,142) -- (64,152) ;
\draw    (34,152) -- (96,152) ;
\draw    (34,152) -- (34,222) ;
\draw    (96,152) -- (96,222) ;
\draw    (166,142) -- (166,152) ;
\draw    (137,152) -- (196,152) ;
\draw    (137,152) -- (137,222) ;
\draw    (196,152) -- (196,222) ;
\draw    (321,122) -- (321,101) ;
\draw    (478,122) -- (478,101) ;
\draw    (583,122) -- (583,101) ;
\draw    (321,101) -- (583,101) ;
\draw    (432,101) -- (432,90) ;
\draw    (64,122) -- (64,101) ;
\draw    (166,122) -- (166,101) ;
\draw    (64,101) -- (166,101) ;
\draw    (116,101) -- (116,90) ;
\draw    (284,40.5) -- (284,51) ;
\draw    (116,51) -- (432,51) ;
\draw    (116,51) -- (116,70) ;
\draw    (432,51) -- (432,70) ;

\draw (275.5,23.5) node [anchor=north west][inner sep=0.75pt]   [align=left] {$\displaystyle v_{1}^{*}$};
\draw (55,229) node [anchor=north west][inner sep=0.75pt]   [align=left] {$\displaystyle \cdots $};
\draw (575,229) node [anchor=north west][inner sep=0.75pt]   [align=left] {$\displaystyle \cdots $};
\draw (470,229) node [anchor=north west][inner sep=0.75pt]   [align=left] {$\displaystyle \cdots $};
\draw (366,229) node [anchor=north west][inner sep=0.75pt]   [align=left] {$\displaystyle \cdots $};
\draw (258,229) node [anchor=north west][inner sep=0.75pt]   [align=left] {$\displaystyle \cdots $};
\draw (157,229) node [anchor=north west][inner sep=0.75pt]   [align=left] {$\displaystyle \cdots $};
\draw (157.5,125) node [anchor=north west][inner sep=0.75pt]   [align=left] {$\displaystyle v_{5}^{*}$};
\draw (423.5,73) node [anchor=north west][inner sep=0.75pt]   [align=left] {$\displaystyle v_{3}^{*}$};
\draw (312.5,125) node [anchor=north west][inner sep=0.75pt]   [align=left] {$\displaystyle v_{6}^{*}$};
\draw (55.5,125) node [anchor=north west][inner sep=0.75pt]   [align=left] {$\displaystyle v_{4}^{*}$};
\draw (469.5,125) node [anchor=north west][inner sep=0.75pt]   [align=left] {$\displaystyle v_{7}^{*}$};
\draw (574.5,125) node [anchor=north west][inner sep=0.75pt]   [align=left] {$\displaystyle v_{8}^{*}$};
\draw (259.5,174) node [anchor=north west][inner sep=0.75pt]   [align=left] {$\displaystyle v_{9}^{*}$};
\draw (363,175) node [anchor=north west][inner sep=0.75pt]   [align=left] {$\displaystyle v_{10}^{*}$};
\draw (642,29) node [anchor=north west][inner sep=0.75pt]   [align=left] {$\displaystyle L_{1}$};
\draw (642,74) node [anchor=north west][inner sep=0.75pt]   [align=left] {$\displaystyle L_{2}$};
\draw (642,126) node [anchor=north west][inner sep=0.75pt]   [align=left] {$\displaystyle L_{3}$};
\draw (642,175) node [anchor=north west][inner sep=0.75pt]   [align=left] {$\displaystyle L_{4}$};
\draw (107.5,73) node [anchor=north west][inner sep=0.75pt]   [align=left] {$\displaystyle v_{2}^{*}$};
\draw (29,228) node [anchor=north west][inner sep=0.75pt]   [align=left] {1};
\draw (123,224) node [anchor=north west][inner sep=0.75pt]   [align=left] {1+$\frac{J}{6}$};
\draw (87,224) node [anchor=north west][inner sep=0.75pt]   [align=left] {$\frac{J}{6}$};
\draw (187,224) node [anchor=north west][inner sep=0.75pt]   [align=left] {$\frac{J}{3}$};
\draw (327,224) node [anchor=north west][inner sep=0.75pt]   [align=left] {1+$\frac{4J}{9}$};
\draw (396,224) node [anchor=north west][inner sep=0.75pt]   [align=left] {$\frac{5J}{9}$};
\draw (224,224) node [anchor=north west][inner sep=0.75pt]   [align=left] {1+$\frac{J}{3}$};
\draw (290,224) node [anchor=north west][inner sep=0.75pt]   [align=left] {$\frac{4J}{9}$};
\draw (499,224) node [anchor=north west][inner sep=0.75pt]   [align=left] {$\frac{7J}{9}$};
\draw (433,224) node [anchor=north west][inner sep=0.75pt]   [align=left] {1+$\frac{5J}{9}$};
\draw (609,227) node [anchor=north west][inner sep=0.75pt]   [align=left] {$J$};
\draw (536,224) node [anchor=north west][inner sep=0.75pt]   [align=left] {1+$\frac{7J}{9}$};

\end{tikzpicture}
}
    \caption{The hierarchical factor structure in the simulation study.}
    \label{fig:sim example}
\end{figure}

We consider 8 simulation settings, given by the combinations of $J = 36, 54$, two sample sizes, $ N = 500, 2000$ and two generating processes of $\Psi^{*}$. For each setting, 100 independent simulations are generated. The results of learning the hierarchical factor structure and estimating the model parameters are shown in 
Tables~\ref{tab:hier std1emc} and \ref{tab:hier std1lmc}. In these tables,  $\bar{K}$ and $\bar{T}$ report the average values of $\hat{K}$ and $\hat{T}$, respectively, and $|\hat{L}_2|$, $|\hat{L}_3|$ and $|\hat{L}_4|$ report the average numbers of factors in $\hat{L}_2$, $\hat{L}_3$ and $\hat{L}_4$, respectively. As shown in Table~\ref{tab:hier std1emc}, the proposed method can accurately recover the true hierarchical factor structure more than 97\% of the time under all the settings, with the highest accuracy of 100\% achieved under the setting with $J= 36$ and heterogeneous diagonal entries in the unique variance matrix.  The MSE of $\hat{\Lambda}$ and $\hat{\Psi}$ show that the loading matrix and the unique variance matrix are accurately estimated when the hierarchical structure is correctly learned.

\begin{table}[tb!]
    \centering
    \caption{The accuracy of the overall estimates of hierarchical structure and parameters.}
    \label{tab:hier std1emc}

    \begin{tabular}{cccccccc}
    \toprule
    $\Psi$ &$J$ & $N$ & $\bar{K}$ &  $\bar{T}$ & EMC & $\text{MSE}_{\hat \Lambda}$ & $\text{MSE}_{\hat \Psi}$\\
    \midrule
    Identity &36&500 & 10.01 & 4.00 & 0.98  &  $2.90\times 10^{-3}$&$1.54\times 10^{-2}$ \\
     && 2000 & 10.04 & 4.00 & 0.97 &$0.74\times 10^{-3}$ & $3.99\times 10^{-3}$ \\
     &54&500 & 10.05 & 4.00 &0.97 & $2.65\times 10^{-3}$  & $6.45\times 10^{-3}$  \\
    &&2000 & 10.02 & 4.00 & 0.99 & $0.66\times 10^{-3}$& $1.63\times 10^{-3}$\\

    Heterogeneous & 36&500 & 10.00 & 4.00 & 1.00  &$3.34\times 10^{-3}$ & $1.45\times 10^{-2}$ \\
     && 2000 & 10.00 & 4.00 & 1.00 & $0.80\times 10^{-3}$ & $3.15\times 10^{-3}$ \\
     &54&500 & 10.01 & 4.00  & 0.99& $2.69\times 10^{-3}$   &  $7.99\times 10^{-3}$ \\
    &&2000 & 10.04 & 4.00 & 0.98 & $0.67\times 10^{-3}$& $2.10\times 10^{-3}$\\
    \bottomrule
    \end{tabular}
    
\end{table}

\begin{table}[tb!]
    \centering
    \caption{The accuracy of the estimated hierarchical structure on each layer.}
    \label{tab:hier std1lmc}
    \begin{tabular}{ccccccccc}
    \toprule
    $\Psi$ &  $J$ & $N$ & $|\hat{L}_2|$ &  $\text{LMC}_{2}$ & $|\hat{L}_3|$ &  $\text{LMC}_{3}$ & $|\hat{L}_4|$ &  $\text{LMC}_{4}$\\
    \midrule
    Identity & 36 & 500 & 2.00 & 1.00 & 4.99 & 0.98 & 2.02 & 0.99 \\
    & & 2000 & 2.00 & 1.00 & 4.97 & 0.97 & 2.07 & 0.97 \\
    &54 & 500 & 2.00 & 1.00 & 4.97 & 0.97  & 2.08 & 0.97 \\
    & & 2000 & 2.00 &  1.00 & 4.99 &  0.99 & 2.03  & 0.99\\
     Heterogeneous & 36& 500 & 2.00  & 1.00  & 5.00   & 1.00 & 2.00 & 1.00   \\
     && 2000 & 2.00  & 1.00  & 5.00   & 1.00 & 2.00 & 1.00  \\
     &54&500 & 2.01  & 0.99  & 5.00 & 0.99  & 2.00 & 1.00   \\
    &&2000 & 2.00 & 0.99  & 4.99 & 0.98  & 2.05 & 0.98 \\
    \bottomrule
    \end{tabular}
\end{table}

\section{Real Data Analysis}\label{sec:real data}
We apply the exploratory hierarchical factor analysis to a personality assessment dataset based on the
International Personality Item Pool (IPIP) NEO 120 personality inventory \citep{johnson2014measuring}. We investigate the structure of the Agreeableness scale based on a sample of 1655 UK participants aged between 30 and 40 years. This scale consists of 24 items, which are designed to measure six facets of Agreeableness, including 
Trust (A1), Morality (A2), Altruism (A3), Cooperation (A4), Modesty (A5), and Sympathy (A6). The responses to all the items are recorded on a 1-5 Likert scale and treated as continuous variables. The reversely worded items have been reversely scored so that a larger score always means a higher level of agreeableness. There is no missing data. 
Detailed descriptions of the items can be found in
the Appendix~\ref{appen:ItemKey}.
The learned hierarchical factor structure, which has four layers and ten factors, is shown in Figure~\ref{fig:real data}, and the estimated loading matrix $\hat{\Lambda}$ is shown in Table~\ref{tab:loading real data}.

\begin{figure}[h]
\centering

\tikzset{every picture/.style={line width=0.75pt}} 

\begin{tikzpicture}[x=0.75pt,y=0.75pt,yscale=-1,xscale=1]

\draw   (389.5,161) -- (410.5,161) -- (410.5,181) -- (389.5,181) -- cycle ;
\draw   (90,82) -- (111,82) -- (111,102) -- (90,102) -- cycle ;
\draw   (40,121) -- (61,121) -- (61,141) -- (40,141) -- cycle ;
\draw   (190,40) -- (211,40) -- (211,60) -- (190,60) -- cycle ;
\draw   (310,161) -- (331,161) -- (331,181) -- (310,181) -- cycle ;
\draw   (350,121) -- (371,121) -- (371,141) -- (350,141) -- cycle ;
\draw   (230,121) -- (251,121) -- (251,141) -- (230,141) -- cycle ;
\draw   (141.5,121) -- (162.5,121) -- (162.5,141) -- (141.5,141) -- cycle ;
\draw   (289,82) -- (310,82) -- (310,102) -- (289,102) -- cycle ;
\draw   (90,121) -- (111,121) -- (111,141) -- (90,141) -- cycle ;
\draw    (200.5,60) -- (200.5,71) ;
\draw    (100.5,71) -- (299.5,71) ;
\draw    (100.5,71) -- (100.5,82) ;
\draw    (299.5,71) -- (299.5,82) ;
\draw    (100.5,102) -- (100.5,121) ;
\draw    (50.5,111) -- (152,111) ;
\draw    (50.5,111) -- (50.5,121) ;
\draw    (152,111) -- (152,121) ;
\draw    (240.5,111) -- (360.5,111) ;
\draw    (240.5,111) -- (240.5,121) ;
\draw    (360.5,111) -- (360.5,121) ;
\draw    (360.5,141) -- (360.5,151) ;
\draw    (320.5,151) -- (400,151) ;
\draw    (320.5,151) -- (320.5,161) ;
\draw    (400,151) -- (400,161) ;
\draw    (299.5,102) -- (299.5,111) ;

\draw (192,43) node [anchor=north west][inner sep=0.75pt]   [align=left] {$\displaystyle \hat{v}_{1}$};
\draw (452,41) node [anchor=north west][inner sep=0.75pt]   [align=left] {$\displaystyle L_{1}$};
\draw (452,83) node [anchor=north west][inner sep=0.75pt]   [align=left] {$\displaystyle L_{2}$};
\draw (452,122) node [anchor=north west][inner sep=0.75pt]   [align=left] {$\displaystyle L_{3}$};
\draw (452,165) node [anchor=north west][inner sep=0.75pt]   [align=left] {$\displaystyle L_{4}$};
\draw (232,124) node [anchor=north west][inner sep=0.75pt]   [align=left] {$\displaystyle \hat{v}_{7}$};
\draw (291,85) node [anchor=north west][inner sep=0.75pt]   [align=left] {$\displaystyle \hat{v}_{3}$};
\draw (352,125) node [anchor=north west][inner sep=0.75pt]   [align=left] {$\displaystyle \hat{v}_{8}$};
\draw (92,124) node [anchor=north west][inner sep=0.75pt]   [align=left] {$\displaystyle \hat{v}_{5}$};
\draw (42,125) node [anchor=north west][inner sep=0.75pt]   [align=left] {$\displaystyle \hat{v}_{4}$};
\draw (312,164) node [anchor=north west][inner sep=0.75pt]   [align=left] {$\displaystyle \hat{v}_{9}$};
\draw (143.5,125) node [anchor=north west][inner sep=0.75pt]   [align=left] {$\displaystyle \hat{v}_{6}$};
\draw (92,85) node [anchor=north west][inner sep=0.75pt]   [align=left] {$\displaystyle \hat{v}_{2}$};
\draw (389,165) node [anchor=north west][inner sep=0.75pt]   [align=left] {$\displaystyle \hat{v}_{10}$};

\end{tikzpicture}
\caption{The hierarchical factor structure from the real data analysis}
\label{fig:real data}
\end{figure}

\begin{table}[ht!]
    \centering
    \caption{The estimated loading matrix $\hat{\Lambda}$ with four layers and ten factors.}
    \label{tab:loading real data}
    \scalebox{0.75}{
    \begin{tabular}{rrrrrrrrrrrr}
    \toprule
    Item & Facet & $F_1$ & $F_2$ &$F_3$ &$F_4$ &$F_5$ &$F_6$ &$F_7$ &$F_8$ &$F_9$ &$F_{10}$ \\
    \midrule
     1 &A1 & 0.47 & 0.14 & 0 & 0.70 & 0 & 0 & 0 & 0 & 0 & 0 \\ 
     2& A1  & 0.36 & 0.23 & 0 & 0.59 & 0 & 0 & 0 & 0 & 0 & 0 \\ 
     3 &A1 &   0.32 & 0.22 & 0 & 0.69 & 0 & 0 & 0 & 0 & 0 & 0 \\ 
     4 &A1 &   0.59 & 0.11 & 0 & 0.64 & 0 & 0 & 0 & 0 & 0 & 0 \\ 
     5 &A2 &   0.44 & 0 & 0.55 & 0 & 0 & 0 & 0.61 & 0 & 0 & 0 \\ 
     6 &A2 &   0.46 & 0 & 0.27 & 0 & 0 & 0 & 0.34 & 0 & 0 & 0 \\ 
     7 &A2 &   0.56 & 0 & 0.42 & 0 & 0 & 0 & 0.61 & 0 & 0 & 0 \\ 
     8 &A2 &   0.45 & 0 & 0.21 & 0 & 0 & 0 & 0 & $-$0.10 & 0.05 & 0 \\ 
    9 &A3 &    0.26 & 0.37 & 0 & 0 & 0.48 & 0 & 0 & 0 & 0 & 0 \\ 
     10 &A3 &   0.26 & 0.54 & 0 & 0 & 0.16 & 0 & 0 & 0 & 0 & 0 \\ 
     11 &A3 &   0.46 & 0.51 & 0 & $-$0.11 & 0 & 0 & 0 & 0 & 0 & 0 \\
     12 &A3 &   0.43 & 0.34 & 0 & 0 & 0.21 & 0 & 0 & 0 & 0 & 0 \\ 
     13 &A4 &   0.21 & 0 & 0.48 & 0 & 0 & 0 & 0 & $-$0.02 & 0 & 0.42 \\ 
     14 &A4 &   0.46 & 0 & 0.14 & 0 & 0 & 0 & 0 & $-$0.15 & 0 & 0.66 \\ 
     15 &A4 &   0.63 & 0 & 0.21 & 0 & 0 & 0 & 0 & $-$0.00 & 0 & 0.48 \\ 
     16 &A4 & 0.57 & 0 & 0.34 & 0 & 0 & 0 & 0 & $-$0.21 & 0 & 0.20 \\ 
     17 &A5 &   0.36 & 0 & 0.43 & 0 & 0 & 0 & 0 & 0.68 & $-$0.06 & 0 \\ 
     18 &A5 &   $-$0.09 & 0 & 0.46 & 0 & 0 & 0 & 0 & 0.70 & 0.48 & 0 \\ 
     19 &A5 &    0.06 & 0 & 0.49 & 0 & 0 & 0 & 0 & 0.86 & 0.41 & 0 \\ 
     20 &A5 &   0.29 & 0 & 0.43 & 0 & 0 & 0 & 0 & 0.15 & 0 & 0.13 \\ 
     21 &A6 &   0.23 & 0.44 & 0 & 0 & 0 & 0.75 & 0 & 0 & 0 & 0 \\ 
     22 &A6 &   0.22 & 0.56 & 0 & 0 & 0 & 0.41 & 0 & 0 & 0 & 0 \\ 
     23 &A6 &   0.41 & 0.57 & 0 & $-$0.04 & 0 & 0 & 0 & 0 & 0 & 0 \\ 
     24 &A6 &   0.34 & 0.40 & 0 & 0 & 0 & 0.38 & 0 & 0 & 0 & 0 \\ 
     \bottomrule
    \end{tabular}
    }
\end{table}

We now examine the learned model. We notice that the loadings on Factor 1 are all positive, except for item 18, which has a small negative loading. Thus, Factor 1 may be interpreted as a general Agreeableness factor. Factor 2 is loaded positively by all items designed to measure the Trust, Altruism, and Sympathy facets. Therefore, it may be interpreted as a higher-order factor of these facets. 
Factors 4--6 are child factors of Factor 2, and based on the loading patterns, they may be interpreted as the Trust, Altruism, and Sympathy factors, respectively. It is worth noting that items 11 (``Am indifferent to the feelings of others") and 23 (``Am not interested in other people’s problems"), which are designed to measure the Altruism and Sympathy facets, now load weakly and negatively on Factor 4 rather than their corresponding factors. 

Factor 3 is another child factor of Factor 1. It is loaded with items designed to measure the facets of Morality, Cooperation, and Modesty. As all the nonzero loadings on Factor 3 are positive, it can be interpreted as 
a higher-order factor of morality, cooperation, and modesty. Factor 7 is the child factor of Factor 3. It is positively loaded by three items designed to measure the Morality facet, and can be interpreted accordingly. Factor 8 is another child factor of Factor 3.  It is loaded positively by all the items designed to measure the Modesty facet 
and negatively, although relatively weakly, by all the items designed to measure the Cooperation facet, and item 8 
(``Obstruct others’ plans") that is designed to measure the Morality facet, but is closely related in concept to cooperation.  
Thus, we can treat Factor 8 as a higher-order factor of modesty and weak aggression (the opposite of cooperation). Finally, Factors 9 and 10 are child factors of Factor 8. Factor 10 may be interpreted as a cooperation factor, while Factor 9 seems to be a weak modesty factor.

Finally, we compare the learned hierarchical factor model with several alternative models based on the Bayesian Information Criterion \citep[BIC;][]{schwarz1978estimating}, including
\begin{enumerate}
    \item (CFA) A six-factor confirmatory factor analysis model with correlated factors. Each factor corresponds to a facet of Agreeableness, loaded by the four items designed to measure this facet. 
    \item (CBF) A confirmatory bi-factor model with one general factor and six group factors, where the group factors are allowed to be correlated. Each group factor corresponds to a facet of Agreeableness, loaded by the four items designed to measure this facet. 
    \item (EBF) An exploratory bi-factor model with one general factor and six group factors, where the group factors are allowed to be correlated. The bi-factor structure is learned using the method proposed in \cite{Qiao_Chen_Ying_2025}. Specifically,   exploratory bi-factor models with $2, 3, \ldots, 12$ group factors are considered, among which the one with six group factors is selected based on the BIC.
     
\end{enumerate}
Table~\ref{tab:hier rd bic} presents the BIC values of all the models, where the model labeled HF is the learned hierarchical factor model. From the results of BIC, the proposed hierarchical factor model fits the data best. Detailed results on the estimated loading matrix and the estimated correlation matrix of the competing models are shown in Appendix~\ref{append:rd compete}.

\begin{table}[ht!]
    \centering
    \caption{The BICs of the hierarchical factor model and the competing models }
    \label{tab:hier rd bic}
    \begin{tabular}{ccccc}
    \toprule
     & HF & CFA & CBF & EBF\\
    \midrule
        BIC & 102,987.54 & 103,841.48 & 103,200.42 & 103,026.10\\
        \bottomrule
    \end{tabular}
    
\end{table}

\section{Discussions} \label{sec:diss}
This paper proposes a divide-and-conquer method with theoretical guarantees for exploring the underlying hierarchical factor structure of the observed data. The method divides the problem into learning the factor structure from the general factor to finer-grained factors. It is computationally efficient, achieved through a greedy search algorithm and an augmented Lagrangian method.  To our knowledge, this is the first statistically consistent method for 
exploratory hierarchical factor analysis that goes beyond the bifactor setting. Our simulation study shows that our method can accurately recover models with up to four factor layers, ten factors, and 54 items under practically reasonable sample sizes, suggesting that it may be suitable for various applications in psychology, education, and related fields. The proposed method is further applied to data from an Agreeableness personality scale, which yields a sensible model with four layers and ten factors that are all psychologically interpretable.  

It is worth noting that the current work assumes that all the factors are orthogonal. 
Mathematically, it is possible to relax this assumption, though certain constraints are still needed. Specifically, two factors need to be orthogonal if one is a descendant of the other. Otherwise, the model is not identifiable due to rotational indeterminacy. For factors without such a relationship, correlations may be allowed. For example, in the exploratory bi-factor analysis in 
\cite{Qiao_Chen_Ying_2025}, which concerns hierarchical factor models with two factor layers,
factors within the second layer are allowed to be correlated. 
However, we should note that relaxing the orthogonality assumption can make the model
less interpretable. Under the orthogonality assumption, the dependence
between two variables is solely due to the shared factors. Such simple interpretations are important when the hierarchical factor model has
a complex structure (e.g., with many factor layers), which is probably why all the existing hierarchical factor models, except for some special bi-factor models,  adopt this orthogonality assumption. Therefore, it may not be worth extending the current theory and method to a more general setting with correlated factors, even though it is possible.

The current method also assumes that a general factor exists and includes it in the first factor layer. However, this may not always be the case. For example, in psychology, there is still a debate about whether a general factor of personality exists \citep[see, e.g.,][]{revelle2013general}.  In cases where we are unsure about the presence of a general factor, the current method can be easily modified to estimate a hierarchical factor model without a general factor, which can be achieved by modifying the first step of Algorithm~\ref{algo:divide and conquer}. 

The current method and asymptotic theory consider a relatively low-dimensional setting where the number of variables $J$ is treated as a constant that does not grow with the sample size. However, in some large-scale settings, $J$ can be on a scale of hundreds or even larger, so it may be better to treat it as a diverging term rather than a fixed constant. In that case, a larger penalty term may be required in the information criterion to account for the larger parameter space, and the asymptotic analysis needs to be modified accordingly.  

Finally, the current work focuses on linear hierarchical factor models, which are suitable for continuous variables. In many applications of hierarchical factor models, we often encounter categorical data (e.g., binary, ordinal, and nominal) that may be better analyzed with non-linear factor models \citep[see, e.g.,][]{chen2020structured}. {We believe it is possible to extend the current framework to the exploratory analysis of non-linear hierarchical factor models. In particular, building upon recent advances in the generalized latent factor model \citep[e.g.,][]{cui2025identifiability}, our approach can be generalized to non-linear hierarchical factor models through likelihood-based estimation, subject to appropriate constraints on both factors and loadings.} 

\clearpage

\vspace{2cm}
\appendix
\noindent
{\Large Appendix}
\numberwithin{equation}{section}
\numberwithin{figure}{section}
\numberwithin{table}{section}

\bigskip

In the Appendix, we provide technical proofs of all theoretical results, additional simulation studies, and further details of the real data analysis presented in the main paper. In particular, Section \ref{appen:proof id} provides the proof of Theorem~\ref{thm:identifiability}, Section~\ref{appen:proof prop C4} proves Proposition~\ref{prop:structure C4}, Section~\ref{append: discussion on condition 3} discusses how Condition~\ref{cond:rank} of Theorem~\ref{thm:identifiability} can be relaxed under a simple hierarchical factor structure, Section~\ref{appen:proof consistency} establishes Theorem~\ref{cor:consistency}, Section~\ref{appen:simulation algo3} presents numerical results demonstrating the convergence of Algorithm~\ref{alg:main} to the global solution from multiple random starting points, Section~\ref{appen:simulation algo3 under} shows the numerical results of Algorithms~\ref{algo:divide and conquer} and \ref{algo:ICB} in learning the hierarchical factor structure when $c_{\max}$ is underspecified, Section~\ref{appen:ItemKey} provides the construct of the data discussed in the real data analysis, and Section~\ref{append:rd compete} presents the numerical results of the alternative models discussed in the real data analysis.

\section{Proof of Theorem~\ref{thm:identifiability}}\label{appen:proof id}
In this section, we give the proof of Theorem~\ref{thm:identifiability}. For simplicity of notation, for any matrix $A\in\mathbb{R}^{m\times n}$, $\mathcal{S}_1\subset\{1,\ldots,m\}$ and $\mathcal{S}_2\subset\{1,\ldots,n\}$, we denote by $A_{[\mathcal{S}_1,:]} = A_{[\mathcal{S}_1,\{1,\ldots,n\}]}$ and $A_{[:,\mathcal{S}_2]} = A_{[\{1,\ldots,m\},\mathcal{S}_2]}$.
\begin{proof}
Suppose that there exists a hierarchical factor model satisfying the constraints C1-C4, and the corresponding loading matrix $\Lambda$ and the unique variance matrix $\Psi$ satisfy $\Sigma = \Lambda\Lambda^{\top} + \Psi$ and $\Sigma = \Sigma^*$. We prove Theorem~\ref{thm:identifiability} by induction on the depth of the hierarchy. It suffices to prove that $\text{Ch}_1 = \text{Ch}^{*}_1$, $v_k = v_{k}^*$ for all $k\in \text{Ch}^{*}_1$ and $\boldsymbol{\lambda}_1 = \boldsymbol{\lambda}^{*}_1$ or $\boldsymbol{\lambda}_1 = -\boldsymbol{\lambda}^{*}_1$ hold, where $v_1,\ldots,v_K$ are the corresponding sets of variables for each factor according to $\Lambda$, $\text{Ch}_1, \ldots, \text{Ch}_K$ are the child factors of each factor according to the hierarchical factor model given $\Lambda$, and $\boldsymbol{\lambda}_1$ and $\boldsymbol{\lambda}^{*}_1$ are the first columns of $\Lambda$ and $\Lambda^*$ respectively.

First, we establish that for each $k\in \text{Ch}_{1}^{*}$, there exits $i\in \text{Ch}_{1}$ such that $v_{k}^{*}\subset v_{i}$. By Condition~\ref{cond:seperate}, we have $\Lambda\Lambda^{\top}= \Lambda^{*}(\Lambda^{*})^{\top}$ and $\Psi = \Psi^{*}$. If $\text{Ch}^{*}_{1} = \emptyset$, the result holds trivially. Otherwise, suppose $\text{Ch}^{*}_{1} \neq \emptyset$. For any $k\in \text{Ch}^{*}_{1}$, define $\mathcal{B}_{k,i} = v^{*}_{k}\cap v_{i}, i\in \text{Ch}_{1}$. \\
If $\text{Ch}^{*}_{k} = \emptyset$, consider the following cases:
\begin{enumerate}
\item We have $|\{i\in \text{Ch}_{1}: |\mathcal{B}_{k,i}|\geq 1 \}| \geq 4$, which implies the existence of four distinct factors $i_1$, $i_2$, $i_3$, $i_4$ such that $v_{i_j}\cap v^{*}_{k}\neq \emptyset$ for $j=1,\ldots, 4$. In this case, choose $j_1 \in \mathcal{B}_{k,i_1}, \ldots, j_4 \in \mathcal{B}_{k,i_4}$. Consider $\Sigma_{[\{j_1,j_2\}, \{j_{3},j_4\}]} = \Sigma^{*}_{[\{j_1,j_2\}, \{j_{3},j_4\}]}$, which is equivalent to
\begin{equation}\label{eq:hier id thm case1 case1}
\Lambda_{[\{j_1,j_2\}, \{1\}]}(\Lambda_{[\{j_3,j_4\}, \{1\}]})^{\top} = \Lambda^{*}_{[\{j_1,j_2\}, \{1,k\}]}(\Lambda^{*}_{[\{j_3,j_4\}, \{1,k\}]})^{\top}.
\end{equation}
Observe that the left-hand side of~\eqref{eq:hier id thm case1 case1} has rank at most 1, whereas, by Condition~\ref{cond:rank}, the right-hand side has rank 2. This contradicts~\eqref{eq:hier id thm case1 case1}. Hence, this case cannot occur.
\item There exist $i_1$ and $i_2 \neq i_1$ such that $|\mathcal{B}_{k,i_{1}}|\geq 2$ and $|\mathcal{B}_{k,i_{2}}|\geq 1$. In this case, choose distinct $j_{1}, j_{2}\in \mathcal{B}_{k,i_{1}}$ and $j_3 \in \mathcal{B}_{k,i_{2}}$. Consider $\Sigma_{[\{j_1, j_2, j_3\}, \{j_{1}, j_2, j_3\}]} = \Sigma^{*}_{[\{j_1, j_2, j_3\}, \{j_{1}, j_2, j_3\}]}$, which is equivalent to
\begin{equation}\label{eq:hier id thm case1 case2}
\Lambda_{[\{j_1, j_2, j_3\},:]}(\Lambda_{[\{j_1, j_2, j_3\},:]})^{\top} = \Lambda^{*}_{[\{j_1, j_2, j_3\},\{1,k\}]}(\Lambda^{*}_{[\{j_1, j_2, j_3\}, \{1,k\}]})^{\top}.
\end{equation}
By Condition~\ref{cond:rank}, the right-hand side of~\eqref{eq:hier id thm case1 case2} has rank 2. Moreover, Condition~\ref{cond:rank} also implies that the submatrix $\Sigma^{*}_{[\{j_1, j_2\}, \{j_{1}, j_2\}]}$ has rank 2, and hence the matrix $\Lambda_{[\{j_1, j_2\},:]}$ must have rank 2 as well. 
However, observe that for any $s\in \{i_2\}\cup D_{i_2}$, $\lambda_{j_1,s} = 0$ and $\lambda_{j_2,s} = 0$, whereas $\Lambda_{[\{j_3\},\{i_2\}\cup D_{i_2}]} \neq \mathbf{0}$. Consequently, $\Lambda_{[\{j_1, j_2, j_3\},:]}$ has rank 3. Thus, the left-hand side of~\eqref{eq:hier id thm case1 case2} has rank 3, which contradicts ~\eqref{eq:hier id thm case1 case2}. Therefore, this case cannot occur.

\item $|v_{k}^{*}|= 3$, and there exist distinct $i_1,i_2,i_3$ such that $|\mathcal{B}_{k,i_{1}}| = |\mathcal{B}_{k,i_{2}}| = |\mathcal{B}_{k,i_{3}}| = 1$. Let $\{j_1\} = \mathcal{B}_{k,i_{1}}$, $\{j_2\} = \mathcal{B}_{k,i_{2}}$, and $\{j_3\} = \mathcal{B}_{k,i_{3}}$. Consider $$\Sigma_{[\{j_1, j_2, j_3\}, \{j_{1}, j_2, j_3\}]} = \Sigma^{*}_{[\{j_1, j_2, j_3\}, \{j_{1}, j_2, j_3\}]},$$ which is equivalent to~\eqref{eq:hier id thm case1 case2}. In this case, the left-hand side of~\eqref{eq:hier id thm case1 case2} has rank 3, whereas, by Condition~\ref{cond:rank}, the right-hand side has rank 2. This contradicts~\eqref{eq:hier id thm case1 case2}. Hence, this case cannot occur.

\item There exists a unique $i\in \text{Ch}_{1}$ such that $\mathcal{B}_{k,i_{1}} = v_{k}^{*}$, which indicates that $v_{k}^{*} \subset v_{i}$. 
\end{enumerate}
When $\text{Ch}^{*}_{k} \neq \emptyset$, consider the following cases:
\begin{enumerate}
\item There exist $s\in \text{Ch}^{*}_{k}$ and $i\in \text{Ch}_{1}$ such that $|\mathcal{B}_{k,i}\cap v_{s}^{*}|\geq 2$. In this case, we claim that
\begin{equation}\label{eq:hier id thm case2 claim1}
|(\cup_{i'\neq i,i'\in \text{Ch}_{1}}\mathcal{B}_{k,i'}) \cap (\cup_{s'\neq s, s'\in \text{Ch}^{*}_{k}}v_{s'}^{*})|\leq 1,
\end{equation}
Otherwise, choose $j_1, j_2 \in \mathcal{B}_{k,i}\cap v_{s}^{*}$ and $j_3, j_4 \in (\cup_{i'\neq i,i'\in \text{Ch}_{1}}  \mathcal{B}_{k,i'}) \cap (\cup_{s'\neq s, s'\in \text{Ch}^{*}_{k}}v_{s'}^{*})$. Consider $\Sigma_{[\{j_1,j_2\}, \{j_{3},j_4\}]} = \Sigma^{*}_{[\{j_1,j_2\}, \{j_{3},j_4\}]}$, which is equivalent to~\eqref{eq:hier id thm case1 case1}. 
The left-hand side of~\eqref{eq:hier id thm case1 case1} has rank 1, whereas by Condition~\ref{cond:rank}, the right-hand side has rank 2. This contradicts~\eqref{eq:hier id thm case1 case1}, and thus the claim in~\eqref{eq:hier id thm case2 claim1} holds.

Now observe that $|v^{*}_{s'}|\geq 3$ for all $s'\neq s, s'\in \text{Ch}^{*}_{k}$. Combined with~\eqref{eq:hier id thm case2 claim1}, $|\mathcal{B}_{k,i}\cap v_{s'}^{*}|\geq 2$ for all $s'\in \text{Ch}^{*}_{k}$. By an analogous argument, we also have, 
\begin{equation}\label{eq:hier id thm case2 claim2}
|(\cup_{i'\neq i,i'\in \text{Ch}_{1}}\mathcal{B}_{k,i'}) \cap v_{s}^{*}|\leq 1,
\end{equation}
holds. Combining~\eqref{eq:hier id thm case2 claim1} with~\eqref{eq:hier id thm case2 claim2} yields
\begin{equation}\label{eq:hier id thm case2 claim3}
|(\cup_{i'\neq i,i'\in \text{Ch}_{1}}\mathcal{B}_{k,i'}) \cap (\cup_{s'\in \text{Ch}^{*}_{k}}v_{s'}^{*})|\leq 2.
\end{equation}
We now analyze the possible values of the left-hand side of~\eqref{eq:hier id thm case2 claim3}. If $|(\cup_{i'\neq i,i'\in \text{Ch}_{1}}\mathcal{B}_{k,i'}) \cap (\cup_{s'\in \text{Ch}^{*}_{k}}v_{s'}^{*})| = 2$, there exists some $s'\neq s$ such that~\eqref{eq:hier id thm case2 claim1} is tight. We choose $j_1, j_2\in \mathcal{B}_{k,i}\cap v_{s}^{*}$, $j_3, j_4\in \mathcal{B}_{k,i}\cap v_{s'}^{*}$, $j_5 \in (\cup_{i'\neq i,i'\in \text{Ch}_{1}}\mathcal{B}_{k,i'}) \cap v_{s}^{*}$ and $j_6\in (\cup_{i'\neq i,i'\in \text{Ch}_{1}}\mathcal{B}_{k,i'}) \cap v_{s'}^{*}$.  Furthermore, when $ \text{Ch}^{*}_{s} \neq \emptyset$, we require that $j_1, j_2$ belong to different child factors of factor $s$ with $j_5$. Similarly, when $ \text{Ch}^{*}_{s'} \neq \emptyset$, we require that $j_3, j_4$ belong to different child factors of factor $s'$ with $j_6$. Such a choice is always possible due to the assumed structure of the hierarchical model. Now consider $\Sigma_{[\{j_1,j_2,j_3,j_4\},\{j_5,j_6\}]} = \Sigma^{*}_{[\{j_1,j_2,j_3,j_4\},\{j_5,j_6\}]}$, which is equivalent to
\begin{equation}\label{eq:hier id thm case2 case1 main1}
\begin{aligned}
&\Lambda_{[\{j_1,j_2,j_3,j_4\},\{1\}]}(\Lambda_{[\{j_5,j_6\},\{1\}]})^{\top} \\
=& \Lambda^{*}_{[\{j_1,j_2,j_3,j_4\},\{1,k,s,s'\}]}(\Lambda^{*}_{[\{j_5,j_6\},\{1,k,s,s'\}]})^{\top},
\end{aligned}
\end{equation}
by the construction of $j_1,\ldots,j_6$. The left-hand side of~\eqref{eq:hier id thm case2 case1 main1} has rank 1. On the other hand, by Condition~\ref{cond:rank}, the matrix $ \Lambda^{*}_{[\{j_1,j_2,j_3,j_4\},\{1,k,s,s'\}]}$ has rank 4, and $\Lambda^{*}_{[\{j_5,j_6\},\{1,k,s,s'\}]}$ has rank 2. By Sylvester's rank inequality~\citep[see, e.g.,][]{horn2012matrix},
\begin{equation*}
\begin{aligned}
 &\text{rank}\big(\Lambda^{*}_{[\{j_1,j_2,j_3,j_4\},\{1,k,s,s'\}]}(\Lambda^{*}_{[\{j_5,j_6\},\{1,k,s,s'\}]})^{\top}\big)\\
\geq &\text{rank}\big(\Lambda^{*}_{[\{j_1,j_2,j_3,j_4\},\{1,k,s,s'\}]}\big) + \text{rank}\big(\Lambda^{*}_{[\{j_5,j_6\},\{1,k,s,s'\}]}\big)-4\\
= &2,
\end{aligned}
\end{equation*}
which contradicts~\eqref{eq:hier id thm case2 case1 main1}. Hence, this case cannot occur.

If $|(\cup_{i'\neq i,i'\in \text{Ch}_{1}}\mathcal{B}_{k,i'}) \cap (\cup_{s'\in \text{Ch}^{*}_{k}}v_{s'}^{*})| = 1$. Without loss of generality, assume $(\cup_{i'\neq i,i'\in \text{Ch}_{1}}\mathcal{B}_{k,i'}) \cap (\cup_{s'\in \text{Ch}^{*}_{k}}v_{s'}^{*}) = \mathcal{B}_{k,i_1} \cap v_{s_1}^{*} = \{j\}$, where $i_1\in  \text{Ch}_{1}, i_1\neq i$ and $s_1\in\text{Ch}^{*}_{k}, s_1\neq s$. Consider $\Sigma_{[v^{*}_{k},v^{*}_{k}]} = \Sigma^{*}_{[v^{*}_{k},v^{*}_{k}]}$, which is equivalent to
\begin{equation}\label{eq:hier id thm case2 case1 main2}
\Lambda_{[v^{*}_{k},:]}(\Lambda_{[v^{*}_{k},:]})^{\top} = \Lambda^{*}_{[v^{*}_{k},\{1,k\}\cup D_{k}^{*}]}(\Lambda^{*}_{[v^{*}_{k},\{1,k\}\cup D_{k}^{*}]})^{\top}.
\end{equation}
By Condition~\ref{cond:rank}, $\Lambda^{*}_{[v^{*}_{k}\setminus\{j\}, \{1,k\}\cup D_{k}^{*}]}$ has rank $2+|D_{k}^{*}|$. Thus, $\Lambda_{[v^{*}_{k}\setminus\{j\}, :]}$ has rank $2+|D_{k}^{*}|$. Since $\Lambda_{[\{j\},\{i_1\}]} \neq 0$ and $\Lambda_{[v_{k}^{*}\setminus\{j\},\{i_1\}]} = \mathbf{0}$, $\Lambda_{[v^{*}_{k}, :]}$ has rank $3+|D_{k}^{*}|$, which contradicts~\eqref{eq:hier id thm case2 case1 main2}. Hence, this case cannot occur.

If $|(\cup_{i'\neq i,i'\in \text{Ch}_{1}}\mathcal{B}_{k,i'}) \cap (\cup_{s'\in \text{Ch}^{*}_{k}}v_{s'}^{*})| = 0$, there exists a unique $i\in \text{Ch}_{1}$ such that $\mathcal{B}_{k,i} = v_{k}^{*}$, which indicates $v_{k}^{*} \subset v_{i}$. 
\item $|\mathcal{B}_{k,i}\cap v_{s}^{*}|\leq 1$ for all $i\in \text{Ch}_{1}$ and $s\in \text{Ch}^{*}_{k}$. If there exist some $i\in \text{Ch}_{1}$ and $s\in \text{Ch}^{*}_{k}$ such that $|\mathcal{B}_{k,i}\cap v_{s}^{*}| = 1$ and $|\mathcal{B}_{k,i}\cap v_{s'}^{*}| = 0$ for all $s'\in \text{Ch}^{*}_{k}, s'\neq s$, assume $\{j\} = \mathcal{B}_{k,i}\cap v_{s}^{*}$. Similar to the proof in~\eqref{eq:hier id thm case2 case1 main2}, the matrices on both sides have unequal ranks. Thus, the assumption does not hold. We assume that there exist $i\in \text{Ch}_{1}$, $s_1\in\text{Ch}^{*}_{k}$ and $s_2\in\text{Ch}^{*}_{k},s_2\neq s_1$ such that $|\mathcal{B}_{k,i}\cap v_{s_1}^{*}|=1$ and $|\mathcal{B}_{k,i}\cap v_{s_2}^{*}|=1$. If there further exists $s_3\in\text{Ch}^{*}_{k}, s_3\neq s_1,s_2$ such that $|\mathcal{B}_{k,i}\cap v_{s_3}^{*}|=0$, we denote by $\{j_1\} = \mathcal{B}_{k,i}\cap v_{s_1}^{*}$ and $\{j_2\} = \mathcal{B}_{k,i}\cap v_{s_2}^{*}$. Consider $\Sigma_{[v_{s_3}^{*},\{j_1,j_2\}]} = \Sigma^{*}_{[v_{s_3}^{*},\{j_1,j_2\}]}$, which is equivalent to
\begin{equation*}
\Lambda_{[v_{s_3}^{*},\{1\}]}(\Lambda_{[\{j_1,j_2\},\{1\}]})^{\top} = \Lambda^{*}_{[v_{s_3}^{*},\{1,k\}]}(\Lambda^{*}_{[\{j_1,j_2\},\{1,k\}]})^{\top}.
\end{equation*}
Noticing that the rank of the matrix on the left side is 1, while according to Condition~\ref{cond:rank}, the rank of the matrix on the right side is 2, the assumption does not hold. 

Thus, for any $i\in\text{Ch}_{1}$, if there exists some $s\in \text{Ch}^{*}_{k}$ such that $|\mathcal{B}_{k,i}\cap v_{s}^{*}| = 1$, then $|\mathcal{B}_{k,i}\cap v_{s}^{*}| = 1$ for all $s\in \text{Ch}^{*}_{k}$, which indicate that $|v_{s}^{*}|$ are the same for $s\in \text{Ch}^{*}_{k}$. If $|\text{Ch}_{k}^{*}|\geq 3$, let $s_1,s_2,s_3\in \text{Ch}_{k}^{*}$ and $i_1,i_2,i_3\in \text{Ch}_{1}$ such that $\{j_1\} = \mathcal{B}_{k,i_1}\cap v_{s_1}^{*}$, $\{j_2\} = \mathcal{B}_{k,i_2}\cap v_{s_1}^{*}$, $\{j_3\} = \mathcal{B}_{k,i_3}\cap v_{s_2}^{*}$, $\{j_4\} = \mathcal{B}_{k,i_3}\cap v_{s_3}^{*}$. Consider $\Sigma_{[\{j_1,j_2\},\{j_3,j_4\}]} = \Sigma^{*}_{[\{j_1,j_2\},\{j_3,j_4\}]}$, which is equivalent to
\begin{equation}
\Lambda_{[\{j_1,j_2\},\{1\}]}(\Lambda_{[\{j_3,j_4\},\{1\}]})^{\top} = \Lambda^{*}_{[\{j_1,j_2\},\{1,k\}]}(\Lambda^{*}_{[\{j_3,j_4\},\{1,k\}]})^{\top}.
\end{equation}
Since the left-hand side has rank 1, while by Condition~\ref{cond:rank}, the right-hand side has rank 2, the assumption does not hold.

Finally, if $|\text{Ch}_{k}^{*}| = 2$, let $\{j_1\} = \mathcal{B}_{k,i_1}\cap v_{s_1}^{*}$, $\{j_2\} = \mathcal{B}_{k,i_1}\cap v_{s_2}^{*}$, $j_3,j_4\in v_{s_1}^{*}, j_3,j_4\neq j_1$ and $j_5,j_6\in v_{s_2}^{*}, j_5,j_6\neq j_2$. Furthermore, when $|\text{Ch}_{s_1}^{*}|\neq 0$, we require that $j_3,j_4$ belong to different child factors of factor $s_1$ with $j_1$. Similarly, when $|\text{Ch}_{s_2}^{*}|\neq 0$, $j_5,j_6$ belong to different child factors of factor $s_2$ with $j_2$. Such a choice is always possible due to the assumed structure of the hierarchical model. Consider $\Sigma_{[\{j_1,j_2\},\{j_3,j_4,j_5,j_6\}]} = \Sigma^{*}_{[\{j_1,j_2\},\{j_3,j_4,j_5,j_6\}]}$, which is equivalent to
\begin{equation}\label{eq:hier id thm case2 worst case}
\begin{aligned}
&\Lambda_{[\{j_1,j_2\},\{1\}]}(\Lambda_{[\{j_3,j_4,j_5,j_6\},\{1\}]})^{\top} \\
=& \Lambda^{*}_{[\{j_1,j_2\},\{1,k,s_1,s_2\}]}(\Lambda^{*}_{[\{j_3,j_4,j_5,j_6\},\{1,k,s_1,s_2\}]})^{\top}.
\end{aligned}
\end{equation}
The left-hand side of~\eqref{eq:hier id thm case2 worst case} has rank 1. On the other hand, by Condition~\ref{cond:rank}, $\Lambda^{*}_{[\{j_3,j_4,j_5,j_6\},\{1,k,s_1,s_2\}]}$ has rank 4 and $\Lambda^{*}_{[\{j_1,j_2\},\{1,k,s_1,s_2\}]}$ has rank 2. By Sylvester's rank inequality,
\begin{equation}
\begin{aligned}
 &\text{rank}\big(\Lambda^{*}_{[\{j_1,j_2\},\{1,k,s_1,s_2\}]}(\Lambda^{*}_{[\{j_3,j_4,j_5,j_6\},\{1,k,s_1,s_2\}]})^{\top}\big)\\
\geq &\text{rank}\big(\Lambda^{*}_{[\{j_1,j_2\},\{1,k,s_1,s_2\}]}\big) + \text{rank}\big(\Lambda^{*}_{[\{j_3,j_4,j_5,j_6\},\{1,k,s_1,s_2\}]}\big)-4\\
= &2,
\end{aligned}
\end{equation}
which contradicts~\eqref{eq:hier id thm case2 worst case}. Thus, the assumption does not hold.
\end{enumerate}

From the previous proof, for any $k\in \text{Ch}^{*}_{1}$, there exists $i\in \text{Ch}_{1}$ such that $v_{k}^{*}\subset v_i$. For any $i\in \text{Ch}_{1}$, define $C_i = \{k\in \text{Ch}^{*}_{1}: v_{k}^{*}\subset v_i\}$. Consider $\Sigma_{[v_i,v_i]} = \Sigma^{*}_{[v_i,v_i]}$, which is equivalent to
\begin{equation*}
 \Lambda_{[v_i,\{1,i\}\cup D_i]}( \Lambda_{[v_i,\{1,i\}\cup D_i]})^{\top} = \Lambda^{*}_{[v_i,\{1\}\cup C_i \cup (\cup_{k\in C_i}D^{*}_k)]}( \Lambda^{*}_{[v_i,\{1\}\cup C_i \cup (\cup_{k\in C_i}D^{*}_k)]})^{\top}.
\end{equation*}
According to Condition~\ref{cond:rank}, the matrix $\Lambda^{*}_{[v_i,\{1\}\cup C_i \cup (\cup_{k\in C_i}D^{*}_k)]}$ has rank $1 + |C_i| + \sum_{k\in C_i} |D^{*}_{k}|$. Thus, we must have $1+|D_i| \geq |C_i| + \sum_{k\in C_i} |D^{*}_{k}|$. Summing both sides over all $i\in\text{Ch}_{1}$, we have
\begin{equation*}
K-1 = \sum_{i\in \text{Ch}_{1}}(1+|D_i|) \geq  \sum_{i\in \text{Ch}_{1}}\left(|C_i| + \sum_{k\in C_i} |D^{*}_{k}|\right) = \sum_{k\in\text{Ch}^{*}_{1}}(1+|D^{*}_{k}|) = K-1.
\end{equation*}
Therefore, 
\begin{equation}\label{eq:hier id thm sum mid}
1+|D_i| = |C_i| + \sum_{k\in C_i} |D^{*}_{k}|,
\end{equation}
for every $i\in \text{Ch}_{1}$. According to Lemma 5.1 of \cite{anderson1956statistical}, there exists an orthogonal matrix $R_i$ such that
\begin{equation}\label{eq:hier id thm sum1}
\Lambda_{[v_i,\{1,i\}\cup D_i]} = \Lambda^{*}_{[v_i,\{1\}\cup C_i \cup (\cup_{k\in C_i}D^{*}_k)]}R_{i}.
\end{equation}
On the other hand, for $i, i' \in \text{Ch}_{1}$, consider $\Sigma_{[v_i, v_{i'}]} = \Sigma^{*}_{[v_i,  v_{i'}]}$, which is equivalent to
\begin{equation}\label{eq:hier id thm sum2}
\Lambda_{[v_i,\{1\}]}(\Lambda_{[v_{i'},\{1\}]})^{\top} = \Lambda^{*}_{[v_i,\{1\}]}(\Lambda^{*}_{[v_{i'},\{1\}]})^{\top}.
\end{equation}
Combining~\eqref{eq:hier id thm sum1} with~\eqref{eq:hier id thm sum2}, either $\Lambda_{[v_i,\{1\}]} = \Lambda^{*}_{[v_i,\{1\}]}$ or $\Lambda_{[v_i,\{1\}]} = -\Lambda^{*}_{[v_i,\{1\}]}$ holds. Without loss of generality, we assume $\Lambda_{[v_i,\{1\}]} = \Lambda^{*}_{[v_i,\{1\}]}$, which further implies $\boldsymbol{\lambda}_1 = \boldsymbol{\lambda}^{*}_1$. 

It remains to show that $|C_i| = 1$ for all $i\in \text{Ch}_{1}$. Suppose, for contradiction, that there exists some $i\in \text{Ch}_{1}$ such that $|C_i| \geq 2$. Since $|D_{i}|\geq 2$, for $s_1, s_2\in \text{Ch}_{i}$, there exist $k_1, k_2 \in C_i$ such that $v_{s_1} \cap v^{*}_{k_1} \neq\emptyset$ and $v_{s_2} \cap v^{*}_{k_2} \neq\emptyset$. Consider $\Sigma_{[v_{s_1} \cap v^{*}_{k_1} , v_{s_2} \cap v^{*}_{k_2} ]} = \Sigma^{*}_{[v_{s_1} \cap v^{*}_{k_1} , v_{s_2} \cap v^{*}_{k_2} ]}$. Combined with $\Lambda_{[v_{s_1} \cap v^{*}_{k_1},\{1\}]} = \Lambda^{*}_{[v_{s_1} \cap v^{*}_{k_1},\{1\}]}$ and $\Lambda_{[v_{s_2} \cap v^{*}_{k_2},\{1\}]} = \Lambda^{*}_{[v_{s_2} \cap v^{*}_{k_2},\{1\}]}$, we have
\begin{equation*}
\Lambda_{[v_{s_1} \cap v^{*}_{k_1},\{i\}]}(\Lambda_{[v_{s_2} \cap v^{*}_{k_2},\{i\}]})^{\top} = \mathbf{0}.
\end{equation*}
Consequently, $\Lambda_{[v_{s_1} \cap v^{*}_{k_1},\{i\}]} = \mathbf{0}$ or $\Lambda_{[v_{s_2} \cap v^{*}_{k_2},\{i\}]} = \mathbf{0}$, which contradicts the definition of $v_{i}$. Thus, $|C_i| = 1$ for all $i\in \text{Ch}_1$. Therefore, we have shown that $\text{Ch}_1 = \text{Ch}^{*}_1$, $v_k = v_{k}^*$ for all $k\in \text{Ch}^{*}_1$. 

Finally, combining $\boldsymbol{\lambda}_1 = \boldsymbol{\lambda}^{*}_1$ with $\Sigma = \Sigma^{*}$, the covariance equality decomposes into $|\text{Ch}^{*}_{1}|$ independent equations 
\begin{equation*}
\Lambda_{[v_{k}^{*},\{k\}\cup D_k]}(\Lambda_{[v_{k}^{*},\{k\}\cup D_k]})^{\top} = \Lambda^{*}_{[v_{k}^{*},\{k\}\cup D^{*}_{k}]}(\Lambda^{*}_{[v_{k}^{*},\{k\}\cup D^{*}_{k}]})^{\top},
\end{equation*}
$k\in \text{Ch}^{*}_1 $. By~\eqref{eq:hier id thm sum mid}, we have $|D_k| = |D^{*}_k|$ for all $k\in \text{Ch}^{*}_1$. Thus, by applying the same argument recursively to factors on the $t$th layer, $t=2,\ldots, T$, we conclude that $\Lambda = \Lambda^{*}Q$ and $\Psi = \Psi^*$ for some sign flip matrix $Q$.
\end{proof}

\section{Proof of Proposition~\ref{prop:structure C4}}\label{appen:proof prop C4}
In this section, we give the proof of Proposition~\ref{prop:structure C4}.
\begin{proof}
Since factor $j$ and its descendant factors construct a hierarchical factor structure that satisfies constraint C1-C4, it suffices to prove that
\begin{equation}\label{eq: proof prop 2 goal}
|v_{1}^{*}|\geq 3+ |D_{1}^{*}|.
\end{equation}

Let $L_{t}$ be the factors that belong to the $t$th layer, $t=1,\ldots,T$. We divide $L_t$ into $L_{t}^{(1)} = \{k\in L_{t}: \mbox{Ch}_{k}^{*} \neq \emptyset \}$ and $L_{t}^{(2)} = \{k\in L_{t}: \mbox{Ch}_{k}^{*} = \emptyset \}$ for $t=2,\ldots,T$ so that $L_{t}^{(1)}\cup L_{t}^{(2)} = L_{t}$ and $L_{t}^{(1)}\cap L_{t}^{(2)} = \emptyset$. By definition, $L_{T}^{(1)} = \emptyset$. By constraint C3, we first have 
\begin{equation}\label{eq:proof prop 2 sum L 1}
|L_{t}^{(1)}| \leq \left\lfloor\frac{1}{2}|L_{t+1}|\right\rfloor = \left\lfloor\frac{1}{2}|L_{t+1}^{(1)}| +  \frac{1}{2}|L_{t+1}^{(2)}|\right\rfloor, t=2,\ldots,T-1.
\end{equation}
Iterating~\eqref{eq:proof prop 2 sum L 1} for $t+1\leq j\leq T-1$ yields
$$
|L_{t}^{(1)}| \leq \sum_{j=t+1}^{T}\frac{1}{2^{j-t}}|L_{j}^{(2)}|.
$$
Consequently,
\begin{equation}\label{eq:proof prop 2 sum D}
\begin{aligned}
|D_{1}^{*}| =& \sum_{t=2}^{T}|L_{t}|\\
=& \sum_{t=2}^{T}|L_{t}^{(1)}| + |L_{t}^{(2)}|\\
\leq & \sum_{t=2}^{T} (|L_{t}^{(2)}| + \sum_{j=t+1}^{T}\frac{1}{2^{j-t}}|L_{j}^{(2)}|)\\
=&\sum_{t=2}^{T}(2-\frac{1}{2^{t-2}})|L_{t}^{(2)}|\\
<&2\sum_{t=2}^{T}|L_{t}^{(2)}|.
\end{aligned}
\end{equation}
On the other hand, constraint C4 implies
\begin{equation}\label{eq:proof prop 2 sum v}
|v_{1}^{*}| \geq 3\sum_{t=2}^{T}|L_{t}^{(2)}|.
\end{equation}
Combining \eqref{eq:proof prop 2 sum D} and \eqref{eq:proof prop 2 sum v}, we have $|v_{1}^{*}|> \frac{2}{3}|D^{*}_{1}|$. In particular, \eqref{eq: proof prop 2 goal} holds when $v_{1}^{*}\geq 9$. When $|v_{1}^{*}| = 7$ or 8, $|D_{1}^{*}|\leq 2$ by constraint C4 and \eqref{eq: proof prop 2 goal} holds. When $3\leq |v_{1}^{*}| \leq 6$, $|D_{1}^{*}|=0$ by constraint C4 and \eqref{eq: proof prop 2 goal} holds.
\end{proof}

\section{Further discussions of Condition~\ref{cond:rank}}\label{append: discussion on condition 3}
In this section, we discuss the identifiability of a bi-factor model with two group factors, which constructs a special case of a two-layer hierarchical factor model. Let $\Lambda^{*}$ and $\Psi^{*}$ be the true loading matrix and the unique variance matrix. Let $v_{1}^{*}, v_{2}^{*}$ and $v_{3}^{*}$ be the sets of variables loading on the general factor (Factor 1) and the two group factors (Factors~2 and~3) such that $v_{1}^{*} = v_{2}^{*}\cup v_{3}^{*}$ and $v_{2}^{*}\cap v_{3}^{*} = \emptyset$. Lemma~\ref{lem:counter} provides a counterexample showing that Conditions~\ref{cond:true para} and the following Condition~\ref{cond: couter sufficient}—the latter being a sufficient condition for Condition~\ref{cond:seperate} under this specific hierarchical structure—are not sufficient to guarantee identifiability of the model. Theorem~\ref{thm: appen bi-factor id} then establishes the identifiability under the additional Condition~\ref{cond: appendix additional}. 

\begin{condition}\label{cond: couter sufficient}
$|v_{g}^{*}|\geq 3$ and $\Lambda^{*}_{[v^{*}_{g},\{1,g\}]}$ is of full rank for $g\in \{2,3\}$. Moreover, there exists $g\in\{2,3\}$ and disjoint partition $E_1$, $E_2$ such that $E_1\cup E_2 = v^{*}_{g}$, $E_1\cap E_2 = \emptyset$, and $\Lambda^{*}_{[E_1,\{1,g\}]}, \Lambda^{*}_{[E_2,\{1,g\}]}$ are of full column rank.
\end{condition}

\begin{lemma}\label{lem:counter}
Suppose that Conditions \ref{cond:true para} and \ref{cond: couter sufficient} hold. There exists another hierarchical factor structure with the loading matrix $\Lambda$ and the unique variance matrix $\Psi$ such that $\Lambda\Lambda^{\top} + \Psi =\Sigma^{*}$.
\end{lemma}
\begin{proof}
It is easy to check that Condition~\ref{cond: couter sufficient} is a sufficient condition for Condition~\ref{cond:seperate}. Thus, we have $\Psi=\Psi^{*}$ and focus on constructing a loading matrix $\Lambda$ that produces the same covariance matrix $\Sigma^*$ but corresponds to a different hierarchical factor structure. The hierarchical factor structure decoded by $\Lambda$ is still a bi-factor structure with two group factors. Let $v_{2}$ and $v_{3}$ be the sets of variables belonging to Factor 2 and 3 according to $\Lambda$.  Moreover, let $\mathcal{B}_{2,2} = v_{2}^{*}\cap v_{2}$, $\mathcal{B}_{2,3} = v_{2}^{*}\cap v_{3}$, $\mathcal{B}_{3,2} = v_{3}^{*}\cap v_{2}$ and $\mathcal{B}_{3,3} = v_{3}^{*}\cap v_{3}$ and assume that $|\mathcal{B}_{i,j}|\neq 0$ for all $i,j\in\{2,3\}$. Now we construct $\Lambda^{*}$ and $\Lambda$ by specifying their nonzero loadings as follows:
\begin{equation}
\begin{aligned}
&\Lambda^{*}_{[\mathcal{B}_{2,2},\{2\}]} = \frac{1}{2}\Lambda^{*}_{[\mathcal{B}_{2,2},\{1\}]}, \Lambda_{[\mathcal{B}_{2,2},\{1\}]} = \Lambda^{*}_{[\mathcal{B}_{2,2},\{1\}]}, \Lambda_{[\mathcal{B}_{2,2},\{2\}]} = \frac{1}{2}\Lambda^{*}_{[\mathcal{B}_{2,2},\{1\}]},\\
&\Lambda^{*}_{[\mathcal{B}_{2,3},\{2\}]} = 2\Lambda^{*}_{[\mathcal{B}_{2,3},\{1\}]}, \Lambda_{[\mathcal{B}_{2,3},\{1\}]} = 2\Lambda^{*}_{[\mathcal{B}_{2,3},\{1\}]}, \Lambda_{[\mathcal{B}_{2,3},\{3\}]} = -\Lambda^{*}_{[\mathcal{B}_{2,3},\{1\}]},\\
&\Lambda^{*}_{[\mathcal{B}_{3,2},\{3\}]} = \frac{1}{2}\Lambda^{*}_{[\mathcal{B}_{3,2},\{1\}]}, \Lambda_{[\mathcal{B}_{3,2},\{1\}]} = \frac{1}{2}\Lambda^{*}_{[\mathcal{B}_{3,2},\{1\}]}, \Lambda_{[\mathcal{B}_{3,2},\{2\}]} = \Lambda^{*}_{[\mathcal{B}_{3,2},\{1\}]},\\
&\Lambda^{*}_{[\mathcal{B}_{3,3},\{3\}]} = -\Lambda^{*}_{[\mathcal{B}_{3,3},\{1\}]}, \Lambda_{[\mathcal{B}_{3,3},\{1\}]} = \Lambda^{*}_{[\mathcal{B}_{3,3},\{1\}]}, \Lambda_{[\mathcal{B}_{3,3},\{3\}]} = \Lambda^{*}_{[\mathcal{B}_{3,3},\{1\}]}.\\
\end{aligned}
\end{equation}

As long as $|\mathcal{B}_{2,2}|\geq 2$ and $|\mathcal{B}_{2,3}|\geq 2$, $\Lambda^{*}$ satisfies Conditions \ref{cond:true para} and \ref{cond: couter sufficient}. However, $\Lambda\Lambda^{\top} = \Lambda^{*}\Lambda^{*\top}$ while $\Lambda$ and $\Lambda^{*}$ produce different hierarchical factor structures.
\end{proof}

To avoid the counterexample raised in Lemma~\ref{lem:counter}, we need the following Condition \ref{cond: appendix additional}.

\begin{condition}\label{cond: appendix additional}
There exists $g\in\{2,3\}$ and $i,j,k \in v_{g}^{*}$ such that $\Lambda^{*}_{[\{i,j\}, \{1,g\}]}$ $\Lambda^{*}_{[\{i,k\}, \{1,g\}]}$ and $\Lambda^{*}_{[\{j,k\}, \{1,g\}]}$ are of full rank.
\end{condition}

\begin{theorem}\label{thm: appen bi-factor id}
Suppose that Conditions~\ref{cond:true para}, \ref{cond: couter sufficient}, and \ref{cond: appendix additional} hold. If there exists some hierarchical factor structure with three factors such that its loading matrix $\Lambda$ and unique variance matrix $\Psi$ satisfy $\Sigma^* = \Lambda\Lambda^{\top} + \Psi$,  there exists some sign flip matrix $Q\in \mathcal{Q}$ such that $\Lambda = \Lambda^{*}Q$, where $\mathcal{Q}$ consists of all the $3\times 3$ diagonal matrix $Q$ whose diagonal entries take values $1$ or $-1$.
\end{theorem}

\begin{proof}

We adopt the notation from the proof of Lemma~\ref{lem:counter}. By Condition~\ref{cond: couter sufficient}, we have $\Lambda\Lambda^{\top} = \Lambda^{*}\Lambda^{*\top}$ and $\Psi = \Psi^{*}$. We consider the following cases:
\begin{enumerate}
\item $|\mathcal{B}_{s_1,s_2}|\neq 0$ for all $s_1,s_2\in\{2,3\}$. 
\item Without loss of generality, $v_2 = v_2^*$ but $v_3 \neq v_3^*$.
\item $v_{2} = v_{2}^{*}$ and $v_{3} = v_{3}^{*}$.
\end{enumerate}

In the first case, since $\Lambda^{*}_{[v^{*}_{2},\{1,2\}]}$ and $\Lambda^{*}_{[v^{*}_{3},\{1,3\}]}$ are of rank-2, all the following submatrices are of rank-1:
\begin{equation*}
\begin{aligned}
&\Lambda^{*}_{[\mathcal{B}_{2,2},\{1,2\}]}, \Lambda^{*}_{[\mathcal{B}_{2,3},\{1,2\}]}, \Lambda^{*}_{[\mathcal{B}_{3,2},\{1,3\}]}, \Lambda^{*}_{[\mathcal{B}_{3,3},\{1,3\}]},\\
&\Lambda_{[\mathcal{B}_{2,2},\{1,2\}]}, \Lambda_{[\mathcal{B}_{2,3},\{1,3\}]}, \Lambda_{[\mathcal{B}_{3,2},\{1,2\}]}, \Lambda_{[\mathcal{B}_{3,3},\{1,3\}]}.
\end{aligned}
\end{equation*}
However, according to Condition~\ref{cond: appendix additional}, there exists at least one of the matrices $\Lambda^{*}_{[\mathcal{B}_{2,2},\{1,2\}]}, \Lambda^{*}_{[\mathcal{B}_{2,3},\{1,2\}]}, \Lambda^{*}_{[\mathcal{B}_{3,2},\{1,3\}]}$ and $\Lambda^{*}_{[\mathcal{B}_{3,3},\{1,3\}]}$ such that it is of rank-2. Thus, the first case is not allowed.

In the second case, we assume $v_{2} = v_{2}^{*}$ while $v_{3} \neq v_{3}^{*}$ without loss of generality. Since $\Lambda_{[v_{2}^{*},\{1,2\}]}\Lambda_{[v_{2}^{*},\{1,2\}]}^{\top} = \Lambda_{[v_{2}^{*},\{1,2\}]}^{*}\Lambda_{[v_{2}^{*},\{1,2\}]}^{*\top}$, there exits some orthogonal rotation matrix $R\in\mathbb{R}^{2\times 2}$ such that
\begin{equation}\label{eq:bi-factor id tm eq 1}
\Lambda_{[v_{2}^{*},\{1,2\}]} = \Lambda_{[v_{2}^{*},\{1,2\}]}^{*}R.
\end{equation}
With $\Lambda_{[v_{2}^{*},\{1\}]}\Lambda_{[\mathcal{B}_{3,3},\{1\}]}^{\top} = \Lambda_{[v_{2}^{*},\{1\}]}^{*}\Lambda_{[\mathcal{B}_{3,3},\{1\}]}^{*\top}$, there exists some constant $a$ such that $\Lambda_{[v_{2}^{*},\{1\}]} = a\Lambda_{[v_{2}^{*},\{1\}]}^{*}$. Combined with \eqref{eq:bi-factor id tm eq 1}, $a=1$ or $-1$ since $\Lambda_{[v_{2}^{*},\{1,2\}]}^{*}$ is of rank-2. Without loss of generality, we assume $\Lambda_{[v_{2}^{*},\{1\}]} = \Lambda_{[v_{2}^{*},\{1\}]}^{*}$ and $\Lambda_{[v_{2}^{*},\{2\}]} = \Lambda_{[v_{2}^{*},\{2\}]}^{*}$ further. Then, consider $\Lambda_{[v_{2}^{*},\{1,2\}]}\Lambda_{[\mathcal{B}_{3,2},\{1,2\}]}^{\top} = \Lambda_{[v_{2}^{*},\{1\}]}^{*}\Lambda_{[\mathcal{B}_{3,2},\{1\}]}^{*\top}$, which leads to $\Lambda_{[\mathcal{B}_{3,2},\{2\}]} = \mathbf{0}$. Thus, the second case is not allowed.

In the third case, similar to the proof in the second case, there exists two orthogonal rotation matrices $R_1, R_2\in\mathbb{R}^{2\times 2}$ and  such that
\begin{equation}\label{eq:bi-factor id tm eq 2}
\Lambda_{[v_{2}^{*},\{1,2\}]} = \Lambda_{[v_{2}^{*},\{1,2\}]}^{*}R_1 \mbox{~and ~} \Lambda_{[v_{3}^{*},\{1,3\}]} = \Lambda_{[v_{3}^{*},\{1,3\}]}^{*}R_2.
\end{equation}
Combined with $\Lambda_{[v_{2}^{*},\{1\}]}\Lambda_{[v_{3}^{*},\{1\}]}^{\top} = \Lambda_{[v_{2}^{*},\{1\}]}^{*}\Lambda_{[v_{3}^{*},\{1\}]}^{*\top}$, there exists some sign flip matrix $Q\in \mathcal{Q}$ such that $\Lambda = \Lambda^{*}Q$.
\end{proof}

\begin{remark}
Theorem~\ref{thm: appen bi-factor id} establishes the identifiability of the bi-factor model with two group factors. Compared to the general hierarchical identifiability result in Theorem~\ref{thm:identifiability}, it requires fewer structural assumptions, but still needs the additional rank condition (Condition~\ref{cond: appendix additional}). The proof of Theorem~\ref{thm: appen bi-factor id} is based on the specific hierarchical structure and we believe the requirement for Condition~\ref{cond:rank} can be simplified based on the true hierarchical factor structure.
\end{remark}

\section{Proof of Theorem~\ref{cor:consistency}}\label{appen:proof consistency}
We first introduce some notations and lemmas needed for the proof of Theorem~\ref{cor:consistency}.  Suppose that $A,\varepsilon\in\mathbb{R}^{m\times n}$. We denote by $\sigma_{1}(A)\geq\ldots\geq\sigma_{\min(m,n)}(A)\geq 0$ are the singular values of $A$, and $U_1,\ldots,U_{\min(m,n)}$ are the corresponding right(left) singular vectors. Similarly, we denote by $\sigma_{1}(A+\varepsilon)\geq\ldots\geq\sigma_{\min(m,n)}(A+\varepsilon)\geq0$ as the singular values of $A+\varepsilon$ and $U_{1}',\ldots,U_{\min(m,n)}'$ the corresponding right(left) singular vectors. We use $\|A\|_{2}$ denote the spectral norm of a matrix $A$.
\begin{lemma}[Weyl's bound, \cite{weyl1912asymptotische}]\label{lem: Weyl}
\begin{equation*}
\max_{1\leq i\leq \min(m, n)}|\sigma_{i}(A)-\sigma_{i}(A+\varepsilon)|\leq \|\varepsilon\|_{2}.
\end{equation*}
\end{lemma}
We further assume that the rank of $A$ is $r$. We denote by $U = (U_{1},\ldots,U_{j})$ and $U' = (U_{1}',\ldots,U_{j}')$, $1\leq j\leq r$. The following Lemma~\ref{lem: wedin} is a modification of Wedin's Theorem \citep{wedin1972perturbation}.
\begin{lemma}\label{lem: wedin}
There exists some orthogonal matrix $R$ such that
\begin{equation*}
\|UR-U'\|_{F}\leq \frac{2^{3/2}r^{1/2}\|\varepsilon\|_{F}}{\delta}.
\end{equation*}
when $\delta = \sigma_{j}(A) - \sigma_{j+1}(A)>0$.
\end{lemma}

\begin{lemma}\label{lem:apprxi anderson}
Given a $J\times K$ dimensional matrix $\Lambda$ following a hierarchical structure that satisfies constraints C1-C4 and a $J\times J$ dimensional diagonal matrix $\Psi = \text{diag}(\psi_1,\ldots,\psi_J)$ with $\psi_j>0, j=1,\ldots,J$. Assume that $\Lambda$ satisfies Condition~\ref{cond:bic correct} and Condition~\ref{cond:hier compact set}. If there exist a series of $J\times K$ dimensional random matrices $\{\hat{\Lambda}_{N}\}_{N=1}^{\infty}$ and a series of $J\times J$ dimensional diagonal random matrices $\{\hat{\Psi}_{N}\}_{N=1}^{\infty}$, where $\hat{\Psi}_{N} = \mbox{diag}(\hat{\psi}_{N,1},\ldots,\hat{\psi}_{N,J})$ with $\hat{\psi}_{N,j}\geq 0, j=1,\ldots,J$, such that $\{\hat{\Lambda}_{N}\}_{N=1}^{\infty}$ satisfies Condition~\ref{cond:hier compact set} and 
\begin{equation}\label{eq:lem approxi anderson main}
\| \hat{\Lambda}_{N}\hat{\Lambda}_{N}^{\top} + \hat{\Psi}_{N} - \Lambda\Lambda^{\top} - \Psi \|_{F} = O_{\mathbb{P}}(1/\sqrt{N}).
\end{equation}
Then we have $\| \hat{\Lambda}_{N}\hat{\Lambda}_{N}^{\top}  - \Lambda\Lambda^{\top}\|_{F} = O_{\mathbb{P}}(1/\sqrt{N})$ and $\| \hat{\Psi}_{N}  - \Psi\|_{F} = O_{\mathbb{P}}(1/\sqrt{N})$.
\end{lemma}

Lemma~\ref{lem:apprxi anderson} is a generalization of Theorem 5.1 in \cite{anderson1956statistical}, and its proof proceeds along the same lines. 
\begin{proof}
For $j=1,\ldots,J$, by Condition~\ref{cond:bic correct}, there exist $E_1, E_2\in \{1,\ldots,J\}\setminus \{j\}$ with $|E_1| = |E_2| = K$ and $E_1 \cap E_2 = \emptyset$ such that $\Lambda_{[E_1,:]}$ and $\Lambda_{[E_2,:]}$ are full-rank matrices. Without loss of generality, we assume that $\Lambda$ and $\hat{\Lambda}_{N}$ can be expressed as
\begin{equation*}
\begin{aligned}
\Lambda =  \left(
    \begin{array}{c}
    \Lambda_1 \\
    \boldsymbol{\lambda}_{j}\\
    \Lambda_2 \\
    \Lambda_3\\
    \end{array}
    \right),
    \end{aligned}
\quad
\begin{aligned}
\hat{\Lambda}_{N} =  \left(
    \begin{array}{c}
    \hat{\Lambda}_{N,1} \\
    \hat{\boldsymbol{\lambda}}_{N,j}\\
    \hat{\Lambda}_{N,2} \\
    \hat{\Lambda}_{N,3}\\
    \end{array}
    \right),
    \end{aligned}
\end{equation*}
where we denote by $\Lambda_1 = \Lambda_{[E_1,:]}$, $\Lambda_2 = \Lambda_{[E_2,:]}$,  $\boldsymbol{\lambda}_{j} = \Lambda_{[\{j\},:]}$ is the $j$th row of $\Lambda$, $\Lambda_3$ consists of the remaining rows in $\Lambda$ with a slight abuse of notation. The blocks $\hat{\Lambda}_{N,1}$, $\hat{\Lambda}_{N,2}$, $\hat{\boldsymbol{\lambda}}_{N,j}$, and $\hat{\Lambda}_{N,3}$ are defined analogously for $\hat{\Lambda}_N$, with the same row partitioning. Thus, we have
\begin{equation*}
\begin{aligned}
\Lambda_{[E_1\cup E_2\cup \{j\},:]}\Lambda^{\top}_{[E_1\cup E_2\cup \{j\},:]} = \left(
    \begin{array}{ccc}
    \Lambda_{1}\Lambda_{1}^{\top} &\Lambda_{1}\boldsymbol{\lambda}_{j}^{\top}  &\Lambda_{1}\Lambda_2^{\top}\\
    \boldsymbol{\lambda}_{j}\Lambda_{1}^{\top}  & \boldsymbol{\lambda}_{j}\boldsymbol{\lambda}_{j}^{\top} &\boldsymbol{\lambda}_{j}\Lambda_2^{\top}\\
    \Lambda_2\Lambda_{1}^{\top}  &  \Lambda_2\boldsymbol{\lambda}_{j}^{\top} & \Lambda_2\Lambda_2^{\top} \\
    \end{array}
    \right),
\end{aligned}
\end{equation*}
and
\begin{equation*}
\begin{aligned}
(\hat{\Lambda}_{N})_{[E_1\cup E_2\cup \{j\},:]}(\hat{\Lambda}_{N})^{\top}_{[E_1\cup E_2\cup \{j\},:]} = \left(
    \begin{array}{ccc}
    \hat{\Lambda}_{N,1}\hat{\Lambda}_{N,1}^{\top} &\hat{\Lambda}_{N,1}\hat{\boldsymbol{\lambda}}_{N,j}^{\top}  &\hat{\Lambda}_{N,1}\hat{\Lambda}_{N,2}^{\top}\\
    \hat{\boldsymbol{\lambda}}_{N,j}\hat{\Lambda}_{N,1}^{\top}  & \hat{\boldsymbol{\lambda}}_{N,j}\hat{\boldsymbol{\lambda}}_{N,j}^{\top} &\hat{\boldsymbol{\lambda}}_{N,j}\hat{\Lambda}_{N,2}^{\top}\\
    \hat{\Lambda}_{N,2}\hat{\Lambda}_{N,1}^{\top}  &  \hat{\Lambda}_{N,2}\hat{\boldsymbol{\lambda}}_{N,j}^{\top} & \hat{\Lambda}_{N,2}\hat{\Lambda}_{N,2}^{\top} \\
    \end{array}
    \right).
\end{aligned}
\end{equation*}
According to~\eqref{eq:lem approxi anderson main}, we have
\begin{equation}\label{eq:lem approxi anderson app eq}
\begin{aligned}
\|\Lambda_{1}\boldsymbol{\lambda}_{j}^{\top} - \hat{\Lambda}_{N,1}\hat{\boldsymbol{\lambda}}_{N,j}^{\top}\| &= O_{\mathbb{P}}(1/\sqrt{N}), \\
\|\Lambda_{2}\boldsymbol{\lambda}_{j}^{\top} - \hat{\Lambda}_{N,2}\hat{\boldsymbol{\lambda}}_{N,j}^{\top}\| &=O_{\mathbb{P}}(1/\sqrt{N}), \\
\|\Lambda_{1}\Lambda_2^{\top}-\hat{\Lambda}_{N,1}\hat{\Lambda}_{N,2}^{\top}\|_{F}&=O_{\mathbb{P}}(1/\sqrt{N}).
\end{aligned}
\end{equation}
Since each of the following $(K+1) \times (K+1)$ matrices has rank at most $K$
\begin{equation*}
\left(
    \begin{array}{cc}
    \Lambda_{1}\boldsymbol{\lambda}_{j}^{\top}  &\Lambda_{1}\Lambda_2^{\top}\\
     \boldsymbol{\lambda}_{j}\boldsymbol{\lambda}_{j}^{\top} &\boldsymbol{\lambda}_{j}\Lambda_2^{\top}\\
    \end{array}
    \right)  \mbox{~~and~~}
    \left(
    \begin{array}{cc}
    \hat{\Lambda}_{N,1}\hat{\boldsymbol{\lambda}}_{N,j}^{\top}  &\hat{\Lambda}_{N,1}\hat{\Lambda}_{N,2}^{\top}\\ \hat{\boldsymbol{\lambda}}_{N,j}\hat{\boldsymbol{\lambda}}_{N,j}^{\top} &\hat{\boldsymbol{\lambda}}_{N,j}\hat{\Lambda}_{N,2}^{\top}\\
    \end{array}
    \right),
\end{equation*}
we have
\begin{equation*}
\begin{aligned}
\det\left(
    \begin{array}{cc}
    \Lambda_{1}\boldsymbol{\lambda}_{j}^{\top}  &\Lambda_{1}\Lambda_2^{\top}\\
     \boldsymbol{\lambda}_{j}\boldsymbol{\lambda}_{j}^{\top} &\boldsymbol{\lambda}_{j}\Lambda_2^{\top}\\
    \end{array}
    \right) 
=\det\left(
    \begin{array}{cc}
    \hat{\Lambda}_{N,1}\hat{\boldsymbol{\lambda}}_{N,j}^{\top}  &\hat{\Lambda}_{N,1}\hat{\Lambda}_{N,2}^{\top}\\ \hat{\boldsymbol{\lambda}}_{N,j}\hat{\boldsymbol{\lambda}}_{N,j}^{\top} &\hat{\boldsymbol{\lambda}}_{N,j}\hat{\Lambda}_{N,2}^{\top}\\
    \end{array}
    \right) = 0.
\end{aligned}
\end{equation*}
Then, we have 
\begin{equation}\label{eq:lem approxi anderson zero eq}
\begin{aligned}
&(-1)^{K}\boldsymbol{\lambda}_{j}\boldsymbol{\lambda}_{j}^{\top}\det(\Lambda_{1}\Lambda_2^{\top}) + f(\Lambda_{1}\boldsymbol{\lambda}_{j}^{\top},\boldsymbol{\lambda}_{j}\Lambda_2^{\top})\\
=& (-1)^{K}\hat{\boldsymbol{\lambda}}_{N,j}\hat{\boldsymbol{\lambda}}_{N,j}^{\top}\det(\hat{\Lambda}_{N,1}\hat{\Lambda}_{N,2}^{\top}) +  f(\hat{\Lambda}_{N,1}\hat{\boldsymbol{\lambda}}_{N,j}^{\top},\hat{\boldsymbol{\lambda}}_{N,j}\hat{\Lambda}_{N,2}^{\top}) \\
=& 0,
\end{aligned}
\end{equation}
where $f(\cdot)$ is a scalar-valued function. Both $f(\cdot)$ and the determinant function $\det(\cdot)$ are Lipschitz continuous with respect to the entries of their matrix arguments, with Lipschitz constants depending only on $K$ and $\tau$. Combined with~\eqref{eq:lem approxi anderson app eq} and~\eqref{eq:lem approxi anderson zero eq}, we have 
\begin{equation*}
\begin{aligned}
&|\boldsymbol{\lambda}_{j}\boldsymbol{\lambda}_{j}^{\top}-\hat{\boldsymbol{\lambda}}_{N,j}\hat{\boldsymbol{\lambda}}_{N,j}^{\top}||\det(\Lambda_{1}\Lambda_2^{\top})|\\
\leq & |f(\Lambda_{1}\boldsymbol{\lambda}_{j}^{\top},\boldsymbol{\lambda}_{j}\Lambda_2^{\top}) - f(\hat{\Lambda}_{N,1}\hat{\boldsymbol{\lambda}}_{N,j}^{\top},\hat{\boldsymbol{\lambda}}_{N,j}\hat{\Lambda}_{N,2}^{\top})| \\
&+|\hat{\boldsymbol{\lambda}}_{N,j}\hat{\boldsymbol{\lambda}}_{N,j}^{\top}||\det(\Lambda_{1}\Lambda_2^{\top}) - \det(\hat{\Lambda}_{N,1}\hat{\Lambda}_{N,2}^{\top})|\\
=& O_{\mathbb{P}}(1/\sqrt{N}).
\end{aligned}
\end{equation*}
Noticing that $|\det(\Lambda_{1}\Lambda_2^{\top})|>0$, we have $|\boldsymbol{\lambda}_{j}\boldsymbol{\lambda}_{j}^{\top}-\hat{\boldsymbol{\lambda}}_{N,j}\hat{\boldsymbol{\lambda}}_{N,j}^{\top}| = O_{\mathbb{P}}(1/\sqrt{N})$. Combined with 
$$
|\boldsymbol{\lambda}_{j}\boldsymbol{\lambda}_{j}^{\top} + \psi_j -\hat{\boldsymbol{\lambda}}_{N,j}\hat{\boldsymbol{\lambda}}_{N,j}^{\top}-\hat{\psi}_{N,j}| = O_{\mathbb{P}}(1/\sqrt{N}),
$$
we have $|\psi_j - \hat{\psi}_{N,j}| = O_{\mathbb{P}}(1/\sqrt{N})$ for $j=1,\ldots,J$. Thus, we have $\| \hat{\Psi}_{N}  - \Psi\|_{F} = O_{\mathbb{P}}(1/\sqrt{N})$ and furthermore we have $\| \hat{\Lambda}_{N}\hat{\Lambda}_{N}^{\top}  - \Lambda\Lambda^{\top}\|_{F} = O_{\mathbb{P}}(1/\sqrt{N})$.
\end{proof}
For Factor $k\in L_{t-1}$, $t \geq 3$, let $\Sigma^{*}_{k,0} := \sum_{i=1}^{k_{t-2}} (\boldsymbol{\lambda}_{i}^{*})_{[v_{k}^{*}]} (\boldsymbol{\lambda}_{i}^{*})_{[v_{k}^{*}]}^\top$ when $k \in \hat{L}_{t-1}$ and $\Sigma_{k}^{*} =(\boldsymbol{\lambda}_{k}^{*})_{[v_{k}^{*}]} (\boldsymbol{\lambda}_{k}^{*})_{[v_{k}^{*}]}^\top + \sum_{i\in D^{*}_{k}} (\boldsymbol{\lambda}_{i}^{*})_{[v_{k}^{*}]} (\boldsymbol{\lambda}_{i}^{*})_{[v_{k}^{*}]}^\top + \Psi^{*}_{[v_{k}^{*},v_{k}^{*}]}$. We further define 
\begin{equation}
\begin{aligned}
\Theta_{k}(c,d) =& \{\Sigma = \Lambda_{k}\Lambda_{k}^{\top} + \Psi_{k} \in \mathbb{R}^{|v_{k}^{*}|\times|v_{k}^{*}|}: \Lambda_{k}\in\mathbb{R}^{|v_{k}^{*}|\times (1+cd)}, \\
&~|\lambda_{k,ij}| \leq \tau, \lambda_{k,ij}\lambda_{k,ij'} = 0,\mbox{~for~} i=1,\ldots,|v_{k}^{*}|, j\in\mathcal{B}_{s}, j'\in \mathcal{B}_{s'}, s\neq s' \\
&\mbox{~and~}\Psi = \mbox{diag}(\psi_{k1},\ldots,\psi_{k|v_{k}^{*}|}) \mbox{~with~}\kappa_1\leq\psi_{ki}\leq\kappa_2 \mbox{~for~} i=1,\ldots,|v_{k}^{*}|\},
\end{aligned}
\end{equation}
where $\mathcal{B}_{s} = 2+(s-1)d,\ldots, 1+sd$ for $s = 1,\ldots,c$, and $\tau$, $\kappa_1$ and $\kappa_2$ are those specified in Condition~\ref{cond:hier compact set}. Given a symmetric positive semi-definite matrix $\tilde{\Sigma}_{k,0}$ serving as an estimator of $\Sigma^{*}_{k,0}$, we define 
\begin{equation}
\hat{\Sigma}_{k} = \argmin_{\Sigma_{k}\in \Theta_{k}(c,d)}l\left( \tilde{\Sigma}_{k,0} + \Sigma_{k}, S_{k}\right).
\end{equation}
The following Lemma~\ref{lem: consistency covariance} and~\ref{lem: convergence rate covariance} establish the consistency and convergence rate of $\hat{\Sigma}_{k}$.

\begin{lemma}[Consistency]\label{lem: consistency covariance}
Suppose $d$ is sufficiently large such that $\Sigma_{k}^{*}\in \Theta_{k}(c,d)$ and $\|\tilde{\Sigma}_{k,0} - \Sigma^{*}_{k,0}\|_{F} = o_{\mathbb{P}}(1)$. If Conditions~\ref{cond:correct pilot} and~\ref{cond:hier compact set} hold, $\hat{\Sigma}_{k}\stackrel{\mathbb{P}}{\to} \Sigma^{*}_{k}$.
\end{lemma}

\begin{proof}
The proof of Lemma~\ref{lem: consistency covariance} follows Theorem 2.1 in \cite{newey1994large}. First, we show that $\Theta_{k}(c,d)$ is a compact set in $\mathbb{R}^{|v_{k}^{*}|\times |v_{k}^{*}|}$. By definition, we directly have that $\Theta_{k}(c,d)$ is a bounded set. To prove that $\Theta_{k}(c,d)$ is also a closed set, we assume $\{\Sigma_{k}^{(n)} = \Lambda_{k}^{(n)}\Lambda_{k}^{(n)\top} + \Psi_{k}^{(n)}\}_{n=1}^{\infty}$ is an arbitrary convergent sequence in $\Theta_{k}(c,d)$. Since $\{\Lambda_{k}^{(n)}\}_{n=1}^{\infty}$ and $\{\Psi_{k}^{(n)}\}_{n=1}^{\infty}$ are bounded sequences, there exists subsequence $\{\Lambda_{k}^{(n_m)}\}_{m=1}^{\infty}$ and $\{\Psi_{k}^{(n_m)}\}_{m=1}^{\infty}$ such that
$$
\lim_{m\to\infty} \Lambda_{k}^{(n_m)} = \Lambda_{k}^{\infty} \mbox{~and~} \lim_{m\to\infty} \Psi_{k}^{(n_m)} = \Psi_{k}^{\infty}.
$$
Since $\lambda_{k,ij}^{n_m}\lambda_{k,ij'}^{n_m} = 0$, $\lim_{m\to\infty}\lambda_{k,ij}^{n_m} = \lambda_{k,ij}^{\infty}$ and $\lim_{m\to\infty}\lambda_{k,ij'}^{n_m} = \lambda_{k,ij'}^{\infty}$, we have $\lambda_{k,ij}^{\infty}\lambda_{k,ij'}^{\infty}=0$ for $i = 1, \ldots, |v_{k}^{*}|$, $j\in \mathcal{B}_{s}$, $j'\in \mathcal{B}_{s'}$, $1\leq s<s'\leq c$. Thus,
$$
\lim_{n\to\infty}\Sigma_{k}^{(n)} = \Lambda_{k}^{\infty}\Lambda_{k}^{\infty\top} + \Psi_{k}^{\infty} \in \Theta_{k}(c,d). 
$$
Then $\Theta_{k}(c,d)$ is a compact set.

Second, let 
\begin{equation}
\begin{aligned}
&a_{k}(x;\Sigma_{k},\tilde{\Sigma}_{k,0}) = \log\det(\tilde{\Sigma}_{k,0}+\Sigma_{k})+ \mbox{tr}\left(x_{[v_{k}^{*}]}x_{[v_{k}^{*}]}^{\top}(\tilde{\Sigma}_{k,0}+\Sigma_{k})^{-1}\right)\\
&M_{k}(\Sigma_{k},\tilde{\Sigma}_{k,0}) = \log\det(\tilde{\Sigma}_{k,0}+\Sigma_{k})+ \mbox{tr}\left(S_{k}(\tilde{\Sigma}_{k,0}+\Sigma_{k})^{-1}\right)\\
&M_{0,k}(\Sigma_{k},\tilde{\Sigma}_{k,0}) = \log\det(\tilde{\Sigma}_{k,0}+\Sigma_{k})+ \mbox{tr}\left((\Sigma_{k,0}^{*}+\Sigma_{k}^{*})(\tilde{\Sigma}_{k,0}+\Sigma_{k})^{-1}\right)\\
\end{aligned}
\end{equation}
We directly have 
\begin{equation}
\hat{\Sigma}_{k} = \argmin_{\Sigma_{k}\in \Theta_{k}(c,d)}M_{k}(\Sigma_{k},\tilde{\Sigma}_{k,0}). 
\end{equation}
Moreover, 
\begin{equation}
\begin{aligned}
&\frac{\partial}{\partial \Sigma_{k}}M_{0,k}(\Sigma_{k},\Sigma^{*}_{k,0}) \\
=& (\Sigma_{k}+\Sigma^{*}_{k,0})^{-1} - (\Sigma_{k}+\Sigma^{*}_{k,0})^{-1}(\Sigma_{k}^{*}+\Sigma^{*}_{k,0})(\Sigma_{k}+\Sigma^{*}_{k,0})^{-1}\\
=&0,
\end{aligned}
\end{equation}
when $\Sigma_{k} = \Sigma^{*}_{k} \in \Theta_{k}(c,d)$. Thus, $M_{0,k}(\Sigma_{k},\Sigma^{*}_{k,0})$ reaches its unique minimum at $\Sigma_{k}^{*}$. 

Third, 
\begin{equation}
\begin{aligned}
&|a_{k}(x;\Sigma_{k},\Sigma^{*}_{k,0})|\\
\leq & |\log\det(\Sigma^{*}_{k,0}+\Sigma_{k})| + \left|\mbox{tr}\left(x_{[v_{k}^{*}]}x_{[v_{k}^{*}]}^{\top}(\Sigma^{*}_{k,0}+\Sigma_{k})^{-1}\right)\right| \\
\leq& |v_{k}^{*}|\max\left(|\log(\sigma_{\min}(\Sigma^{*}_{k,0}+\Sigma_{k}))|,|\log(\sigma_{\max}(\Sigma^{*}_{k,0}+\Sigma_{k}))|\right) + \frac{\left\|x_{[v_{k}^{*}]}\right\|^{2}}{\sigma_{\min}(\Sigma^{*}_{k,0}+\Sigma_{k})}\\
\leq& |v_{k}^{*}|\max\left(|\log\kappa_1|,\left|\log\left(|v_{k}^{*}|\left((1+cd)^{2}+K^{2}\right)\tau^{2}+\kappa_2\right)\right|\right)+ \frac{1}{\kappa_1}\left\|x_{[v_{k}^{*}]}\right\|^{2}.
\end{aligned}
\end{equation}
Since $\mathbb{E}\left(\left\|x_{[v_{k}^{*}]}\right\|^{2}\right)<\infty$, by Lemma 2.4 of \cite{newey1994large}
\begin{equation}\label{eq:lemma consistency uniform convergence}
\begin{aligned}
\sup_{\Sigma_{k}\in \Theta_{k}(c,d)} |M_{k}(\Sigma_{k},\Sigma^{*}_{k,0})-M_{0,k}(\Sigma_{k},\Sigma^{*}_{k,0})|\stackrel{\mathbb{P}}{\to}0.
\end{aligned}
\end{equation}

Now we have
\begin{equation}\label{eq:lemma consistency decompose}
\begin{aligned}
&M_{0,k}(\hat{\Sigma}_{k},\Sigma^{*}_{k,0})\\
=& M_{0,k}(\Sigma^{*}_{k},\Sigma^{*}_{k,0}) +(M_{0,k}(\hat{\Sigma}_{k},\Sigma^{*}_{k,0})-M_{0,k}(\hat{\Sigma}_{k},\tilde{\Sigma}_{k,0})) + (M_{0,k}(\hat{\Sigma}_{k},\tilde{\Sigma}_{k,0})\\
&-M_{k}(\hat{\Sigma}_{k},\tilde{\Sigma}_{k,0})) + (M_{k}(\hat{\Sigma}_{k},\tilde{\Sigma}_{k,0})-M_{k}(\Sigma^{*}_{k},\tilde{\Sigma}_{k,0}))+(M_{k}(\Sigma^{*}_{k},\tilde{\Sigma}_{k,0})\\
&-M_{0,k}(\Sigma^{*}_{k},\tilde{\Sigma}_{k,0})) + (M_{0,k}(\Sigma_{k}^{*},\tilde{\Sigma}_{k,0})-M_{0,k}(\Sigma^{*}_{k},\Sigma^{*}_{k,0}))\\
\leq & M_{0,k}(\Sigma^{*}_{k},\Sigma^{*}_{k,0}) + 2\sup_{\Sigma_{k}\in \Theta_{k}(c,d)}|M_{0,k}(\Sigma_{k},\Sigma^{*}_{k,0})-M_{0,k}(\Sigma_{k},\tilde{\Sigma}_{k,0})| \\
&+ 2\sup_{\Sigma_{k}\in \Theta_{k}(c,d)}|M_{0,k}(\Sigma_{k},\tilde{\Sigma}_{k,0})-M_{k}(\Sigma_{k},\tilde{\Sigma}_{k,0}) - M_{0,k}(\Sigma_{k},\Sigma^{*}_{k,0})+M_{k}(\Sigma_{k},\Sigma^{*}_{k,0})| \\
&+ 2\sup_{\Sigma_{k}\in \Theta_{k}(c,d)}|M_{k}(\Sigma_{k},\Sigma^{*}_{k,0})-M_{0,k}(\Sigma_{k},\Sigma^{*}_{k,0})|.
\end{aligned}
\end{equation}
For arbitrary $\Sigma_{k}\in \Theta_{k}(c,d)$, according to Taylor's expansion there exists some $\eta\in (0,1)$ such that
\begin{equation}\label{eq:lemma consistency uniform gap 1}
\begin{aligned}
&|M_{0,k}(\Sigma_{k},\Sigma^{*}_{k,0})-M_{0,k}(\Sigma_{k},\tilde{\Sigma}_{k,0})|\\
\leq& \left|\mbox{tr}\left((\tilde{\Sigma}_{k,0}-\Sigma^{*}_{k,0})\left(\Sigma_{k}+(1-\eta)\Sigma^{*}_{k,0}+\eta\tilde{\Sigma}_{k,0}\right)^{-1}(\Sigma^{*}_{k,0}+\Sigma^{*}_{k})\Big(\Sigma_{k}+(1-\eta)\Sigma^{*}_{k,0}\right.\right.\\
&\left.\left.\left.+\eta\tilde{\Sigma}_{k,0}\right)^{-1}\right)\right|+\left|\mbox{tr}\left((\tilde{\Sigma}_{k,0}-\Sigma^{*}_{k,0})\left((1-\eta)\Sigma^{*}_{k,0}+\eta\tilde{\Sigma}_{k,0}\right)^{-1}\right)\right| \\
\leq& \frac{|v_{k}^{*}|K^{2}\tau^{2}+\kappa_2 + \kappa_1}{\kappa_{1}^{2}}\|\tilde{\Sigma}_{k,0}-\Sigma^{*}_{k,0}\|_{F},
\end{aligned}
\end{equation}
where the last inequality follows Ruhe’s trace inequality \citep{ruhe1970perturbation}. Similarly, with probability approaching 1 as $N$ grows to infinity, 
\begin{equation}\label{eq:lemma consistency uniform gap 2}
\begin{aligned}
&|M_{k}(\Sigma_{k},\Sigma^{*}_{k,0})-M_{k}(\Sigma_{k},\tilde{\Sigma}_{k,0})|\\
\leq& \left|\mbox{tr}\left((\tilde{\Sigma}_{k,0}-\Sigma^{*}_{k,0})\left(\Sigma_{k}+(1-\eta)\Sigma^{*}_{k,0}+\eta\tilde{\Sigma}_{k,0}\right)^{-1}S_{k}\left(\Sigma_{k}+(1-\eta)\Sigma^{*}_{k,0}+\eta\tilde{\Sigma}_{k,0}\right)^{-1}\right)\right|\\
&+\left|\mbox{tr}\left((\tilde{\Sigma}_{k,0}-\Sigma^{*}_{k,0})\left(\Sigma_{k}+(1-\eta)\Sigma^{*}_{k,0}+\eta\tilde{\Sigma}_{k,0}\right)^{-1}\right)\right| \\
\leq& \frac{2\left(|v_{k}^{*}|K^{2}\tau^{2}+\kappa_{2}\right)+\kappa_{1}}{\kappa_{1}^{2}}\|\tilde{\Sigma}_{k,0}-\Sigma^{*}_{k,0}\|_{F},
\end{aligned}
\end{equation}
With \eqref{eq:lemma consistency uniform gap 1} and \eqref{eq:lemma consistency uniform gap 2}
\begin{equation}\label{eq:lemma consistency uniform op 1}
\begin{aligned}
\sup_{\Sigma_{k}\in \Theta_{k}(c,d)}|M_{0,k}(\Sigma_{k},\Sigma^{*}_{k,0})-M_{0,k}(\Sigma_{k},\tilde{\Sigma}_{k,0})| =  o_{\mathbb{P}}(1), 
\end{aligned}
\end{equation}
and
\begin{equation}\label{eq:lemma consistency uniform op 2}
\begin{aligned}
&\sup_{\Sigma_{k}\in \Theta_{k}(c,d)}|M_{0,k}(\Sigma_{k},\tilde{\Sigma}_{k,0})-M_{k}(\Sigma_{k},\tilde{\Sigma}_{k,0}) - M_{0,k}(\Sigma_{k},\Sigma^{*}_{k,0})+M_{k}(\Sigma_{k},\Sigma^{*}_{k,0})|\\
=&  o_{\mathbb{P}}(1).
\end{aligned}
\end{equation}
For arbitrary $\epsilon>0$, let 
$$\Delta(\epsilon) = \inf_{\Sigma_{k}\in \Theta_{k}(c,d),\|\Sigma_{k}-\Sigma_{k}^{*}\|_{F}\geq \epsilon}M_{0,k}(\Sigma_{k},\Sigma^{*}_{k,0})-M_{0,k}(\Sigma^{*}_{k},\Sigma^{*}_{k,0})>0.$$ Combined with \eqref{eq:lemma consistency uniform convergence}, \eqref{eq:lemma consistency decompose}, \eqref{eq:lemma consistency uniform op 1} and \eqref{eq:lemma consistency uniform op 2}, with probability approaching 1 as $N$ grows to infinity,
$$
M_{0,k}(\hat{\Sigma}_{k},\Sigma^{*}_{k,0}) < M_{0,k}(\Sigma^{*}_{k},\Sigma^{*}_{k,0}) + \Delta(\epsilon),
$$
which indicates $\|\hat{\Sigma}_{k}-\Sigma^{*}_{k}\|_{F}<\epsilon$. Thus, $\hat{\Sigma}_{k}\stackrel{\mathbb{P}}{\to} \Sigma^{*}_{k}$.
\end{proof}

\begin{lemma}[Convergence rate]\label{lem: convergence rate covariance}
Suppose $d$ is sufficiently large such that $\Sigma_{k}^{*}\in \Theta_{k}(c,d)$ and $\|\tilde{\Sigma}_{k,0} - \Sigma^{*}_{k,0}\|_{F} = O_{\mathbb{P}}(1/\sqrt{N})$. If Conditions~\ref{cond:correct pilot} and~\ref{cond:hier compact set} hold, $\|\hat{\Sigma}_{k}- \Sigma^{*}_{k}\|_{F} = O_{\mathbb{P}}(1/\sqrt{N})$.
\end{lemma}

\begin{proof}
Consider
\begin{equation}
\begin{aligned}
&M_{k}(\hat{\Sigma}_{k},\tilde{\Sigma}_{k,0}) - M_{k}(\Sigma^{*}_{k},\Sigma^{*}_{k,0}) \\
=& M_{0,k}(\hat{\Sigma}_{k},\tilde{\Sigma}_{k,0}) - M_{0,k}(\Sigma^{*}_{k},\Sigma^{*}_{k,0}) + M_{k}(\hat{\Sigma}_{k},\tilde{\Sigma}_{k,0}) - M_{0,k}(\hat{\Sigma}_{k},\tilde{\Sigma}_{k,0}) \\
& - M_{k}(\Sigma^{*}_{k},\Sigma^{*}_{k,0}) + M_{0,k}(\Sigma^{*}_{k},\Sigma^{*}_{k,0}).
\end{aligned}
\end{equation}
Let $\Delta_{\hat{\Sigma}_{k},\tilde{\Sigma}_{k,0}} = \hat{\Sigma}_{k} + \tilde{\Sigma}_{k,0} -\Sigma^{*}_{k} - \Sigma^{*}_{k,0}$. By Taylor's expansion, there exists some $\eta\in(0,1)$ such that
\begin{equation}
\begin{aligned}
&M_{0,k}(\hat{\Sigma}_{k},\tilde{\Sigma}_{k,0}) - M_{0,k}(\Sigma^{*}_{k},\Sigma^{*}_{k,0}) \\
=& \frac{1}{2}\mbox{tr}\left(\Delta_{\hat{\Sigma}_{k},\tilde{\Sigma}_{k,0}}\left(\Sigma^{*}_{k} + \Sigma^{*}_{k,0} + \eta\Delta_{\hat{\Sigma}_{k},\tilde{\Sigma}_{k,0}}\right)^{-1}\Delta_{\hat{\Sigma}_{k},\tilde{\Sigma}_{k,0}}\left(\Sigma^{*}_{k} + \Sigma^{*}_{k,0} + \eta\Delta_{\hat{\Sigma}_{k},\tilde{\Sigma}_{k,0}}\right)^{-1} \right.\\
&\left. \left(2\left(\Sigma^{*}_{k} + \Sigma^{*}_{k,0}\right)\left(\Sigma^{*}_{k} + \Sigma^{*}_{k,0} + \eta\Delta_{\hat{\Sigma}_{k},\tilde{\Sigma}_{k,0}}\right)^{-1}-\mathbf{I}\right)\right).
\end{aligned}
\end{equation}
For simplicity of the notations, let 
\begin{equation*}
\begin{aligned}
&\Delta_1 = \Delta_{\hat{\Sigma}_{k},\tilde{\Sigma}_{k,0}}\left(\Sigma^{*}_{k} + \Sigma^{*}_{k,0} + \eta\Delta_{\hat{\Sigma}_{k},\tilde{\Sigma}_{k,0}}\right)^{-1}\Delta_{\hat{\Sigma}_{k},\tilde{\Sigma}_{k,0}},\\
&\Delta_2 = \left(\Sigma^{*}_{k} + \Sigma^{*}_{k,0} + \eta\Delta_{\hat{\Sigma}_{k},\tilde{\Sigma}_{k,0}}\right)^{-1} \left(2\left(\Sigma^{*}_{k} + \Sigma^{*}_{k,0}\right)\left(\Sigma^{*}_{k} + \Sigma^{*}_{k,0} + \eta\Delta_{\hat{\Sigma}_{k},\tilde{\Sigma}_{k,0}}\right)^{-1}-\mathbf{I}\right)
\end{aligned}
\end{equation*}
According to Lemma~\ref{lem: consistency covariance}, $\hat{\Sigma}_{k}\stackrel{\mathbb{P}}{\to} \Sigma^{*}_{k}$. Combined with Lemma~\ref{lem: Weyl}, with probability approaching 1 as $N$ grows to infinity, 
\begin{equation}
\begin{aligned}
&\frac{\|\Delta_{\hat{\Sigma}_{k},\tilde{\Sigma}_{k,0}}\|_{F}^{2}}{|v_{k}^{*}|K^{2}\tau^{2}+\kappa_2}\\
\leq&\frac{1}{2}\sigma_{\min}\left(\left(\Sigma^{*}_{k} + \Sigma^{*}_{k,0}\right)^{-1}\right)\|\Delta_{\hat{\Sigma}_{k},\tilde{\Sigma}_{k,0}}\|_{F}^{2} \\
\leq& \mbox{tr}(\Delta_1) \\
\leq&2\sigma_{\max}\left(\left(\Sigma^{*}_{k} + \Sigma^{*}_{k,0}\right)^{-1}\right)\|\Delta_{\hat{\Sigma}_{k},\tilde{\Sigma}_{k,0}}\|_{F}^{2} \\
\leq&2\|\Delta_{\hat{\Sigma}_{k},\tilde{\Sigma}_{k,0}}\|_{F}^{2}/\kappa_{1},
\end{aligned}
\end{equation}
and
\begin{equation}
\begin{aligned}
&\sigma_{\min}(\Delta_2) \geq \frac{1}{2}\sigma_{\min}\left(\left(\Sigma^{*}_{k} + \Sigma^{*}_{k,0}\right)^{-1}\right) \geq \frac{1}{|v_{k}^{*}|K^{2}\tau^{2}+\kappa_2}, \\
&\sigma_{\max}(\Delta_2)\leq 2\sigma_{\max}\left(\left(\Sigma^{*}_{k} + \Sigma^{*}_{k,0}\right)^{-1}\right) \leq \frac{2}{\kappa_1}.
\end{aligned}
\end{equation}
By the Ruhe’s trace inequality, we have 
\begin{equation}\label{eq:lem rate final bound 1}
\begin{aligned}
\|\Delta_{\hat{\Sigma}_{k},\tilde{\Sigma}_{k,0}}\|_{F}^{2} = O_{\mathbb{P}}\left(M_{0,k}(\hat{\Sigma}_{k},\tilde{\Sigma}_{k,0}) - M_{0,k}(\Sigma^{*}_{k},\Sigma^{*}_{k,0})\right).
\end{aligned}
\end{equation}
Next, by Taylor's expansion, there exists some $\eta\in (0,1)$ such that 
\begin{equation}
\begin{aligned}
&M_{k}(\hat{\Sigma}_{k},\tilde{\Sigma}_{k,0}) - M_{0,k}(\hat{\Sigma}_{k},\tilde{\Sigma}_{k,0}) - M_{k}(\Sigma^{*}_{k},\Sigma^{*}_{k,0}) + M_{0,k}(\Sigma^{*}_{k},\Sigma^{*}_{k,0}) \\
=& \mbox{tr}\left(\Delta_{\hat{\Sigma}_{k},\tilde{\Sigma}_{k,0}} \left(\Sigma^{*}_{k} + \Sigma^{*}_{k,0} + \eta\Delta_{\hat{\Sigma}_{k},\tilde{\Sigma}_{k,0}}\right)^{-1}\left(S_{k} - \Sigma^{*}_{[v_{k}^{*},v_{k}^{*}]}\right)\right.\\ &\left.\left(\Sigma^{*}_{k}+ \Sigma^{*}_{k,0}+ \eta\Delta_{\hat{\Sigma}_{k},\tilde{\Sigma}_{k,0}}\right)^{-1}\right).
\end{aligned}
\end{equation}
Combined with Condition~\ref{cond:correct pilot}, Lemma~\ref{lem: Weyl} and the Ruhe’s trace inequality, with probability approaching 1 as $N$ grows to infinity, 
\begin{equation}\label{eq:lem rate final bound 2}
\begin{aligned}
&\left|M_{k}(\hat{\Sigma}_{k},\tilde{\Sigma}_{k,0}) - M_{0,k}(\hat{\Sigma}_{k},\tilde{\Sigma}_{k,0}) - M_{k}(\Sigma^{*}_{k},\Sigma^{*}_{k,0}) + M_{0,k}(\Sigma^{*}_{k},\Sigma^{*}_{k,0})\right| \\
=& O_{\mathbb{P}}(\|\Delta_{\hat{\Sigma}_{k},\tilde{\Sigma}_{k,0}}\|_{F}/\sqrt{N}).
\end{aligned}
\end{equation}
Similarly, let $
\Delta_{\Sigma^{*}_{k},\tilde{\Sigma}_{k,0}} = \Sigma^{*}_{k}  + \tilde{\Sigma}_{k,0} - \Sigma^{*}_{k} - \Sigma^{*}_{k,0} =  \tilde{\Sigma}_{k,0} - \Sigma^{*}_{k,0}$. We have 
\begin{equation}\label{eq:lem rate main bound 2}
\begin{aligned}
M_{k}(\Sigma^{*}_{k},\tilde{\Sigma}_{k,0}) - M_{k}(\Sigma^{*}_{k},\Sigma^{*}_{k,0}) &= O_{\mathbb{P}}(\|\Delta_{\Sigma^{*}_{k},\tilde{\Sigma}_{k,0}}\|_{F}^{2}) + O_{\mathbb{P}}(\|\Delta_{\Sigma^{*}_{k},\tilde{\Sigma}_{k,0}}\|_{F}/\sqrt{N})\\
& = O_{\mathbb{P}}(1/N)
\end{aligned}
\end{equation}
Thus, we have 
\begin{equation}
\begin{aligned}
0\leq& M_{0,k}(\hat{\Sigma}_{k},\tilde{\Sigma}_{k,0}) - M_{0,k}(\Sigma^{*}_{k},\Sigma^{*}_{k,0}) \\
\leq & M_{k}(\hat{\Sigma}_{k},\tilde{\Sigma}_{k,0}) - M_{k}(\Sigma^{*}_{k},\Sigma^{*}_{k,0}) + O_{\mathbb{P}}\left(\|\Delta_{\hat{\Sigma}_{k},\tilde{\Sigma}_{k,0}}\|_{F}/\sqrt{N}\right)\\
\leq& M_{k}(\Sigma^{*}_{k},\tilde{\Sigma}_{k,0}) - M_{k}(\Sigma^{*}_{k},\Sigma^{*}_{k,0}) + O_{\mathbb{P}}\left(\|\Delta_{\hat{\Sigma}_{k},\tilde{\Sigma}_{k,0}}\|_{F}/\sqrt{N}\right)\\
=& O_{\mathbb{P}}(1/N) + O_{\mathbb{P}}\left(\|\Delta_{\hat{\Sigma}_{k},\tilde{\Sigma}_{k,0}}\|_{F}/\sqrt{N}\right).
\end{aligned}
\end{equation}
Combined with \eqref{eq:lem rate final bound 1}, we have $\|\Delta_{\hat{\Sigma}_{k},\tilde{\Sigma}_{k,0}}\|_{F}^{2} = O_{\mathbb{P}}(1/N)+ O_{\mathbb{P}}(\|\Delta_{\hat{\Sigma}_{k},\tilde{\Sigma}_{k,0}}\|_{F}/\sqrt{N})$, which leads to 
$
\|\Delta_{\hat{\Sigma}_{k},\tilde{\Sigma}_{k,0}}\|_{F} = O_{\mathbb{P}}(1/\sqrt{N})
$. Furthermore, we have
$$
\|\hat{\Sigma}_{k} - \Sigma^{*}_{k}\|_{F} \leq \|\Delta_{\hat{\Sigma}_{k},\tilde{\Sigma}_{k,0}}\|_{F} + \|\tilde{\Sigma}_{k,0} - \Sigma^{*}_{k,0}\|_{F} = O_{\mathbb{P}}(1/\sqrt{N}).
$$
\end{proof}

When the true hierarchical factor structure is known, the estimates of the loading matrix and the unique variance matrix are defined as follows:
\begin{equation}\label{eq:confirmatory estimator definition}
\begin{aligned}
\hat{\Lambda},\hat{\Psi} = \argmin_{\Lambda,\Psi}~&l(\Lambda\Lambda^{\top}+\Psi;S)\\
\mbox{s.t.~} & |\lambda_{ik}|\leq \tau \mbox{~and~} \lambda_{jk}=0 \mbox{~for~} k=1,\ldots,K, i\in v^{*}_{k}, j\notin v^{*}_{k},\\
& \Psi = \mbox{diag}(\psi_1,\ldots,\psi_{J}), \kappa_1\leq|\psi_j|\leq \kappa_2, j=1,\ldots,J.
\end{aligned}
\end{equation}

\begin{lemma}\label{lem: confirmatory rate of convergence}
Suppose that the hierarchical factor structure is known. If Conditions~\ref{cond:true para}, \ref{cond:rank}, \ref{cond:bic correct}, \ref{cond:correct pilot} and \ref{cond:hier compact set} hold, we have 
\begin{equation}
\|\hat{\Lambda} - \Lambda^{*}\hat{Q}\|_{F} =  O_{\mathbb{P}}(1/\sqrt{N}) \mbox{~and~} \|\hat{\Psi}-\Psi^{*}\|_{F} = O_{\mathbb{P}}(1/\sqrt{N}),
\end{equation}
where $\hat{Q}$ is the diagonal matrix with diagonal entries consisting of the signs of the corresponding entries of $\hat{\Lambda}^{\top}\Lambda^*$ defined in Theorem~\ref{cor:consistency}.
\end{lemma}

\begin{proof}
Similar to the proof of Lemma~\ref{lem: convergence rate covariance}, we have $\|\hat{\Lambda}\hat{\Lambda}^{\top}+ \hat{\Psi}-\Sigma^{*}\|_{F} = O_{\mathbb{P}}(1/\sqrt{N})$. Furthermore, according to Lemma~\ref{lem:apprxi anderson}, we have $\|\hat{\Lambda}\hat{\Lambda}^{\top}-\Lambda^{*}\Lambda^{*\top}\|_{F} = O_{\mathbb{P}}(1/\sqrt{N})$ and $\|\hat{\Psi}-\Psi^{*}\|_{F} = O_{\mathbb{P}}(1/\sqrt{N})$.

To prove that $\|\hat{\Lambda} - \Lambda^{*}\hat{Q}\|_{F}= O_{\mathbb{P}}(1/\sqrt{N})$, we first show that there exists some orthogonal rotation matrix $R$ such that $\|\hat{\Lambda} - \Lambda^{*}R\|_{F}= O_{\mathbb{P}}(1/\sqrt{N})$. Second, we show that $\|\hat{\boldsymbol{\lambda}}_1 - \boldsymbol{\lambda}_{1}^{*}\mbox{sign}(\hat{\boldsymbol{\lambda}}_{1}^{\top}\boldsymbol{\lambda}_{1}^{*})\|_{F}= O_{\mathbb{P}}(1/\sqrt{N})$. Third, we conclude the proof by recursively applying the same argument to the factors in the $t$th layer, $t = 2,\ldots, T$.

Let $$\Lambda^{*} = U^{*}\mbox{diag}\left(\sigma_{1}(\Lambda^{*}),\ldots,\sigma_{K}(\Lambda^{*})\right)V^{*\top}$$ be the singular value decomposition of $\Lambda^{*}$ and $$\hat{\Lambda} = \hat{U}\mbox{diag}\left(\sigma_{1}(\hat{\Lambda}),\ldots,\sigma_{K}(\hat{\Lambda})\right)\hat{V}^{\top}$$ be the singular value decomposition of $\hat{\Lambda}$. Then $\Lambda^{*}\Lambda^{*\top} = U^{*}\mbox{diag}\left(\sigma^{2}_{1}(\Lambda^{*}),\ldots,\sigma^{2}_{K}(\Lambda^{*})\right)U^{*\top}$ and $\hat{\Lambda}\hat{\Lambda}^{\top} =  \hat{U}\mbox{diag}\left(\sigma_{1}^{2}(\hat{\Lambda}),\ldots,\sigma_{K}^{2}(\hat{\Lambda})\right)\hat{U}^{\top}$.
By Lemma~\ref{lem: Weyl}, $|\sigma^{2}_{i}(\Lambda^{*}) - \sigma_{i}^{2}(\hat{\Lambda})| = O_{\mathbb{P}}(1/\sqrt{N})$ for $i=1,\ldots, K$, which further leads to $|\sigma_{i}(\Lambda^{*}) - \sigma_{i}(\hat{\Lambda})| = O_{\mathbb{P}}(1/\sqrt{N})$ for all $i$. By Lemma~\ref{lem: wedin}, there exits some orthogonal rotation matrix $\tilde{R}$ such that 
$\|\hat{U} - U^{*}\tilde{R}\|_{F} = O_{\mathbb{P}}(1/\sqrt{N})$. Moreover, $\tilde{R}$ satisfies $\tilde{R}\mbox{diag}\left(\sigma_{1}(\Lambda^{*}),\ldots,\sigma_{K}(\Lambda^{*})\right) = \mbox{diag}\left(\sigma_{1}(\Lambda^{*}),\ldots,\sigma_{K}(\Lambda^{*})\right)\tilde{R}$ with probability approaching 1 as $N$ grows to infinity. Taking $R = V^{*}\tilde{R}\hat{V}^{\top}$, we have
\begin{equation}\label{eq:confirmatory rate convergence rotation gap}
\begin{aligned}
&\|\hat{\Lambda}-\Lambda^{*}R\|_{F}\\
=&\left\|\hat{U}\mbox{diag}\left(\sigma_{1}(\hat{\Lambda}),\ldots,\sigma_{K}(\hat{\Lambda})\right)\hat{V}^{\top} - U^{*}\mbox{diag}\left(\sigma_{1}(\Lambda^{*}),\ldots,\sigma_{K}(\Lambda^{*})\right)\tilde{R}\hat{V}^{\top}\right\|_{F} \\
=&\|\hat{U}\mbox{diag}\left(\sigma_{1}(\hat{\Lambda}),\ldots,\sigma_{K}(\hat{\Lambda})\right) - U^{*}\tilde{R}\mbox{diag}\left(\sigma_{1}(\Lambda^{*}),\ldots,\sigma_{K}(\Lambda^{*})\right)\|_{F} \\
\leq&  \left\|\hat{U}\left(\mbox{diag}\left(\sigma_{1}(\hat{\Lambda}),\ldots,\sigma_{K}(\hat{\Lambda})\right) - \mbox{diag}\left(\sigma_{1}(\Lambda^{*}),\ldots,\sigma_{K}(\Lambda^{*})\right)\right)\right\|_{F}+ \big\|(\hat{U}-U^{*}\tilde{R})\\
&\mbox{diag}\left(\sigma_{1}(\Lambda^{*}),\ldots,\sigma_{K}(\Lambda^{*})\right)\big\|_{F} \\
=&O_{\mathbb{P}}(1/\sqrt{N}).
\end{aligned}
\end{equation}
For $i,j\in\text{Ch}_{1}^{*}$, $i\neq j$, by Lemma~\ref{lem: wedin} and 
$$\left\|\hat{\Lambda}_{[v^{*}_{i}, \{1\}]}\hat{\Lambda}_{[v^{*}_{j}, \{1\}]}^{\top} - \Lambda^{*}_{[v^{*}_{i}, \{1\}]}(\Lambda^{*}_{[v^{*}_{j}, \{1\}]})^{\top} \right\|_{F} = O_{\mathbb{P}}(1/\sqrt{N}),$$
we have
\begin{equation}
\left\|\frac{\hat{\Lambda}_{[v_{i}^{*} ,\{1\}]}}{\big\|\hat{\Lambda}_{[v_{i}^{*},\{1\}]} \big\|}  -  \frac{\Lambda^{*}_{[v_{i}^{*} ,\{1\}]}}{\big\|\Lambda^{*}_{[v_{i}^{*} ,\{1\}]} \big\|}\mbox{sign}\left(\hat{\Lambda}_{[v_{i}^{*} ,\{1\}]}^{\top}\Lambda^{*}_{[v_{i}^{*} ,\{1\}]}\right)\right\| = O_{\mathbb{P}}(1/\sqrt{N}).
\end{equation}
Then, we further have
\begin{equation}\label{eq:confirm first column positive gap final}
\left\|\frac{\hat{\Lambda}_{[v_{j}^{*} ,\{1\}]}}{\big\|\hat{\Lambda}_{[v_{j}^{*},\{1\}]} \big\|}  -  \frac{\Lambda^{*}_{[v_{j}^{*} ,\{1\}]}}{\big\|\Lambda^{*}_{[v_{j}^{*} ,\{1\}]} \big\|}\mbox{sign}\left(\hat{\Lambda}_{[v_{i}^{*} ,\{1\}]}^{\top}\Lambda^{*}_{[v_{i}^{*} ,\{1\}]}\right)\right\| = O_{\mathbb{P}}(1/\sqrt{N})
\end{equation}
for all $j\in\text{Ch}_{1}^{*}$, which also leads to the fact that $\mbox{sign}\left(\hat{\Lambda}_{[v_{i}^{*} ,\{1\}]}^{\top}\Lambda^{*}_{[v_{i}^{*} ,\{1\}]}\right) = \mbox{sign}(\hat{\boldsymbol{\lambda}}_{1}^{\top}\boldsymbol{\lambda}_{1}^{*})$ with probability approaching 1 as $N$ grows to infinity. According to~\eqref{eq:confirmatory rate convergence rotation gap}, for each $i\in\text{Ch}_{1}^{*}$, we have
\begin{equation*}
    \left\|\hat{\Lambda}_{[v_{i}^{*},\{1\}]} - \Lambda^{*}_{[v_{i}^{*},\{1,i\}\cup D^{*}_{i}]}R_{[\{1,i\}\cup D^{*}_{i},\{1\}]}\right\|_{F} = O_{\mathbb{P}}(1/\sqrt{N}).
\end{equation*}
Let $$P_{i} = \frac{\Lambda^{*}_{[v_{i}^{*},\{1\}]}\big(\Lambda^{*}_{[v_{i}^{*},\{1\}]}\big)^{\top}}{\big(\Lambda^{*}_{[v_{i}^{*},\{1\}]}\big)^{\top}\Lambda^{*}_{[v_{i}^{*},\{1\}]}}$$ and $\Lambda^{*}_{\text{Proj},i} = (\mathbf{I}-P_{i})\Lambda^{*}_{[v_{i}^{*},\{i\}\cup D^{*}_{i}]}$. By Condition~\ref{cond:rank}, $\sigma_{1+|D^{*}_{i}|}\big(\Lambda^{*}_{\text{Proj},i}\big) >0$. We have
\begin{equation*}
\begin{aligned}
&\left\|\hat{\Lambda}_{[v_{i}^{*},\{1\}]}-\Lambda^{*}_{[v_{i}^{*},\{1,i\}\cup D^{*}_{i}]}R_{[\{1,i\}\cup D^{*}_{i},\{1\}]}\right\| \\
\geq& \left\|\frac{\big\|\hat{\Lambda}_{[v_{i}^{*} ,\{1\}]}\big\|}{\big\|\Lambda^{*}_{[v_{i}^{*} ,\{1\}]}\big\|}\Lambda^{*}_{[v_{i}^{*} ,\{1\}]}\mbox{sign}(\hat{\boldsymbol{\lambda}}_{1}^{\top}\boldsymbol{\lambda}_{1}^{*}) - \Lambda^{*}_{[v_{i}^{*},\{1,i\}\cup D^{*}_{i}]}R_{[\{1,i\}\cup D^{*}_{i},\{1\}]}\right\|\\
& - \big\|\hat{\Lambda}_{[v_{i}^{*} ,\{1\}]}\big\| \left\|\frac{\hat{\Lambda}_{[v_{i}^{*} ,\{1\}]}}{\big\|\hat{\Lambda}_{[v_{i}^{*},\{1\}]} \big\|} -  \frac{\Lambda^{*}_{[v_{i}^{*} ,\{1\}]}}{\big\|\Lambda^{*}_{[v_{i}^{*} ,\{1\}]} \big\|}\mbox{sign}(\hat{\boldsymbol{\lambda}}_{1}^{\top}\boldsymbol{\lambda}_{1}^{*})\right\|\\
\geq& \big\|R_{[\{i\}\cup D^{*}_{i},\{1\}]}\big\|\sigma_{1+|D_{i}^{*}|}\big(\Lambda^{*}_{\text{Proj},i}\big) + O_{\mathbb{P}}(1/\sqrt{N}).
\end{aligned}
\end{equation*}
Thus, we have $\big\|R_{[\{i\}\cup D^{*}_{i},\{1\}]}\big\| = O_{\mathbb{P}}(1/\sqrt{N})$ for all $i\in \text{Ch}^{*}_{i}$, which leads to $\big\|R_{[\{2,\ldots,K\},\{1\}]}\big\| = O_{\mathbb{P}}(1/\sqrt{N})$ and $\big|R_{[\{1\},\{1\}]}-\mbox{sign}(\hat{\boldsymbol{\lambda}}_{1}^{\top}\boldsymbol{\lambda}_{1}^{*})\big| = O_{\mathbb{P}}(1/\sqrt{N})$. We then have
\begin{equation}\label{eq:hier algo proof rot final eq}
\begin{aligned}
&\big\|\hat{\boldsymbol{\lambda}}_{1} -\boldsymbol{\lambda}_{1}^{*}\mbox{sign}(\hat{\boldsymbol{\lambda}}_{1}^{\top}\boldsymbol{\lambda}_{1}^{*})\big\| \\
\leq & \big\|\hat{\boldsymbol{\lambda}}_{1} - \Lambda^{*}R_{[:,\{1\}]}\big\| + \big|R_{[\{1\},\{1\}]}-\mbox{sign}(\hat{\boldsymbol{\lambda}}_{1}^{\top}\boldsymbol{\lambda}_{1}^{*})\big|\big\|\boldsymbol{\lambda}_{1}^{*}\big\| \\
&+ \big\|R_{[\{2,\ldots,K\},\{1\}]}\big\|\big\|\Lambda^{*}_{[:,\{2,\ldots,K\}]}\big\|_{F} \\
=& O_{\mathbb{P}}(1/\sqrt{N}).
\end{aligned}
\end{equation}
Finally, with~\eqref{eq:hier algo proof rot final eq}, we have $$\|\hat{\Lambda}_{[v_{i}^{*},\{i\}\cup D^{*}_{i}]}\hat{\Lambda}^{\top}_{[v_{i}^{*},\{i\}\cup D^{*}_{i}]}-\Lambda^{*}_{[v_{i}^{*},\{i\}\cup D^{*}_{i}]}\Lambda^{*\top}_{[v_{i}^{*},\{i\}\cup D^{*}_{i}]}\|_{F} = O_{\mathbb{P}}(1/\sqrt{N}),$$
for all $i\in \mbox{Ch}_{1}^{*}$. Then, Lemma~\ref{lem: confirmatory rate of convergence} can be proved by recursively applying the same argument to the factors in the $t$th layer, $t = 2,\ldots, T$.
\end{proof}
We now give the proof of Theorem~\ref{cor:consistency}.
\begin{proof}
The proof follows in a recursive manner.  We first prove that with probability approaching 1 as $N$ grows to infinity, 
\begin{equation}\label{eq: theorem 2 goal factor 1}
\hat{\text{Ch}}_1 = \text{Ch}_{1}^{*}, \mbox{~and~}\hat{v}_i = v_{i}^{*} \mbox{~for all~} i\in \text{Ch}_{1}^{*},
\end{equation}
and as a by-product, $\|\tilde{\boldsymbol{\lambda}}_{1} - \boldsymbol{\lambda}_{1}^{*}\mbox{sign}(\boldsymbol{\lambda}_{1}^{*\top}\tilde{\boldsymbol{\lambda}}_{1})\| = O_{\mathbb{P}}(1/\sqrt{N})$, which further implies $\|\tilde{\boldsymbol{\lambda}}_{1}\tilde{\boldsymbol{\lambda}}_{1}^{\top} - \boldsymbol{\lambda}_{1}^{*}\boldsymbol{\lambda}_{1}^{*\top}\|_{F} = O_{\mathbb{P}}(1/\sqrt{N})$. Then Theorem~\ref{cor:consistency} is proved by applying the same argument to the factors in the $t$th layer, $t=2,\ldots, T$, in conjunction with Lemma~\ref{lem: confirmatory rate of convergence}.

For simplicity of the notation, we denote $c^{*} = |\mbox{Ch}_{1}^{*}|$. When $c^{*} = 0$, the proof of \eqref{eq: theorem 2 goal factor 1} is trivial. When $c^{*} \geq 2$, the proof of \eqref{eq: theorem 2 goal factor 1} consists of two main steps:
\begin{enumerate}
    \item For sufficiently large $d$ such that $\Sigma^{*}\in \Theta_{1}(c,d)$, let $\bar{\Lambda}_{1,c}$ and $\bar{\Psi}_{1,c}$ be the estimates according to \eqref{opt:alm aug bbf} and $v^{1,c}_{1},\ldots, v^{1,c}_{c}$ be the sets of variables belonging to the child factors of factor 1 decoded by $\bar{\Lambda}_{1,c}$. The possible configurations for $v^{1,c}_{1},\ldots, v^{1,c}_{c}$ are:
    \begin{enumerate}
        \item[A.] For each $k\in \mbox{Ch}_{1}^{*}$, there exists some $s\in\{1,\ldots,c\}$ such that $v_{k}^{*}\subset v^{1,c}_{s}$.
        \item[B.] There exists some $j\in \{1,\ldots,J\}$ such that $$(v^{1,c}_{1},v^{1,c}_{2}) = (\{1,\ldots,J\}\setminus\{j\},\{j\}) \mbox{~and~} v^{1,c}_{s} = \emptyset \mbox{~for~} s>2.$$
    \end{enumerate}

    \item $c^{*} = \argmin_{c=0,2,\ldots,c_{\max}}\tilde{\mbox{IC}}_{1,c}$ and $\tilde{d}_{s}^{c^{*}} = 1+ |D_{1+s}^{*}|$ for $s=1,\ldots,c^{*}$ with probability approaching 1 as $N$ grows to infinity.
\end{enumerate}

Given $v^{1,c}_{1},\ldots, v^{1,c}_{c}$, we prove the first part by showing that for arbitrary $\Lambda\in \tilde{\mathcal{A}}^{1}(c,d_1\ldots,d_c)$ and $\Psi$, there exists some constant $C>0$ depending only on $\Lambda^{*}$ such that $\|\Lambda\Lambda^{\top}+\Psi -\Sigma^{*}\|_{F} \geq C$ if $v^{1,c}_{1},\ldots, v^{1,c}_{c}$ are not in case A or B. Let $\mathcal{B}_{i,s} = v^{*}_{i} \cap v_{s}^{1,c^{*}}$ for $i=2,\ldots,1+c^*$ and $s=1,\ldots,c$. We first claim that such a constant $C$ exists if there exists $i\in\{2,\ldots,1+c^*\}$ such that the following cases do not hold: (1) $\mathcal{B}_{i,s} = v^{*}_{i}$ for some $s\in\{1,\ldots,c\}$ and (2) $v^{*}_{i} = \mathcal{B}_{i,s_1} \cup \mathcal{B}_{i,s_2}$, $\mathcal{B}_{i,s_1}, \mathcal{B}_{i,s_2}\neq\emptyset$ for $s_1, s_2\in\{1,\ldots,c\}$ and $|v^{1,c}_{s_2}| = 1$. We then claim that such a constant $C$ exists if the second case holds for some $i$ but case B does not hold.

Now we give the proof of the first claim. Let $\Sigma = \Lambda\Lambda^{\top}+\Psi$. For $i=2,\ldots,1+c^*$, consider the following cases where $\text{Ch}_{i}^{*} = \emptyset$:
\begin{enumerate}
    \item $|\{s : |\mathcal{B}_{i,s}|\geq 1 \}| \geq 4$. Let $s_1,\ldots,s_4 \in \{1,\ldots,c\}$ such that $|\mathcal{B}_{i,s_1}|\geq 1, \ldots, |\mathcal{B}_{i,s_4}|\geq 1$ and $j_1 \in \mathcal{B}_{i,s_1}, \ldots, j_4 \in \mathcal{B}_{i,s_4}$. We have
    \begin{equation*}
    \Sigma_{[\{j_1,j_2\}, \{j_{3},j_4\}]} = \Lambda_{[\{j_1,j_2\},\{1\}]}(\Lambda_{[\{j_3,j_4\},\{1\}]})^{\top},
    \end{equation*}
    has rank 1, while by Condition~\ref{cond:rank},
    \begin{equation*}
    \Sigma^{*}_{[\{j_1,j_2\}, \{j_{3},j_4\}]} = \Lambda^{*}_{[\{j_1,j_2\},\{1,i\}]}(\Lambda^{*}_{[\{j_3,j_4\},\{1,i\}]})^{\top},
    \end{equation*}
    has rank 2. By Lemma~\ref{lem: Weyl}, we have
    \begin{equation}\label{eq:hier algo 2*2 block}
    \begin{aligned}
    &\|\Sigma-\Sigma^{*}\|_{F} \\
    \geq &  \big\|\Sigma_{[\{j_1,j_2\}, \{j_{3},j_4\}]}-\Sigma^{*}_{[\{j_1,j_2\}, \{j_{3},j_4\}]} \big\|_{F}\\
    \geq & \sigma_{2}\big(\Lambda^{*}_{[\{j_1,j_2\},\{1,i\}]}(\Lambda^{*}_{[\{j_1,j_2\},\{1,i\}]})^{\top}\big)\\
    > &0.
    \end{aligned}
    \end{equation}
    \item There exists some $1\leq s \leq c$ such that $|\mathcal{B}_{i,s}|\geq 2$ and $|v^{*}_{i}\setminus \mathcal{B}_{i,s}|\geq 2$. In this case, choose $j_1, j_2 \in \mathcal{B}_{i,s}$ and $j_3, j_4 \in v^{*}_{i}\setminus \mathcal{B}_{i,s}$,~\eqref{eq:hier algo 2*2 block} also holds.
    \item There exists some $1\leq s \leq c$ such $|\mathcal{B}_{i,s}| = 1$ and $|v^{1,c}_{s}| > 1$. Let $\{j\} = \mathcal{B}_{i,s}$ and we have
    \begin{equation}\label{eq:hier algo proof no child case3 eq main}
    \begin{aligned}
    &\|\Sigma - \Sigma^*\|_{F} \\
    \geq & \frac{1}{\sqrt{2}}\left( \big\|\Sigma_{[v_{i}^{*}\setminus \{j\},\{j\}]} - \Sigma^{*}_{[v_{i}^{*}\setminus \{j\},\{j\}]} \big\| +  \big\|\Sigma_{[v_{i}^{*}\setminus \{j\},v^{1,c}_{s}\setminus\{j\}]} - \Sigma^{*}_{[v_{i}^{*}\setminus \{j\},v^{1,c}_{s}\setminus\{j\}]} \big\|_{F}\right).
    \end{aligned}
    \end{equation}
    Notice that $$\Sigma_{[v_{i}^{*}\setminus \{j\},v^{1,c}_{s}\setminus\{j\}]} = \Lambda_{[v_{i}^{*}\setminus \{j\},\{1\}]}(\Lambda_{[v^{1,c}_{s}\setminus\{j\},\{1\}]})^{\top}
    ,$$ and $$\Sigma^{*}_{[v_{i}^{*}\setminus \{j\},v^{1,c}_{s}\setminus\{j\}]} = \Lambda^{*}_{[v_{i}^{*}\setminus \{j\},\{1\}]}(\Lambda^{*}_{[v^{1,c}_{s}\setminus\{j\},\{1\}]})^{\top}.$$ We denote by $$\delta =  \big\|\Sigma_{[v_{i}^{*}\setminus \{j\},v^{1,c}_{s}\setminus\{j\}]} - \Sigma^{*}_{[v_{i}^{*}\setminus \{j\},v^{1,c}_{s}\setminus\{j\}]} \big\|_{F}. $$ By Lemma~\ref{lem: wedin}, 
    \begin{equation}\label{eq:hier algo proof sine1}
    \left\|\frac{\Lambda_{[v_{i}^{*}\setminus \{j\},\{1\}]}}{ \big\|\Lambda_{[v_{i}^{*}\setminus \{j\},\{1\}]}  \big\|}  -  \frac{\Lambda^{*}_{[v_{i}^{*}\setminus \{j\},\{1\}]}}{ \big\|\Lambda^{*}_{[v_{i}^{*}\setminus \{j\},\{1\}]}  \big\|}\right\|\leq \frac{2^{3/2}\delta}{ \big\|\Lambda^{*}_{[v_{i}^{*}\setminus \{j\},\{1\}]} \big\|  \big\|\Lambda^{*}_{[v^{1,c}_{s}\setminus\{j\},\{1\}]} \big\|},
    \end{equation}
    or
    \begin{equation*}
    \left\|\frac{\Lambda_{[v_{i}^{*}\setminus \{j\},\{1\}]}}{ \big\|\Lambda_{[v_{i}^{*}\setminus \{j\},\{1\}]}  \big\|}  +  \frac{\Lambda^{*}_{[v_{i}^{*}\setminus \{j\},\{1\}]}}{ \big\|\Lambda^{*}_{[v_{i}^{*}\setminus \{j\},\{1\}]}  \big\|}\right\|\leq \frac{2^{3/2}\delta}{ \big\|\Lambda^{*}_{[v_{i}^{*}\setminus \{j\},\{1\}]} \big\|  \big\|\Lambda^{*}_{[v^{1,c}_{s}\setminus\{j\},\{1\}]} \big\|}
    \end{equation*}
    holds. Without loss of generality, we assume that~\eqref{eq:hier algo proof sine1} holds. On the other hand, notice that $$\Sigma_{[v_{i}^{*}\setminus \{j\},\{j\}]} = \lambda_{j,1}\Lambda_{[v_{i}^{*}\setminus \{j\},\{1\}]},$$ and $$\Sigma^{*}_{[v_{i}^{*}\setminus \{j\},\{j\}]} = \lambda_{j,1}^{*}\Lambda^{*}_{[v_{i}^{*}\setminus \{j\},\{1\}]} + \lambda_{j,i}^{*}\Lambda^{*}_{[v_{i}^{*}\setminus \{j\},\{i\}]}.$$
    Let $$P_{i} = \frac{\Lambda^{*}_{[v_{i}^{*}\setminus \{j\},\{1\}]}\big(\Lambda^{*}_{[v_{i}^{*}\setminus \{j\},\{1\}]}\big)^{\top}}{\big(\Lambda^{*}_{[v_{i}^{*}\setminus \{j\},\{1\}]}\big)^{\top}\Lambda^{*}_{[v_{i}^{*}\setminus \{j\},\{1\}]}}$$ and $\boldsymbol{\mu} = (\mathbf{I}-P_{i})\Lambda^{*}_{[v_{i}^{*}\setminus \{j\},\{j\}]}$. 
    According to Condition~\ref{cond:rank}, $\boldsymbol{\mu}\neq\mathbf{0}$. According to Condition~\ref{cond:hier compact set}, we have
    \begin{equation}~\label{eq:hier algo proof no child case3 eq main2}
    \begin{aligned}
    & \big\|\Sigma_{[v_{i}^{*}\setminus \{j\},\{j\}]} - \Sigma^{*}_{[v_{i}^{*}\setminus \{j\},\{j\}]}\big\| \\
    = & \big\|\lambda_{j,1}\Lambda_{[v_{i}^{*}\setminus \{j\},\{1\}]} -\lambda_{j,1}^{*}\Lambda^{*}_{[v_{i}^{*}\setminus \{j\},\{1\}]} - \lambda_{j,i}^{*}\Lambda^{*}_{[v_{i}^{*}\setminus \{j\},\{i\}]} \big\| \\
    \geq& |\lambda_{j,i}^{*}|\|\boldsymbol{\mu}\| - |\lambda_{j,1}|\left\|\Lambda_{[v_{i}^{*}\setminus \{j\},\{1\}]} - \frac{\big\|\Lambda_{[v_{i}^{*}\setminus \{j\},\{1\}]}\big\|}{\big\|\Lambda^{*}_{[v_{i}^{*}\setminus \{j\},\{1\}]} \big\|}\Lambda^{*}_{[v_{i}^{*}\setminus \{j\},\{1\}]}\right\| \\
    \geq& |\lambda_{j,i}^{*}|\|\boldsymbol{\mu}\| - \frac{2^{3/2}\tau^{2}\delta(|v_{i}^{*}|-1)^{1/2}}{\big\|\Lambda^{*}_{[v_{i}^{*}\setminus \{j\},\{1\}]}\big\| \big\|\Lambda^{*}_{[v^{1,c}_{s}\setminus\{j\},\{1\}]}\big\|}.
    \end{aligned}
    \end{equation}
    Combining~\eqref{eq:hier algo proof no child case3 eq main} and~\eqref{eq:hier algo proof no child case3 eq main2}, we have
    \begin{equation*}
    \|\Sigma-\Sigma^{*}\|_{F}\geq\min\left(\frac{\sqrt{2}}{4},\frac{\big\|\Lambda^{*}_{[v_{i}^{*}\setminus \{j\},\{1\}]}\big\| \big\|\Lambda^{*}_{[v^{1,c}_{s}\setminus\{j\},\{1\}]}\big\|}{8\tau^{2}(|v_{i}^{*}|-1)^{1/2}}\right)|\lambda_{j,i}^{*}|\|\boldsymbol{\mu}\|>0.
    \end{equation*}

    \item $|v_{i}^{*}| = \cup_{k=1,2,3}\mathcal{B}_{i,s_{k}}$ with $\{j_{k}\} = \mathcal{B}_{i,s_{k}}$ for $k=1, 2, 3$. If there exists some $k$ such that $|v^{1,c}_{s_{k}}|>1$, with a similar argument from \eqref{eq:hier algo proof no child case3 eq main} to \eqref{eq:hier algo proof no child case3 eq main2}, we have
    \begin{equation*}
    \|\Sigma-\Sigma^{*}\|_{F}\geq\min\left(\frac{\sqrt{2}}{4},\frac{\big\|\Lambda^{*}_{[v_{i}^{*}\setminus \{j_{k}\},\{1\}]}\big\| \big\|\Lambda^{*}_{[v^{1,c}_{s_{k}}\setminus\{j_{k}\},\{1\}]}\big\|}{8\tau^{2}(|v_{i}^{*}|-1)^{1/2}}\right)|\lambda_{j_{k},i}^{*}|\|\boldsymbol{\mu}\|>0,
    \end{equation*}
    where $\boldsymbol{\mu}$ is defined similarly in \eqref{eq:hier algo proof no child case3 eq main2}. Otherwise, $\{j_{k}\} = v^{1,c}_{s_{k}}$ for $k=1, 2, 3$. Consider $i' \in \text{Ch}^{*}_{i}$ and $i'\neq i$. We have
    \begin{equation}
    \begin{aligned}
    &\|\Sigma - \Sigma^*\|_{F} \\
    \geq & \frac{1}{\sqrt{2}}\left( \big\|\Sigma_{[v_{i}^{*}\setminus\{j_k\},\{j_k\}]} - \Sigma^{*}_{[v_{i}^{*}\setminus\{j_k\},\{j_k\}]} \big\| +  \big\|\Sigma_{[v_{i}^{*},v_{i'}^{*}]]} - \Sigma^{*}_{[v_{i}^{*},v_{i'}^{*}]]} \big\|_{F}\right).
    \end{aligned}
    \end{equation}
    With a similar argument from \eqref{eq:hier algo proof no child case3 eq main} to \eqref{eq:hier algo proof no child case3 eq main2}, we have
    \begin{equation*}
    \|\Sigma-\Sigma^{*}\|_{F}\geq\min\left(\frac{\sqrt{2}}{4},\frac{\big\|\Lambda^{*}_{[v_{i}^{*}\setminus \{j_{k}\},\{1\}]}\big\| \big\|\Lambda^{*}_{[v^{*}_{i'},\{1\}]}\big\|}{8\tau^{2}(|v_{i}^{*}|-1)^{1/2}}\right)|\lambda_{j_{k},i}^{*}|\|\boldsymbol{\mu}_{k}\|>0,
    \end{equation*}
    where $\boldsymbol{\mu}_{k}$s are defined similarly in \eqref{eq:hier algo proof no child case3 eq main2} for $k=1, 2, 3$.

    \item The rest of the cases are included in the two cases of our first claim.
\end{enumerate}
When $\mbox{Ch}_{i}^{*} \neq \emptyset$, consider the following cases:
\begin{enumerate}
\item There exist $k\in \text{Ch}^{*}_{i}$ and $s = 1,\ldots,c$ such that $|\mathcal{B}_{i,s}\cap v_{k}^{*}|\geq 2$. If we further have 
\begin{equation*}
|(\cup_{1\leq s'\leq c, s'\neq s}\mathcal{B}_{i,s'}) \cap (\cup_{k'\neq k, k'\in \text{Ch}^{*}_{i}}v_{k'}^{*})|\geq 2,
\end{equation*}
choose $j_1, j_2 \in \mathcal{B}_{i,s}\cap v_{k}^{*}$ and $j_3, j_4 \in (\cup_{1\leq s'\leq c^{*}, s'\neq s} \mathcal{B}_{i,s'}) \cap (\cup_{k'\neq k, k'\in \text{Ch}^{*}_{1}}v_{k'}^{*})$. We have
    \begin{equation*}
    \Sigma_{[\{j_1,j_2\}, \{j_{3},j_4\}]} = \Lambda_{[\{j_1,j_2\},\{1\}]}(\Lambda_{[\{j_3,j_4\},\{1\}]})^{\top},
    \end{equation*}
    which has rank 1, while by Condition~\ref{cond:rank}
    \begin{equation*}
    \Sigma^{*}_{[\{j_1,j_2\}, \{j_{3},j_4\}]} = \Lambda^{*}_{[\{j_1,j_2\},\{1,i\}]}(\Lambda^{*}_{[\{j_3,j_4\},\{1,i\}]})^{\top},
    \end{equation*}
    has rank 2. By Lemma~\ref{lem: Weyl}, we also have
    \begin{equation*}
    \begin{aligned}
    &\|\Sigma-\Sigma^{*}\|_{F} \\
    \geq & \big\|\Sigma_{[\{j_1,j_2\}, \{j_{3},j_4\}]}-\Sigma^{*}_{[\{j_1,j_2\}, \{j_{3},j_4\}]}\big\|_{F}\\
    \geq & \sigma_{2}\big(\Lambda^{*}_{[\{j_1,j_2\},\{1,i\}]}(\Lambda^{*}_{[\{j_1,j_2\},\{1,i\}]})^{\top}\big)\\
    > &0.
    \end{aligned}
    \end{equation*}

If 
\begin{equation}\label{eq:hier algo thm case2 claim1}
|(\cup_{1\leq s'\leq c, s'\neq s}\mathcal{B}_{i,s'}) \cap (\cup_{k'\neq k, k'\in \text{Ch}^{*}_{i}}v_{k'}^{*})|\leq 1,
\end{equation}

since $|v^{*}_{k'}|\geq 3$ for $k'\neq k, k'\in \text{Ch}^{*}_{i}$, by~\eqref{eq:hier algo thm case2 claim1} we also have $|\mathcal{B}_{i,s}\cap v_{k'}^{*}|\geq 2$ for all $k'\in \text{Ch}^{*}_{i}$. Similar to~\eqref{eq:hier algo thm case2 claim1}, we have
\begin{equation}\label{eq:hier algo thm case2 claim2}
|(\cup_{1\leq s'\leq c, s'\neq s}\mathcal{B}_{i,s'}) \cap v_{k}^{*}|\leq 1.
\end{equation}
Combining~\eqref{eq:hier algo thm case2 claim1} and~\eqref{eq:hier algo thm case2 claim2}, we have
\begin{equation*}
|(\cup_{1\leq s'\leq c, s'\neq s}\mathcal{B}_{i,s'}) \cap (\cup_{k'\in \text{Ch}^{*}_{i}}v_{k'}^{*})|\leq 2.
\end{equation*}
First, if $|(\cup_{1\leq s'\leq c, s'\neq s}\mathcal{B}_{i,s'}) \cap (\cup_{k'\in \text{Ch}^{*}_{i}}v_{k'}^{*})| = 2$, we denote by $k'\neq k$ such that~\eqref{eq:hier algo thm case2 claim1} is tight. Choose $j_1, j_2\in \mathcal{B}_{i,s}\cap v_{k}^{*}$, $j_3, j_4\in \mathcal{B}_{i,s}\cap v_{k'}^{*}$, $j_5 \in (\cup_{1\leq s'\leq c, s'\neq s}\mathcal{B}_{i,s'}) \cap v_{k}^{*}$ and $j_6\in (\cup_{1\leq s'\leq c, s'\neq s}\mathcal{B}_{i,s'}) \cap v_{k'}^{*}$. Furthermore, we require that when $ \text{Ch}^{*}_{k} \neq \emptyset$, $j_1, j_2$ belong to different child factors of factor $k$ with $j_5$ and when $ \text{Ch}^{*}_{k'} \neq \emptyset$, $j_3, j_4$ belong to different child factors of factor $k'$ with $j_6$. Such a choice is always possible due to the assumed structure of the hierarchical model. It is easy to check that
    \begin{equation*}
    \Sigma_{[\{j_1,j_2,j_3,j_4\},\{j_5,j_6\}]} = \Lambda_{[\{j_1,j_2,j_3,j_4\},\{1\}]}(\Lambda_{[\{j_5,j_6\},\{1\}]})^{\top}
    \end{equation*}
    has rank 1. On the other hand,
    \begin{equation*}
    \Sigma^{*}_{[\{j_1,j_2,j_3,j_4\},\{j_5,j_6\}]} = \Lambda^{*}_{[\{j_1,j_2,j_3,j_4\},\{1,i,k,k'\}]}(\Lambda^{*}_{[\{j_5,j_6\},\{1,i,k,k'\}]})^{\top}.
    \end{equation*}
    According to Condition~\ref{cond:rank}, the rank of $\Lambda^{*}_{[\{j_1,j_2,j_3,j_4\},\{1,i,k,k'\}]}$ is 4 and the rank of $\Lambda^{*}_{[\{j_5,j_6\},\{1,i,k,k'\}]}$ is 2. By Sylvester's rank inequality,
    \begin{equation*}
    \begin{aligned}
     &\text{rank}\big(\Lambda^{*}_{[\{j_1,j_2,j_3,j_4\},\{1,i,k,k'\}]}(\Lambda^{*}_{[\{j_5,j_6\},\{1,i,k,k'\}]})^{\top}\big)\\
    \geq &\text{rank}\big(\Lambda^{*}_{[\{j_1,j_2,j_3,j_4\},\{1,i,k,k'\}]}\big) + \text{rank}\big(\Lambda^{*}_{[\{j_5,j_6\},\{1,i,k,k'\}]}\big)-4\\
    = &2.
    \end{aligned}
    \end{equation*}
    Thus, by Lemma~\ref{lem: Weyl},
    \begin{equation*}
    \begin{aligned}
    &\|\Sigma-\Sigma^{*}\|_{F} \\
    \geq & \big\|\Sigma_{[\{j_1,j_2,j_3,j_4\},\{j_5,j_6\}]}-\Sigma^{*}_{[\{j_1,j_2,j_3,j_4\},\{j_5,j_6\}]}\big\|_{F}\\
    \geq & \sigma_{2}\big(\Lambda^{*}_{[\{j_1,j_2,j_3,j_4\},\{1,i,k,k'\}]}(\Lambda^{*}_{[\{j_5,j_6\},\{1,i,k,k'\}]})^{\top}\big)\\
    > &0.
    \end{aligned}
    \end{equation*}

Second, if $|(\cup_{1\leq s'\leq c, s'\neq s}\mathcal{B}_{i,s'}) \cap (\cup_{k'\in \text{Ch}^{*}_{i}}v_{k'}^{*})| = 1$, let $(\cup_{1\leq s'\leq c, s'\neq s}\mathcal{B}_{i,s'})\cap (\cup_{k'\in \text{Ch}^{*}_{i}}v_{k'}^{*}) = \mathcal{B}_{i,s_1} \cap v_{k_1}^{*} = \{j\}$ without loss of generality. When $|v^{1,c}_{s_1}|=1$, the second case of our first claim holds. Otherwise, it is easy to check that
    \begin{equation}\label{eq:hier algo proof with child case12 eq main}
    \begin{aligned}
    &\|\Sigma - \Sigma^*\|_{F} \\
    \geq & \frac{1}{\sqrt{2}}\left(\big\|\Sigma_{[v_{i}^{*}\setminus \{j\},\{j\}]} - \Sigma^{*}_{[v_{i}^{*}\setminus \{j\},\{j\}]}\big\| + \big\|\Sigma_{[v_{i}^{*}\setminus \{j\},v^{1,c}_{s_1}\setminus\{j\}]} - \Sigma^{*}_{[v_{i}^{*}\setminus \{j\},v^{1,c}_{s_1}\setminus\{j\}]}\big\|_{F}\right).
    \end{aligned}
    \end{equation}
    Notice that $$\Sigma_{[v_{i}^{*}\setminus \{j\},v^{1,c}_{s_1}\setminus\{j\}]} = \Lambda_{[v_{i}^{*}\setminus \{j\},\{1\}]}(\Lambda_{[v^{1,c}_{s_1}\setminus\{j\},\{1\}]})^{\top}
    ,$$ while $$\Sigma^{*}_{[v_{i}^{*}\setminus \{j\},v^{1,c}_{s_1}\setminus\{j\}]} = \Lambda^{*}_{[v_{i}^{*}\setminus \{j\},\{1\}]}(\Lambda^{*}_{[v^{1,c}_{s_1}\setminus\{j\},\{1\}]})^{\top}.$$
    We denote by $\delta = \big\|\Sigma_{[v_{i}^{*}\setminus \{j\},v^{1,c}_{s_1}\setminus\{j\}]} - \Sigma^{*}_{[v_{i}^{*}\setminus \{j\},v^{1,c}_{s_1}\setminus\{j\}]}\big\|_{F} $. By Lemma~\ref{lem: wedin}, either
    \begin{equation}\label{eq:hier algo proof with child sine1}
    \left\|\frac{\Lambda_{[v_{i}^{*}\setminus \{j\},\{1\}]}}{\big\|\Lambda_{[v_{i}^{*}\setminus \{j\},\{1\}]} \big\|}  -  \frac{\Lambda^{*}_{[v_{i}^{*}\setminus \{j\},\{1\}]}}{\big\|\Lambda^{*}_{[v_{i}^{*}\setminus \{j\},\{1\}]} \big\|}\right\|\leq \frac{2^{3/2}\delta}{\big\|\Lambda^{*}_{[v_{i}^{*}\setminus \{j\},\{1\}]}\big\| \big\|\Lambda^{*}_{[v^{1,c}_{s_1}\setminus\{j\},\{1\}]}\big\|},
    \end{equation}
    or
    \begin{equation*}
    \left\|\frac{\Lambda_{[v_{i}^{*}\setminus \{j\},\{1\}]}}{\big\|\Lambda_{[v_{i}^{*}\setminus \{j\},\{1\}]} \big\|}  +  \frac{\Lambda^{*}_{[v_{i}^{*}\setminus \{j\},\{1\}]}}{\big\|\Lambda^{*}_{[v_{i}^{*}\setminus \{j\},\{1\}]} \big\|}\right\|\leq \frac{2^{3/2}\delta}{\big\|\Lambda^{*}_{[v_{i}^{*}\setminus \{j\},\{1\}]}\big\| \big\|\Lambda^{*}_{[v^{1,c}_{s_1}\setminus\{j\},\{1\}]}\big\|}.
    \end{equation*}
    holds. Without loss of generality, we assume that~\eqref{eq:hier algo proof with child sine1} holds. On the other hand, notice that
    $$
    \Sigma_{[v_{i}^{*}\setminus \{j\},\{j\}]} = \lambda_{j,1}\Lambda_{[v_{i}^{*}\setminus \{j\},\{1\}]},
    $$
    and $$\Sigma^{*}_{[v_{i}^{*}\setminus \{j\},\{j\}]} = \Lambda^{*}_{[v_{i}^{*}\setminus \{j\},\{1,i\}\cup D_{i}^{*}]}(\Lambda^{*}_{[\{j\},\{1,i\}\cup D_{i}^{*}]})^{\top}.$$

    Let $$P_{i} = \frac{\Lambda^{*}_{[v_{i}^{*}\setminus \{j\},\{1\}]}\big(\Lambda^{*}_{[v_{i}^{*}\setminus \{j\},\{1\}]}\big)^{\top}}{\big(\Lambda^{*}_{[v_{i}^{*}\setminus \{j\},\{1\}]}\big)^{\top}\Lambda^{*}_{[v_{i}^{*}\setminus \{j\},\{1\}]}}$$ and $\Lambda^{*}_{\text{Proj},i} = (\mathbf{I}-P_{i})\Lambda^{*}_{[v_{i}^{*}\setminus\{j\},\{i\}\cup D^{*}_{i}]}$. By Condition~\ref{cond:rank}, $\sigma_{1+|D^{*}_{i}|}\big(\Lambda^{*}_{\text{Proj},i}\big) >0$. By Condition~\ref{cond:hier compact set}, 
    \begin{equation}~\label{eq:hier algo proof with child case12 eq main2}
    \begin{aligned}
    & \big\|\Sigma_{[v_{i}^{*}\setminus \{j\},\{j\}]} - \Sigma^{*}_{[v_{i}^{*}\setminus \{j\},\{j\}]}\big\| \\
    \geq& - |\lambda_{j,1}|\left\|\Lambda_{[v_{i}^{*}\setminus \{j\},\{1\}]} - \frac{\big\|\Lambda_{[v_{i}^{*}\setminus \{j\},\{1\}]}\big\|}{\big\|\Lambda^{*}_{[v_{i}^{*}\setminus \{j\},\{1\}]} \big\|}\Lambda^{*}_{[v_{i}^{*}\setminus \{j\},\{1\}]}\right\| \\
    &+\big\|\Lambda^{*}_{[\{j\},\{i\}\cup D_{i}^{*}]}\big\|\sigma_{1+|D^{*}_{i}|}\big(\Lambda^{*}_{\text{Proj},i}\big)\\
    \geq& \big\|\Lambda^{*}_{[\{j\},\{i\}\cup D_{i}^{*}]}\big\|\sigma_{1+|D^{*}_{i}|}\big(\Lambda^{*}_{\text{Proj},i}\big) - \frac{2^{3/2}\tau^{2}\delta(|v_{i}^{*}|-1)^{1/2}}{\big\|\Lambda^{*}_{[v_{i}^{*}\setminus \{j\},\{1\}]}\big\| \big\|\Lambda^{*}_{[v^{1,c}_{s_1}\setminus\{j\},\{1\}]}\big\|}.
    \end{aligned}
    \end{equation}
    Combining~\eqref{eq:hier algo proof with child case12 eq main} and~\eqref{eq:hier algo proof with child case12 eq main2}, we have
    \begin{equation}~\label{eq:hier algo proof with child case12 eq main3}
    \begin{aligned}
    &\|\Sigma-\Sigma^{*}\|_{F}\\
    \geq&\min\left(\frac{\sqrt{2}}{4},\frac{\big\|\Lambda^{*}_{[v_{i}^{*}\setminus \{j\},\{1\}]}\big\| \big\|\Lambda^{*}_{[v^{1,c}_{s}\setminus\{j\},\{1\}]}\big\|}{8\tau^{2}(|v_{i}^{*}|-1)^{1/2}}\right)\big\|\Lambda^{*}_{[\{j\},\{i\}\cup D_{i}^{*}]}\big\|\sigma_{1+|D^{*}_{i}|}\big(\Lambda^{*}_{\text{Proj},i}\big)\\
    >&0.
    \end{aligned}
    \end{equation}
    
Finally, when $|(\cup_{1\leq s'\leq c^{*}, s'\neq s}\mathcal{B}_{i,s'}) \cap (\cup_{k'\in \text{Ch}^{*}_{i}}v_{k'}^{*})| = 0$, the first case of our first claim holds. 
\item $|\mathcal{B}_{i,s}\cap v_{k}^{*}|\leq 1$ for all $1\leq s\leq c$ and $k\in \text{Ch}^{*}_{i}$. First, consider the case when there exist some $1\leq s\leq c$ such that $|\mathcal{B}_{i,s}\cap v_{k_1}^{*}| = 1$ and $|\mathcal{B}_{i,s}\cap v_{k_2}^{*}| = 1$ for some $k_1, k_2 \in \text{Ch}^{*}_{i}$. we denote $\{j_1\} = \mathcal{B}_{i,s}\cap v_{k_1}^{*}$ and $\{j_2\} = \mathcal{B}_{i,s}\cap v_{k_2}^{*}$. Moreover, choose $j_3, j_4 \in v_{k_1}^{*}$ and $j_3, j_4 \in v_{k_2}^{*}$. Furthermore, if $|\text{Ch}_{k_1}^{*}|\neq 0$, we further require that $j_3,j_4$ belong to different child factors of factor $k_1$ with $j_1$. Similarly, if $|\text{Ch}_{k_2}^{*}|\neq 0$, $j_5,j_6$ belong to different child factor of factor $k_2$ with $j_2$. Such a choice is always possible due to the assumed structure of the hierarchical model. It is easy to check that
    \begin{equation*}
    \Sigma_{[\{j_1,j_2\},\{j_3,j_4,j_5,j_6\}]} = \Lambda_{[\{j_1,j_2\},\{1\}]}(\Lambda_{[\{j_3,j_4,j_5,j_6\},\{1\}]})^{\top},
    \end{equation*}
    has rank 1. By Condition~\ref{cond:rank}, $\Lambda^{*}_{[\{j_3,j_4,j_5,j_6\},\{1,i,k_1,k_2\}]}$ has rank 4 and $\Lambda^{*}_{[\{j_1,j_2\},\{1,i,k_1,k_2\}]}$ has rank 2. By Sylvester's rank inequality,
\begin{equation*}
\begin{aligned}
 &\text{rank}\big(\Lambda^{*}_{[\{j_1,j_2\},\{1,i,k_1,k_2\}]}(\Lambda^{*}_{[\{j_3,j_4,j_5,j_6\},\{1,i,k_1,k_2\}]})^{\top}\big)\\
\geq &\text{rank}\big(\Lambda^{*}_{[\{j_1,j_2\},\{1,i,k_1,k_2\}]}\big) + \text{rank}\big(\Lambda^{*}_{[\{j_3,j_4,j_5,j_6\},\{1,i,k_1,k_2\}]}\big)-4\\
= &2.
\end{aligned}
\end{equation*}
By Lemma~\ref{lem: Weyl},
    \begin{equation}
    \begin{aligned}
    &\|\Sigma-\Sigma^{*}\|_{F} \\
    \geq & \big\|\Sigma_{[\{j_1,j_2\},\{j_3,j_4,j_5,j_6\}]}-\Sigma^{*}_{[\{j_1,j_2\},\{j_3,j_4,j_5,j_6\}]}\big\|_{F}\\
    \geq & \sigma_{2}\big(\Lambda^{*}_{[\{j_1,j_2\},\{1,i,k_1,k_2\}]}(\Lambda^{*}_{[\{j_3,j_4,j_5,j_6\},\{1,i,k_1,k_2\}]})^{\top}\big)\\
    > &0.
    \end{aligned}
    \end{equation}
Second, for each $1\leq s\leq c$, if $|\mathcal{B}_{i,s}\cap v_{k}^{*}|= 1$ for some $k\in\text{Ch}_{i}^{*}$, $|\mathcal{B}_{i,s}\cap v_{k'}^{*}|= 0$ for all $k'\in\text{Ch}_{i}^{*}$, $k'\neq k$, which indicates $|\mathcal{B}_{i,s}\cap v_{i}^{*}|\leq 1$ for $1\leq s\leq c$. Since $|v_{i}^{*}|\geq 7$ by constraint C4, choose $s_1, s_2, s_3, s_4$ such that $\{j_k\} = \mathcal{B}_{i,s_k}\cap v_{i}^{*}$ for $k=1,\ldots,4$ . Moreover, we require that $j_1, j_2$ and $j_3, j_4$ belong to different child factors of Factor $i$. We have
    \begin{equation*}
    \Sigma_{[\{j_1,j_2\}, \{j_{3},j_4\}]} = \Lambda_{[\{j_1,j_2\},\{1\}]}(\Lambda_{[\{j_3,j_4\},\{1\}]})^{\top},
    \end{equation*}
    has rank 1, while by Condition~\ref{cond:rank},
    \begin{equation*}
    \Sigma^{*}_{[\{j_1,j_2\}, \{j_{3},j_4\}]} = \Lambda^{*}_{[\{j_1,j_2\},\{1,i\}]}(\Lambda^{*}_{[\{j_3,j_4\},\{1,i\}]})^{\top},
    \end{equation*}
    has rank 2. By Lemma~\ref{lem: Weyl}, we have
    \begin{equation*}
    \begin{aligned}
    &\|\Sigma-\Sigma^{*}\|_{F} \\
    \geq &  \big\|\Sigma_{[\{j_1,j_2\}, \{j_{3},j_4\}]}-\Sigma^{*}_{[\{j_1,j_2\}, \{j_{3},j_4\}]} \big\|_{F}\\
    \geq & \sigma_{2}\big(\Lambda^{*}_{[\{j_1,j_2\},\{1,i\}]}(\Lambda^{*}_{[\{j_3,j_4\},\{1,i\}]})^{\top}\big)\\
    \geq &0.
    \end{aligned}
    \end{equation*}

\end{enumerate}

Now the first claim is proved, and we focus on our second claim. We assume that there exist $i_{1}\in\{2,\ldots, 1+c^{*}\}$ and $s_1, s_2 \in \{1,\ldots,c\}$ and $s_1 \neq s_2$ that satisfy $v_{i_1}^{*} = \mathcal{B}_{i_1,s_1}\cup \mathcal{B}_{i_1,s_2}$ and $v_{s_2}^{1,c} = \{j_1\}$ for some $j_1\in\{1,\ldots,J\}$. Furthermore, for each $i_2 \in\{2,\ldots,1+c^{*}\}$ and $i_2\neq i_1$, we denote by $v_{i_2}^{*} = \mathcal{B}_{i_2,s_3}\cup \mathcal{B}_{i_2,s_4}$ for some $s_3, s_4 \in \{1,\ldots,c\}$ that satisfy $\mathcal{B}_{i_2,s_4} = \emptyset$ or $\mathcal{B}_{i_2,s_4} = v^{1,c}_{s_4} = \{j_2\}$ for some $j_2\in\{1,\ldots,J\}$. We will first show that $s_3 = s_1$ for all $i_2\neq i_1$ and second, show that $\mathcal{B}_{i_2,s_4}= \emptyset$ for all $i_2\neq i_1$, which finally leads to case B.

First, when $s_3\neq s_1$ for some $i_2$, it is easy to check that 
\begin{equation}\label{eq:new claim 2 eq main1}
    \begin{aligned}
    &\|\Sigma - \Sigma^*\|_{F} \\
    \geq & \frac{1}{\sqrt{2}}\left(\big\|\Sigma_{[v_{i_1}^{*}\setminus \{j_1\},\{j_1\}]} - \Sigma^{*}_{[v_{i_1}^{*}\setminus \{j_1\},\{j_1\}]}\big\| + \big\|\Sigma_{[v_{i}^{*}\setminus \{j_1\},\mathcal{B}_{i_2,s_3}]} - \Sigma^{*}_{[v_{i}^{*}\setminus \{j_1\},\mathcal{B}_{i_2,s_3}]}\big\|_{F}\right).
    \end{aligned}
    \end{equation}
Similarly to the proof in \eqref{eq:hier algo proof with child case12 eq main} to \eqref{eq:hier algo proof with child case12 eq main3}, we have
\begin{equation}~\label{eq:new claim 2 eq main1 result1}
    \begin{aligned}
    &\|\Sigma-\Sigma^{*}\|_{F}\\
    \geq&\min\left(\frac{\sqrt{2}}{4},\frac{\big\|\Lambda^{*}_{[v_{i_1}^{*}\setminus \{j_1\},\{1\}]}\big\| \big\|\Lambda^{*}_{[\mathcal{B}_{i_2,s_3},\{1\}]}\big\|}{8\tau^{2}(|v_{i}^{*}|-1)^{1/2}}\right)\big\|\Lambda^{*}_{[\{j_1\},\{i_1\}\cup D_{i_1}^{*}]}\big\|\sigma_{1+|D^{*}_{i_1}|}\big(\Lambda^{*}_{\text{Proj},i_1}\big)\\
    >&0.
    \end{aligned}
    \end{equation}
Second, when $\mathcal{B}_{i_2,s_4} \neq \emptyset$ for some $i_2$
\begin{equation}\label{eq:new claim 2 eq main2}
    \begin{aligned}
    &\|\Sigma - \Sigma^*\|_{F} \\
    \geq & \frac{1}{\sqrt{2}}\left(\big\|\Sigma_{[v_{i_1}^{*}\setminus \{j_1\},\{j_1\}]} - \Sigma^{*}_{[v_{i_1}^{*}\setminus \{j_1\},\{j_1\}]}\big\| + \big\|\Sigma_{[v_{i}^{*}\setminus \{j_1\},\mathcal{B}_{i_2,s_4}]} - \Sigma^{*}_{[v_{i}^{*}\setminus \{j_1\},\mathcal{B}_{i_2,s_4}]}\big\|_{F}\right).
    \end{aligned}
    \end{equation}
Again, similarly to the proof in \eqref{eq:hier algo proof with child case12 eq main} to \eqref{eq:hier algo proof with child case12 eq main3}, we have
\begin{equation}~\label{eq:new claim 2 eq main2 result1}
    \begin{aligned}
    &\|\Sigma-\Sigma^{*}\|_{F}\\
    \geq&\min\left(\frac{\sqrt{2}}{4},\frac{\big\|\Lambda^{*}_{[v_{i_1}^{*}\setminus \{j_1\},\{1\}]}\big\| \big\|\Lambda^{*}_{[\mathcal{B}_{i_2,s_4},\{1\}]}\big\|}{8\tau^{2}(|v_{i}^{*}|-1)^{1/2}}\right)\big\|\Lambda^{*}_{[\{j_1\},\{i_1\}\cup D_{i_1}^{*}]}\big\|\sigma_{1+|D^{*}_{i_1}|}\big(\Lambda^{*}_{\text{Proj},i_1}\big)\\
    >&0.
    \end{aligned}
    \end{equation}
Now we have finished the first part of our proof.

For the second part, we mainly focus on case A and omit the proof when $v_{1}^{1,c},\ldots, v_{c}^{1,c}$ are included in case B for two reasons: (1) case B does not satisfy constraint C4 and will never be selected in our algorithms and (2) by a similar argument below, the information criterion brought by such case will be strictly larger than the optimal solution with probability approaching 1 as $N$ grows to infinity. We also assume that $d_{\max}$ is sufficiently large to avoid further discussions.

Now, we focus on case A and we only discuss the case when $v^{1,c}_{s}$ are nonempty for $s=1, \ldots, c$. First, we show that when $c=c^{*}$ in case A, 
\begin{equation}\label{eq:target information criterion}
\tilde{\mbox{IC}}_{1,c^{*}} = \sum_{k\in \text{Ch}_{1}^{*}}\left(|v_{k}^{*}|(|D_{k}^{*}|+1)-|D_{k}^{*}|(|D_{k}^{*}|+1)/2\right)\log N + O_{\mathbb{P}}(1).
\end{equation}
In such a case, we have $v^{1,c^{*}}_{s} = v^{*}_{1+s}$ for $s=1, \ldots, c^{*}$. We claim that Step 6 of Algorithm~\ref{algo:ICB} outputs $\tilde{d}_{s}^{c^*} = 1 + |D_{1+s}^{*}|$ for $s=1,\ldots,c^*$ with probability approaching 1 as $N$ grows to infinity. When $s=1$, for $d_1\geq 1+|D_{2}^{*}|$, let $\underline{\Lambda}_{d_1}$ and $\underline{\Psi}_{d_1}$ be the solution to 
\begin{equation}\label{eq: bic the first node}
\tilde{\mbox{IC}}_{1}(c,d_1,\min(|v_{3}^{*}|,d),\ldots,\min(|v_{1+c^*}^{*}|,d)).
\end{equation}
Similar to the proof of Lemma~\ref{lem: convergence rate covariance}, $\big\|\underline{\Lambda}_{d_1}\underline{\Lambda}_{d_1}^{\top}+\underline{\Psi}_{d_1}-\Sigma^*\big\|_{F} = O_{\mathbb{P}}(1/\sqrt{N})$, and we further have
\begin{equation}\label{eq:thm algo ch1 exlicit}
\begin{aligned}
&\tilde{\mbox{IC}}_{1}(c,d_1,\min(|v_{3}^{*}|,d),\ldots,\min(|v_{c^*}^{*}|,d)) \\
=&l\big(\underline{\Lambda}_{d_1}\underline{\Lambda}_{d_1}^{\top}+\underline{\Psi}_{d_1};S\big) + p_{1}\big(\underline{\Lambda}_{d_1}\big)\log N\\
=& O_{\mathbb{P}}(1) + \big(|v_{2}^{*}|d_1 - d_1(d_1-1)/2 \big)\log N+ \sum_{2\leq s \leq c^*} \big(|v_{s+1}^{*}|d_{s}-d_s(d_s-1)/2 \big)\log N,
\end{aligned}
\end{equation}
where we define $d_s = \min(|v_{1+s}^{*}|,d)$, $s=2,\ldots,c^*$ for simplicity. Noticing that the third term of~\eqref{eq:thm algo ch1 exlicit} is independent of the choice of $d_1$ and the second term is strictly increasing with respect to $d_1$ for $1+|D_{2}^{*}|\leq d_1 \leq \min(|v_{2}^{*}|,d)$, we then have
\begin{equation}\label{eq:thm hier algo d1 eq1}
1+|D^{*}_{2}| = \argmin_{1+|D^{*}_{2}|\leq d_1 \leq \min(|v_{2}^{*}|,d)}\tilde{\mbox{IC}}_{1}(c,d_1,\min(|v_{3}^{*}|,d),\ldots,\min(|v_{c^*}^{*}|,d)),
\end{equation}
with probability approaching 1 as $N$ grows to infinity.

When $d_1 < 1+|D^{*}_{2}|$, for any $\Lambda \in \tilde{\mathcal{A}}^{1}(c^*,d_1,\min(|v_{3}^{*}|,d),\ldots,\min(|v_{1+c^*}^{*}|,d))$ and $\Psi$, we denote by $\Sigma = \Lambda\Lambda^{\top} + \Psi$. According to Condition~\ref{cond:hier always select correct number}, there exist $E_1, E_2 \subset v_{2}^{*}$ with $|E_1| = 2 + |D_{2}^{*}|$, $|E_2| = 1 + |D_{2}^{*}|$ and $E_1\cap E_2 = \emptyset$ such that $\Lambda^{*}_{[E_1, \{1,2\}\cup D^{*}_{2}]}$ and $\Lambda^{*}_{[E_2, \{2\}\cup D^{*}_{2}]}$ are of full rank. We further denote by $B_1 = \{2,\ldots,1+d_1\}$. First we have
\begin{equation}\label{eq:hier algo proof bic d1 is correct main!}
\|\Sigma-\Sigma^{*}\|_{F}\geq  \frac{1}{\sqrt{2}}\left(\big\|\Sigma_{[v_{2}^{*},v_{i}^{*}]}-\Sigma^{*}_{[v_{2}^{*},v_{i}^{*}]} \big\|_{F} + \big\|\Sigma_{[E_1,E_2]}-\Sigma^{*}_{[E_1, E_2]} \big\|_{F}\right),
\end{equation}
for any $i = 3,\ldots,1+c^*$.We denote by $\delta = \big\|\Sigma_{[v_{2}^{*},v_{i}^{*}]}-\Sigma^{*}_{[v_{2}^{*},v_{i}^{*}]} \big\|_{F}$. Notice that
\begin{equation*}
\Sigma_{[v_{2}^{*},v_{i}^{*}]} = \Lambda_{[v_{2}^{*},\{1\}]}(\Lambda_{[v_{i}^{*},\{1\}]})^{\top},
\end{equation*}
and
\begin{equation*}
\Sigma^{*}_{[v_{2}^{*},v_{i}^{*}]} = \Lambda^{*}_{[v_{2}^{*},\{1\}]}(\Lambda^{*}_{[v_{i}^{*},\{1\}]})^{\top}.
\end{equation*}
According to Lemma~\ref{lem: wedin}, either
\begin{equation}\label{eq:hier algo proof d1 diag sine1}
    \left\|\frac{\Lambda_{[v_{2}^{*},\{1\}]}}{ \big\|\Lambda_{[v_{2}^{*},\{1\}]}  \big\|}  -  \frac{\Lambda^{*}_{[v_{2}^{*},\{1\}]}}{ \big\|\Lambda^{*}_{[v_{2}^{*},\{1\}]}  \big\|}\right\|\leq \frac{2^{3/2}\delta}{ \big\|\Lambda^{*}_{[v_{2}^{*},\{1\}]} \big\|  \big\|\Lambda^{*}_{[v_{i}^{*},\{1\}]} \big\|},
    \end{equation}
or
\begin{equation*}
     \left\|\frac{\Lambda_{[v_{2}^{*},\{1\}]}}{ \big\|\Lambda_{[v_{2}^{*},\{1\}]}  \big\|}  +  \frac{\Lambda^{*}_{[v_{2}^{*},\{1\}]}}{ \big\|\Lambda^{*}_{[v_{2}^{*},\{1\}]}  \big\|}\right\|\leq \frac{2^{3/2}\delta}{ \big\|\Lambda^{*}_{[v_{2}^{*},\{1\}]} \big\|  \big\|\Lambda^{*}_{[v_{i}^{*},\{1\}]} \big\|},
    \end{equation*}
holds. Without loss of generality, we assume that~\eqref{eq:hier algo proof d1 diag sine1} holds. On the other hand, notice that
\begin{equation}\label{eq:hier algo proof E1E2 diag main1}
\begin{aligned}
&\Sigma_{[E_1,E_2]} - \Sigma^{*}_{[E_1,E_2]}\\
=& \Lambda_{[E_1,\{1\}]}(\Lambda_{[E_2,\{1\}]})^{\top} + \Lambda_{[E_1, B_1]}(\Lambda_{[E_2, B_1]})^{\top} - \Lambda^{*}_{[E_1,\{1\}]}(\Lambda^{*}_{[E_2,\{1\}]})^{\top} \\
&- \Lambda^{*}_{[E_1,\{2\}\cup D_{2}^{*}]}(\Lambda^{*}_{[E_2,\{2\}\cup D_{2}^{*}]})^{\top}\\
=& \Lambda_{[E_1,\{1\}]}(\Lambda_{[E_2,\{1\}]})^{\top} -\frac{\big\|\Lambda_{[v_{2}^{*},\{1\}]}  \big\|^{2}}{\big\|\Lambda^{*}_{[v_{2}^{*},\{1\}]}  \big\|^{2}} \Lambda^{*}_{[E_1,\{1\}]}(\Lambda^{*}_{[E_2,\{1\}]})^{\top} \\
&+ \Lambda_{[E_1, B_1]}(\Lambda_{[E_2, B_1]})^{\top} - \left(1-\frac{\big\|\Lambda_{[v_{2}^{*},\{1\}]}  \big\|^{2}}{\big\|\Lambda^{*}_{[v_{2}^{*},\{1\}]}  \big\|^{2}}\right)\Lambda^{*}_{[E_1,\{1\}]}(\Lambda^{*}_{[E_2,\{1\}]})^{\top}\\
&-\Lambda^{*}_{[E_1,\{2\}\cup D_{2}^{*}]}(\Lambda^{*}_{[E_2,\{2\}\cup D_{2}^{*}]})^{\top}.
\end{aligned}
\end{equation}
Combined with~\eqref{eq:hier algo proof d1 diag sine1}, we have
\begin{equation}\label{eq:hier algo proof E1E2 diag main2}
\begin{aligned}
&\left\|\Lambda_{[E_1,\{1\}]}(\Lambda_{[E_2,\{1\}]})^{\top} -\frac{\big\|\Lambda_{[v_{2}^{*},\{1\}]}  \big\|^{2}}{\big\|\Lambda^{*}_{[v_{2}^{*},\{1\}]}  \big\|^{2}}\Lambda^{*}_{[E_1,\{1\}]}(\Lambda^{*}_{[E_2,\{1\}]})^{\top}\right\|_{F}\\
\leq&\left\|\left(\Lambda_{[E_1,\{1\}]}-\frac{\big\|\Lambda_{[v_{2}^{*},\{1\}]}  \big\|}{\big\|\Lambda^{*}_{[v_{2}^{*},\{1\}]}  \big\|}\Lambda^{*}_{[E_1,\{1\}]}\right)(\Lambda_{[E_2,\{1\}]})^{\top}\right\|_{F} \\
&+ \frac{\big\|\Lambda_{[v_{2}^{*},\{1\}]}  \big\|}{\big\|\Lambda^{*}_{[v_{2}^{*},\{1\}]}  \big\|}\left\|\Lambda^{*}_{[E_1,\{1\}]}\left(\Lambda_{[E_2,\{1\}]}-\frac{\big\|\Lambda_{[v_{2}^{*},\{1\}]}  \big\|}{\big\|\Lambda^{*}_{[v_{2}^{*},\{1\}]}  \big\|}\Lambda^{*}_{[E_2,\{1\}]}\right)^{\top}\right\|_{F}\\
\leq & \frac{2^{3/2}\delta\big\|\Lambda_{[v_{2}^{*},\{1\}]}  \big\|}{ \big\|\Lambda^{*}_{[v_{2}^{*},\{1\}]} \big\|  \big\|\Lambda^{*}_{[v_{i}^{*},\{1\}]} \big\|}\left( \big\|\Lambda_{[E_2,\{1\}]}  \big\| + \frac{\big\|\Lambda_{[v_{2}^{*},\{1\}]}  \big\|}{\big\|\Lambda^{*}_{[v_{2}^{*},\{1\}]}  \big\|}\big\|\Lambda^{*}_{[E_1,\{1\}]}  \big\|\right)\\
\leq &\frac{2^{5/2}\tau^{2}|v_{2}^{*}|\delta}{\big\|\Lambda^{*}_{[v_{2}^{*},\{1\}]} \big\|  \big\|\Lambda^{*}_{[v_{i}^{*},\{1\}]} \big\|}.
\end{aligned}
\end{equation}
We denote by
\begin{equation*}
\Lambda_{E}^{\top} = \big(\Lambda^{*}_{[E_1, \{1,2\}\cup D_{2}^*]}\big)^{-1} \Lambda_{[E_1, B_1]}(\Lambda_{[E_2, B_1]})^{\top},
\end{equation*}
whose rank is at most $d_1<1+|D_{2}^{*}|$, and 
\begin{equation*}
\begin{aligned}
\Lambda^{*}_{E} = \left(\begin{array}{cc}
\left(1-\frac{\big\|\Lambda_{[v_{2}^{*},\{1\}]}  \big\|^{2}}{\big\|\Lambda^{*}_{[v_{2}^{*},\{1\}]}  \big\|^{2}}\right)\Lambda^{*}_{[E_2,\{1\}]}, & \Lambda^{*}_{[E_2, \{2\}\cup D_{2}^*]}
\end{array}
\right). 
\end{aligned}
\end{equation*}
By Condition~\ref{cond:hier always select correct number}, $\Lambda^{*}_{[E_2, \{2\}\cup D_{2}^*]}$ has rank $1+|D_{2}^*|$. Thus, by Lemma~\ref{lem: Weyl}
\begin{equation}\label{eq:hier algo proof E1E2 diag main3}
\begin{aligned}
&\left\|\Lambda_{E} - \Lambda_{E}^{*}\right\|_{F}\\
\geq & \left\|\big(\Lambda_{E}\big)_{[:,B_1]} - \big(\Lambda_{E}^{*}\big)_{[:,B_1]}\right\|_{F} \\
\geq & \sigma_{1+|D_{2}^*|}\big(\Lambda^{*}_{[E_2, \{2\}\cup D_{2}^*]}\big).
\end{aligned}
\end{equation}
Combined with~\eqref{eq:hier algo proof E1E2 diag main1},~\eqref{eq:hier algo proof E1E2 diag main2} and~\eqref{eq:hier algo proof E1E2 diag main3}, we have
\begin{equation}~\label{eq:hier algo proof E1E2 diag main4}
\begin{aligned}
&\left\|\Sigma_{[E_1,E_2]} - \Sigma^{*}_{[E_1,E_2]}\right\|_{F}\\
\geq & \sigma_{2+|D_{2}^*|}\big(\Lambda^{*}_{[E_1, \{1,2\}\cup D_{2}^*]}\big)\sigma_{1+|D_{2}^*|}\big(\Lambda^{*}_{[E_2, \{2\}\cup D_{2}^*]}\big)-\frac{2^{5/2}\tau^{2}|v_{2}^{*}|\delta}{\big\|\Lambda^{*}_{[v_{2}^{*},\{1\}]} \big\|  \big\|\Lambda^{*}_{[v_{i}^{*},\{1\}]} \big\|}.
\end{aligned}
\end{equation}
Combined with~\eqref{eq:hier algo proof bic d1 is correct main!} we further have
\begin{equation*}
\begin{aligned}
&\|\Sigma-\Sigma^{*}\|_{F}\\
\geq & \min\left(\frac{\sqrt{2}}{4},\frac{\big\|\Lambda^{*}_{[v_{2}^{*},\{1\}]} \big\|\big\|\Lambda^{*}_{[v_{i}^{*},\{1\}]} \big\|}{16\tau^{2}|v_{2}^{*}|}\right)\sigma_{2+|D_{2}^*|}\big(\Lambda^{*}_{[E_1, \{1,2\}\cup D_{2}^*]}\big)\sigma_{1+|D_{2}^*|}\big(\Lambda^{*}_{[E_2, \{2\}\cup D_{2}^*]}\big).
\end{aligned}
\end{equation*}
Thus, the derived information criterion satisfies $$
\tilde{\mbox{IC}}_{1}(c,d_1,\min(|v_{3}^{*}|,d),\ldots,\min(|v_{c^*}^{*}|,d))= O_{\mathbb{P}}(N).$$ Thus, with probability approaching 1 as $N$ grows to infinity, we have
\begin{equation}\label{eq:thm hier algo d1 eq2}
1+|D^{*}_{2}| = \argmin_{1\leq d_1 \leq 1+|D^{*}_{2}|}\tilde{\mbox{IC}}_{1}(c,d_1,\min(|v_{3}^{*}|,d),\ldots,\min(|v_{c^*}^{*}|,d)).
\end{equation}
Combining ~\eqref{eq:thm hier algo d1 eq1} with~\eqref{eq:thm hier algo d1 eq2}, we have $\tilde{d}_{1}^{c^*} = 1+ |D^{*}_{2}|$. Similarly, we have $\tilde{d}_{s}^{c^*} = 1+ |D^{*}_{1+s}|$, for $s=1,\ldots,c^*$. Then we have
\begin{equation*}
\begin{aligned}
&\tilde{\mbox{IC}}_{1}(c^{*},1+ |D^{*}_{2}|,\ldots,1+ |D^{*}_{1+c^*}|)\\
=& \sum_{k\in \text{Ch}_{1}^{*}}\left(|v_{k}^{*}|(|D_{k}^{*}|+1)-|D_{k}^{*}|(|D_{k}^{*}|+1)/2\right)\log N + O_{\mathbb{P}}(1),
\end{aligned}
\end{equation*}
and ~\eqref{eq:target information criterion} holds.

Second, when $c<c^{*}$ in case A. We will show that the $\tilde{d}_{s}^{c}$ given by Step 6 of Algorithm~\ref{algo:ICB} satisfies $\tilde{d}_{s}^{c} = \sum_{v_{i}^{*} \subset v_{s}^{1,c}} 1 + |D_{i}^{*}|$ for $s=1,\ldots,c$ with probability approaching 1 as $N$ grows to infinity. \\
For $s=1$, when $d_1\geq \sum_{v_{i}^{*} \subset v_{s}^{1,c}} 1 + |D_{i}^{*}|$, let  $\underline{\Lambda}_{d_1}$ and $\underline{\Psi}_{d_1}$ be the solution to 
\begin{equation*}
\tilde{\mbox{IC}}_{1}(c,d_1,\min(|v_{2}^{1,c}|,d),\ldots,\min(|v_{c}^{1,c}|,d)).
\end{equation*}
Similarly to Lemma~\ref{lem: convergence rate covariance}, we have $\big\|\underline{\Lambda}_{d_1}\underline{\Lambda}_{d_1}^{\top}+\underline{\Psi}_{d_1}-\Sigma^*\big\|_{F} = O_{\mathbb{P}}(1/\sqrt{N})$ and by Taylor's expansion, we have 
\begin{equation}\label{eq:hier algo proof under case ch1 exlicit}
\begin{aligned}
&\tilde{\mbox{IC}}_{1}(c,d_1,\min(|v_{2}^{1,c}|,d),\ldots,\min(|v_{c}^{1,c}|,d)) \\
=&l\big(\underline{\Lambda}_{d_1}\underline{\Lambda}_{d_1}^{\top}+\underline{\Psi}_{d_1};S\big) + p_{1}\big(\underline{\Lambda}_{d_1}\big)\log N\\
=& O_{\mathbb{P}}(1) + \big(|v_{1}^{1,c}|d_1 - d_1(d_1-1)/2\big)\log N+ \sum_{2\leq s \leq c}\big(|v_{s}^{1,c}|d_{s}-d_s(d_s-1)\big)\log N,
\end{aligned}
\end{equation}
where we denoted by $d_s = \min(|v_{s}^{1,c}|,d)$, $s=2,\ldots,c$ for simplicity. Notice that the third term in~\eqref{eq:hier algo proof under case ch1 exlicit} is independent of the choice of $d_1$ and the second term is strictly increasing with respect to $d_1$ when $\sum_{v_{i}^{*} \subset v_{1}^{1,c}} 1 + |D_{i}^{*}|\leq d_1\leq \min(|v_{1}^{1,c}|,d)$. Thus, with probability approaching 1, as $N$ grows to infinity, we have
\begin{equation}\label{eq:thm hier algo under case d1 eq1}
\begin{aligned}
&\sum_{v_{i}^{*} \subset v_{1}^{1,c}} 1 + |D_{i}^{*}| \\
=& \argmin_{\sum_{v_{i}^{*} \subset v_{1}^{1,c}} 1 + |D_{i}^{*}|\leq d_1 \leq \min(|v_{1}^{1,c}|,d)}\tilde{\mbox{IC}}_{1}(c,d_1,\min(|v_{2}^{1,c}|,d),\ldots,\min(|v_{c}^{1,c}|,d)).
\end{aligned}
\end{equation}
When $d_1 < \sum_{v_{i}^{*} \subset v_{1}^{1,c}} 1 + |D_{i}^{*}|$, similar to the proof in~\eqref{eq:hier algo proof bic d1 is correct main!}-\eqref{eq:hier algo proof E1E2 diag main4}, we have
\begin{equation}\label{eq:thm hier algo under case d1 eq2}
\sum_{v_{i}^{*} \subset v_{1}^{1,c}} 1 + |D_{i}^{*}| = \argmin_{1\leq d_1 \leq \sum_{v_{i}^{*} \subset v_{1}^{1,c}} 1 + |D_{i}^{*}|}\tilde{\mbox{IC}}_{1}(c,d_1,\min(|v_{2}^{1,c}|,d),\ldots,\min(|v_{c}^{1,c}|,d)),
\end{equation}
with probability approaching $1$ as $N$ grows to infinity. Combining~\eqref{eq:thm hier algo under case d1 eq1} with~\eqref{eq:thm hier algo under case d1 eq2}, we have $\tilde{d}_{1}^{c} = \sum_{v_{i}^{*} \subset v_{1}^{1,c}} (1 + |D_{i}^{*}|)$. Similarly, we also have $\tilde{d}_{s}^{c} = \sum_{v_{i}^{*} \subset v_{s}^{1,c}} (1 + |D_{i}^{*}|), s=1,\ldots,c$. However, it is obvious that
\begin{equation*}
\begin{aligned}
\sum_{s=1}^{c}\big(|v_{s}^{1,c}|\tilde{d}_{s}^{c} - \tilde{d}_{s}^{c}(\tilde{d}_{s}^{c}-1)/2\big)
>  \sum_{i\in \text{Ch}^{*}_{i}}\big(|v_{s}^{*}|(|D_{s}^{*}|+1)-|D_{s}^{*}|(|D_{s}^{*}|+1)/2\big),
\end{aligned}
\end{equation*}
when $\tilde{d}_{s}^{c} = \sum_{v_{i}^{*} \subset v_{s}^{1,c}} (1 + |D_{i}^{*}|), s=1,\ldots,c$. Thus, with probability approaching 1 as $N$ grows to infinity, the derived $\tilde{\mbox{IC}}_{1}(c,\tilde{d}_{1}^{c}, \ldots, \tilde{d}_{c}^{c})$ is larger than~\eqref{eq:target information criterion}.

Finally, when $v_{1}^{1,c},\ldots, v_{c}^{1,c}$ are not included in case A or B, the derived information criterion is strictly larger than \eqref{eq:target information criterion} with probability approaching 1 as $N$ grows to infinity by the first part of our proof. Thus, the second part is proved.

At the end of the proof, we conclude that with the same argument in Lemma~\ref{lem: confirmatory rate of convergence}, $\|\tilde{\boldsymbol{\lambda}}_{1} - \boldsymbol{\lambda}_{1}^{*}\mbox{sign}(\boldsymbol{\lambda}_{1}^{*\top}\tilde{\boldsymbol{\lambda}}_{1})\| = O_{\mathbb{P}}(1/\sqrt{N})$, which indicating $\|\tilde{\boldsymbol{\lambda}}_{1}\tilde{\boldsymbol{\lambda}}_{1}^{\top} - \boldsymbol{\lambda}_{1}^{*}\boldsymbol{\lambda}_{1}^{*\top}\|_{F} = O_{\mathbb{P}}(1/\sqrt{N})$. Then Theorem~\ref{cor:consistency} is proved by applying the same argument to the factors in the $t$th layer, $t=2,\ldots, T$, together with Lemma~\ref{lem: confirmatory rate of convergence}.
\end{proof}

\section{Simulation studies for Algorithm \ref{alg:main} with correctly estimated number of child factors}\label{appen:simulation algo3}
As discussed in Section \ref{sec:computation}, Algorithm \ref{alg:main} may converge only to a local optimum, and the local solution may not satisfy constraint C4. In this section, we examine the performance of Algorithm \ref{alg:main} to find a global optimum and decode the structure of the child factors of Factor k given $c = |\mbox{Ch}_{k}^{*}|$ with multiple random starts in detail. We consider the hierarchical structure shown in Figure~\ref{fig:sim under-estimated c} with $J\in\{24,36\}$, $v^{*}_{1} = \{1,\ldots,J\}$, $v^{*}_{2} = \{1,\ldots,J/3\}$, $v^{*}_{3} = \{1+J/3,\ldots, 2J/3\}$, $v^{*}_{3} = \{1+2J/3,\ldots, J\}$, $v^{*}_5 =  \{1,\ldots,J/6\}$ and $v^{*}_6 = \{1+J/6,\ldots,J/3\}$.

\begin{figure}[ht!]
\centering

\tikzset{every picture/.style={line width=0.75pt}} 

\begin{tikzpicture}[x=0.75pt,y=0.75pt,yscale=-1,xscale=1]

\draw   (173.5,71) -- (194.5,71) -- (194.5,91) -- (173.5,91) -- cycle ;
\draw   (284.5,19) -- (305.5,19) -- (305.5,39) -- (284.5,39) -- cycle ;
\draw   (227,120) -- (248,120) -- (248,140) -- (227,140) -- cycle ;
\draw   (330.5,71) -- (351.5,71) -- (351.5,91) -- (330.5,91) -- cycle ;
\draw   (120.5,120) -- (141.5,120) -- (141.5,140) -- (120.5,140) -- cycle ;
\draw   (435.5,71) -- (456.5,71) -- (456.5,91) -- (435.5,91) -- cycle ;
\draw   (251.5,170) -- (286.5,170) -- (286.5,211) -- (251.5,211) -- cycle ;
\draw    (131,140) -- (131,150) ;
\draw    (101,150) -- (163,150) ;
\draw    (101,150) -- (101,171) ;
\draw    (163,150) -- (163,171) ;
\draw    (206,150) -- (269,150) ;
\draw    (237.5,140) -- (237.5,150) ;
\draw    (206,150) -- (206,171) ;
\draw    (269,150) -- (269,171) ;
\draw    (184,91) -- (184,101) ;
\draw    (131,101) -- (237.5,101) ;
\draw    (131,101) -- (131,120) ;
\draw    (237.5,101) -- (237.5,120) ;
\draw    (311,101) -- (311,171) ;
\draw    (311,101) -- (373,101) ;
\draw    (373,101) -- (373,171) ;
\draw    (341,91) -- (341,101) ;
\draw    (416,101) -- (478,101) ;
\draw    (446,91) -- (446,101) ;
\draw    (416,101) -- (416,171) ;
\draw    (478,100) -- (478,171) ;
\draw    (184,71) -- (184,50) ;
\draw    (341,71) -- (341,50) ;
\draw    (446,71) -- (446,50) ;
\draw    (184,50) -- (446,50) ;
\draw    (295,50) -- (295,39) ;
\draw   (144,171) -- (179,171) -- (179,212) -- (144,212) -- cycle ;
\draw   (188.5,170) -- (223.5,170) -- (223.5,211) -- (188.5,211) -- cycle ;
\draw   (83.5,170) -- (118.5,170) -- (118.5,211) -- (83.5,211) -- cycle ;
\draw   (293.5,170) -- (328.5,170) -- (328.5,211) -- (293.5,211) -- cycle ;
\draw   (355.5,170) -- (390.5,170) -- (390.5,211) -- (355.5,211) -- cycle ;
\draw   (460.5,170) -- (495.5,170) -- (495.5,211) -- (460.5,211) -- cycle ;
\draw   (398.5,170) -- (433.5,170) -- (433.5,211) -- (398.5,211) -- cycle ;

\draw (437,185) node [anchor=north west][inner sep=0.75pt]   [align=left] {$\displaystyle \cdots $};
\draw (332,185) node [anchor=north west][inner sep=0.75pt]   [align=left] {$\displaystyle \cdots $};
\draw (228,185) node [anchor=north west][inner sep=0.75pt]   [align=left] {$\displaystyle \cdots $};
\draw (121,185) node [anchor=north west][inner sep=0.75pt]   [align=left] {$\displaystyle \cdots $};
\draw (286.5,22) node [anchor=north west][inner sep=0.75pt]   [align=left] {$\displaystyle v_{1}^{*}$};
\draw (175.5,74) node [anchor=north west][inner sep=0.75pt]   [align=left] {$\displaystyle v_{2}^{*}$};
\draw (332.5,74) node [anchor=north west][inner sep=0.75pt]   [align=left] {$\displaystyle v_{3}^{*}$};
\draw (437.5,74) node [anchor=north west][inner sep=0.75pt]   [align=left] {$\displaystyle v_{4}^{*}$};
\draw (122.5,123) node [anchor=north west][inner sep=0.75pt]   [align=left] {$\displaystyle v_{5}^{*}$};
\draw (229,123) node [anchor=north west][inner sep=0.75pt]   [align=left] {$\displaystyle v_{6}^{*}$};
\draw (190.5,175) node [anchor=north west][inner sep=0.75pt]   [align=left] {1+$\displaystyle \frac{J}{6}$};
\draw (96,183) node [anchor=north west][inner sep=0.75pt]   [align=left] {1};
\draw (155,175) node [anchor=north west][inner sep=0.75pt]   [align=left] {$\displaystyle \frac{J}{6}$};
\draw (262.5,175) node [anchor=north west][inner sep=0.75pt]   [align=left] {$\displaystyle \frac{J}{3}$};
\draw (296,175) node [anchor=north west][inner sep=0.75pt]   [align=left] {1+$\displaystyle \frac{J}{3}$};
\draw (363.5,175) node [anchor=north west][inner sep=0.75pt]   [align=left] {$\displaystyle \frac{2J}{3}$};
\draw (397.5,175) node [anchor=north west][inner sep=0.75pt]   [align=left] {1+$\displaystyle \frac{2J}{3}$};
\draw (473,183) node [anchor=north west][inner sep=0.75pt]   [align=left] {$\displaystyle J$};

\end{tikzpicture}
\caption{The hierarchical factor structure in the simulation studies of Section~\ref{appen:simulation algo3}.}
    \label{fig:sim under-estimated c}
\end{figure}

In the data generation model, $\Lambda^*$ is generated by
\begin{equation}
\begin{aligned}
\lambda_{jk}^{*} = \left\{
    \begin{array}{l}
        u_{jk} \mbox{~~if~~} k=1; \\
        0 \mbox{~~if~~} k>1 ,j \notin v_{k}^{*}; \\
        (1-2x_{jk})u_{jk} \mbox{~~if~~} k>1, j\in v_{k}^{*},
    \end{array}
\right.
\end{aligned}
\end{equation}
for $j = 1,\ldots,J$, and $k = 1,\ldots, K$. Here, $u_{jk}$s are i.i.d., following a Uniform$(0.5,2)$ distribution and $x_{jk}$s are i.i.d., following 
a Bernoulli$(0.5)$ distribution. $\Psi^{*}$ is either an identity matrix or $\Psi^{*} = \mbox{diag}(\psi_{1}^{*2},\ldots,\psi_{J}^{*2})$ with $\psi_{j}^{*}, j=1,\ldots,J$ i.i.d following a Uniform$(0.5,1.5)$ distribution.

Let $\hat{\Lambda}$ and $\hat{\Psi}$ be the estimates of $\Lambda^*$ and $\Psi^*$ given by Algorithm 3 and $\{\hat{v}_{1+i}\}_{i=1}^{c}$ be the estimated set of variables belonging to child factors of factor 1. To define a global optimal solution to the optimization problem in \eqref{opt:alm aug bbf}, we consider the ideal case when $S = \Sigma^{*}$. It is easy to notice that the objective function 
\begin{equation*}
\begin{aligned}
&\tilde{l}\left( \Lambda\Lambda^{\top} + \Psi, \Sigma^{*}\right) \\
=& \log(\det(\Lambda\Lambda^\top + \Psi)) + \textnormal{tr}(\Sigma^{*} (\Lambda\Lambda^\top + \Psi)^{-1}) - \log(\det(\Sigma^{*})) - J
\end{aligned}
\end{equation*}
reach its global minimum at 0. Thus, for each optimization result from a random starting point, we define the following criterion
\begin{enumerate}
    \item GS(Global Solution): a binary variable equal to 1 if $|\tilde{l}(\hat{\Lambda}\hat{\Lambda}^{\top} + \hat{\Psi}, \Sigma^*)| <\delta$ and 0 otherwise, where $\delta$ is a tolerance parameter.
    \item CR(Correctness Rate): a binary variable equal to 1 if $\{\hat{v}_{1+i}\}_{i=1}^{c} = \{v^{*}_{2},v^{*}_{3},v^{*}_{4}\} $ and 0 otherwise.
\end{enumerate}

We apply Algorithm~\ref{alg:main} with $c=3$ and $d=5$ and further denote $\hat{v}_{2}$, $\hat{v}_{3}$, and $\hat{v}_{4}$ as the estimated set of variables belonging to the child factors of factor 1. In this simulation study, we consider 4 simulation settings, given by the combinations of $J = 24, 36$ and two generation processes of $\Psi$. For each setting, 100 independent simulations are generated, and in each simulation, we use 100 random starting points with the tolerance parameter $\delta = 10^{-4}$. The numerical results are given in Table~\ref{tab:appen sim right c}. As shown in Table~\ref{tab:appen sim right c}, when $J=24$, around 57\% of the random starting points converge to a global optimum and 15\% of the estimation results correctly decode the underlying hierarchical factor structure. When $J = 36$, there exists a decrease in both GS and CR, with around 38\% and 11\% of the random starting points converging to a global optimum, respectively.

\begin{table}[htb!]
    \centering
    \caption{The mean value and standard deviation of GS and CR in the simulation study.}
    \label{tab:appen sim right c}

    \begin{tabular}{cccc}
    \toprule
    $\Psi^{*}$ & $J$ & GS &  CR  \\
    \midrule
    Identity &24 & $57.55_{(16.32)}$ & $15.42_{(5.39)}$ \\
     &36 &$37.52_{(12.11)}$ & $11.19_{(4.23)}$ \\
    Heterogeneous&24 & $56.48_{(15.61)}$& $14.88_{(5.48)}$\\
      & 36 & $39.49_{(12.29)}$ & $11.58_{(4.18)}$ \\
    \bottomrule
    \end{tabular}
    
\end{table}

\begin{remark}
We emphasize that when the optimization problem~\eqref{opt:alm aug bbf} reaches a global solution, the estimated sets of variable $\hat{v}_2$, $\hat{v}_3$, $\hat{v}_4$ are not necessarily equal to $v^{*}_{2},v^{*}_{3},v^{*}_{4}$. In the current setting, the following configurations can yield equivalent covariance structures while satisfying the constraints of the optimization problem:
\begin{enumerate}
\item[A.] $\hat{v}_2, \hat{v}_3$, $\hat{v}_{4}$ are equal to $\{1,\ldots,2J/3\}$, $\{1+2J/3,\ldots,J\}$, $\emptyset$.
\item[B.] $\hat{v}_2, \hat{v}_3 $, $\hat{v}_{4}$ are equal to $\{1,\ldots,J/3,1+2J/3,\ldots,J\}$, $\{1+J/3,\ldots,2J/3\}$, $\emptyset$. 
\item[C.] $\hat{v}_2, \hat{v}_3 $, $\hat{v}_{4}$ are equal to $\{1,\ldots,J/3\}$, $\{1+J/3,\ldots,J\}$, $\emptyset$. 
\item[D.] $\hat{v}_2, \hat{v}_3$, $\hat{v}_{4}$ are equal to $\{1,\ldots,J\}$, $\emptyset$, $\emptyset$. 
\item[E.] $\hat{v}_2, \hat{v}_3$, $\hat{v}_{4}$ are equal to $\{1,\ldots,J\}\setminus \{i\}$, $\{i\}$, $\emptyset$ for some $i\in\{1,\ldots,J\}$.
\end{enumerate}
These cases correspond precisely to the cases discussed in the proof of Theorem~\ref{cor:consistency}. To be more exact, case A, B, C are the cases when two of $v^{*}_{2},v^{*}_{3},v^{*}_{4}$ are merged into one set, and Case D is the case when $v^{*}_{2},v^{*}_{3},v^{*}_{4}$ are merged. Case E constructs the following parametric space for the loading matrix $\Lambda$: 
\begin{equation*}
\begin{aligned}
\{\Lambda\in \mathbb{R}^{J\times 16}: &~\lambda_{jk} = 0 \mbox{~for~} j\neq i, 7\leq k\leq 16 \mbox{~and~} \lambda_{ik} = 0 \\
&\mbox{~for~} k=2,\ldots,6,12,\ldots,16\}.
\end{aligned}
\end{equation*}
Given an arbitrary $\Lambda^{*}\in\mathbb{R}^{J\times6}$ and unique variance matrix $\Psi^{*}$, we construct the loading matrix $\tilde{\Lambda}$ and unique variance matrix $\tilde{\Psi} = \mbox{diag}(\tilde{\psi}_{1},\ldots,\tilde{\psi}_{J})$ belonging to the parametric space defined in Case E such that $\tilde{\Lambda}\tilde{\Lambda}^{\top} +  \tilde{\Psi}= \Lambda^{*}\Lambda^{*\top} + \Psi^{*}$ as follows:
\begin{enumerate}
\item Let $R\in\mathbb{R}^{6\times 6}$ be an orthogonal matrix such that $R_{[\{1,\ldots,6\},\{1\}\}]}^{\top} = \frac{\Lambda^{*}_{[\{i\},\{1,\ldots,6\}]}}{\left\|\Lambda^{*}_{[\{i\},\{1,\ldots,6\}]}\right\|}$. 
\item Let $\tilde{\Lambda}_{[\{1,\ldots,J\}\setminus\{i\},\{1,\ldots,6\}]} = \Lambda^{*}_{[\{1,\ldots,J\}\setminus\{i\},\{1,\ldots,6\}]}R$ and $\tilde{\Lambda}_{[\{i\},\{1\}]} = \left\|\Lambda^{*}_{[\{i\},\{1,\ldots,6\}]}\right\|$.
\item Let $\tilde{\Lambda}_{[\{i\},\{7,\ldots,11\}]}$ be an arbitrary vector such that $\left\|\tilde{\Lambda}_{[\{i\},\{7,\ldots,11\}]}\right\|^{2}< \psi_{i}^{*}$, $\tilde{\psi}_{j} = \psi^{*}_{j}$ for $j\neq i$ and $\tilde{\psi}_{i} = \psi^{*}_{i} - \left\|\tilde{\Lambda}_{[\{i\},\{7,\ldots,11\}]}\right\|^{2}$.
\end{enumerate}
This construction shows that Case E also yields a global minimizer of the objective function. However, all cases A-E have at least one empty set among $\hat{v}_2, \hat{v}_3$ and $\hat{v}_{4}$. Since our goal is to recover the structure of three non-empty child factors of factor 1, such solutions violate the intention of the modeling and are excluded in Steps 5–8 of Algorithm~\ref{algo:ICB}.
\end{remark}

\section{Simulation studies for underestimated number of child factors}\label{appen:simulation algo3 under}
In this section, we examine the performance of Algorithm~\ref{algo:divide and conquer} and \ref{algo:ICB} when $c_{\max}$, the upper bound for the possible number of child factors of each factor, is underestimated. We adopt the same hierarchical structure and data generation model used in Section~\ref{appen:simulation algo3}. As illustrated in Figure~\ref{fig:sim under-estimated c}, $c_{\max}$ should be at least 3. However, in this simulation study, we deliberately set $c_{\max}=2$ and $d_{\max}=5$ when  applying Algorithms~\ref{algo:divide and conquer} and \ref{algo:ICB}. In this simulation study, we consider 8 simulation settings, given by the combinations of $J = 24, 36$, two sample sizes $N = 500, 2000$ and two generation processes of $\Psi$ used in Section~\ref{appen:simulation algo3}. For each setting, we generate the loading matrix and the unique variance matrix once, and then 100 independent simulations are generated. 
\begin{figure}[ht!]
\centering

\tikzset{every picture/.style={line width=0.75pt}} 

\begin{tikzpicture}[x=0.75pt,y=0.75pt,yscale=-1,xscale=1]

\draw   (304.5,40) -- (325.5,40) -- (325.5,60) -- (304.5,60) -- cycle ;
\draw    (204,71) -- (424,71) ;
\draw    (315,71) -- (315,60) ;
\draw   (193.5,92) -- (214.5,92) -- (214.5,112) -- (193.5,112) -- cycle ;
\draw   (247,141) -- (268,141) -- (268,161) -- (247,161) -- cycle ;
\draw   (140.5,141) -- (161.5,141) -- (161.5,161) -- (140.5,161) -- cycle ;
\draw   (271.5,191) -- (306.5,191) -- (306.5,232) -- (271.5,232) -- cycle ;
\draw    (151,161) -- (151,171) ;
\draw    (121,171) -- (183,171) ;
\draw    (121,171) -- (121,192) ;
\draw    (183,171) -- (183,192) ;
\draw    (226,171) -- (289,171) ;
\draw    (257.5,161) -- (257.5,171) ;
\draw    (226,171) -- (226,192) ;
\draw    (289,171) -- (289,192) ;
\draw    (204,112) -- (204,122) ;
\draw    (151,122) -- (257.5,122) ;
\draw    (151,122) -- (151,141) ;
\draw    (257.5,122) -- (257.5,141) ;
\draw    (204,92) -- (204,71) ;
\draw   (164,192) -- (199,192) -- (199,233) -- (164,233) -- cycle ;
\draw   (208.5,191) -- (243.5,191) -- (243.5,232) -- (208.5,232) -- cycle ;
\draw   (103.5,191) -- (138.5,191) -- (138.5,232) -- (103.5,232) -- cycle ;
\draw   (413.5,92) -- (434.5,92) -- (434.5,112) -- (413.5,112) -- cycle ;
\draw   (467,141) -- (488,141) -- (488,161) -- (467,161) -- cycle ;
\draw   (360.5,141) -- (381.5,141) -- (381.5,161) -- (360.5,161) -- cycle ;
\draw   (491.5,191) -- (526.5,191) -- (526.5,232) -- (491.5,232) -- cycle ;
\draw    (371,161) -- (371,171) ;
\draw    (341,171) -- (403,171) ;
\draw    (341,171) -- (341,192) ;
\draw    (403,171) -- (403,192) ;
\draw    (446,171) -- (509,171) ;
\draw    (477.5,161) -- (477.5,171) ;
\draw    (446,171) -- (446,192) ;
\draw    (509,171) -- (509,192) ;
\draw    (424,112) -- (424,122) ;
\draw    (371,122) -- (477.5,122) ;
\draw    (371,122) -- (371,141) ;
\draw    (477.5,122) -- (477.5,141) ;
\draw    (424,92) -- (424,71) ;
\draw   (384,192) -- (419,192) -- (419,233) -- (384,233) -- cycle ;
\draw   (428.5,191) -- (463.5,191) -- (463.5,232) -- (428.5,232) -- cycle ;
\draw   (323.5,191) -- (358.5,191) -- (358.5,232) -- (323.5,232) -- cycle ;

\draw (306.5,43) node [anchor=north west][inner sep=0.75pt]   [align=left] {$\displaystyle v_{1}^{*}$};
\draw (248,206) node [anchor=north west][inner sep=0.75pt]   [align=left] {$\displaystyle \cdots $};
\draw (141,206) node [anchor=north west][inner sep=0.75pt]   [align=left] {$\displaystyle \cdots $};
\draw (195.5,95) node [anchor=north west][inner sep=0.75pt]   [align=left] {$\displaystyle v_{2}^{*}$};
\draw (142.5,144) node [anchor=north west][inner sep=0.75pt]   [align=left] {$\displaystyle v_{5}^{*}$};
\draw (249,144) node [anchor=north west][inner sep=0.75pt]   [align=left] {$\displaystyle v_{6}^{*}$};
\draw (211,197) node [anchor=north west][inner sep=0.75pt]   [align=left] {1+$\displaystyle \frac{J}{6}$};
\draw (116,203.5) node [anchor=north west][inner sep=0.75pt]   [align=left] {1};
\draw (174,197) node [anchor=north west][inner sep=0.75pt]   [align=left] {$\displaystyle \frac{J}{6}$};
\draw (282.5,197) node [anchor=north west][inner sep=0.75pt]   [align=left] {$\displaystyle \frac{J}{3}$};
\draw (468,206) node [anchor=north west][inner sep=0.75pt]   [align=left] {$\displaystyle \cdots $};
\draw (362,206) node [anchor=north west][inner sep=0.75pt]   [align=left] {$\displaystyle \cdots $};
\draw (416,96) node [anchor=north west][inner sep=0.75pt]   [align=left] {$\displaystyle v_{f}$};
\draw (362.5,144) node [anchor=north west][inner sep=0.75pt]   [align=left] {$\displaystyle v_{3}^{*}$};
\draw (469,144) node [anchor=north west][inner sep=0.75pt]   [align=left] {$\displaystyle v_{4}^{*}$};
\draw (427.5,197) node [anchor=north west][inner sep=0.75pt]   [align=left] {1+$\displaystyle \frac{2J}{3}$};
\draw (326,197) node [anchor=north west][inner sep=0.75pt]   [align=left] {1+$\displaystyle \frac{J}{3}$};
\draw (392,197) node [anchor=north west][inner sep=0.75pt]   [align=left] {$\displaystyle \frac{2J}{3}$};
\draw (502.5,204) node [anchor=north west][inner sep=0.75pt]   [align=left] {$\displaystyle J$};

\end{tikzpicture}
\caption{The hierarchical factor structure learned with underestimated $c_{\max}$.}
    \label{fig:sim structure under-estimated c}
\end{figure}

Figure~\ref{fig:sim structure under-estimated c} displays the most frequently estimated hierarchical structure, which is selected in more than 60\% of the 100 replications across all settings. As shown, Algorithms~\ref{algo:divide and conquer} and~\ref{algo:ICB} recover a correctly specified but less parsimonious representation of the true hierarchy. To be more exact, a redundant factor, whose sets of variables $v_{f} = v_{3}^{*}\cup v_{4}^{*}$, is learned due to the choice of $c_{\max} = 2$ in the current simulation settings.

\section{Real Data Analysis: Agreeableness Scale Item Key}\label{appen:ItemKey}

\begin{table}[htb!]
    \centering
    \caption{Agreeableness Item Key }
    \label{tab:ItemKey}
    \scalebox{1}{
    \begin{tabular}{clll}
    \toprule
    Item & Sign & Facet & Item detail\\
    \midrule
        1&+ & Trust(A1) & Trust others. \\ 
        2&+  & Trust(A1) & Believe that others have good intentions. \\ 
        3&+  & Trust(A1) & Trust what people say. \\ 
        4&$-$  & Trust(A1) & Distrust people. \\ 
        5&$-$  & Morality(A2) & Use others for my own ends. \\ 
        6&$-$  & Morality(A2) & Cheat to get ahead. \\ 
        7&$-$  & Morality(A2) & Take advantage of others. \\ 
        8&$-$  & Morality(A2) & Obstruct others' plans. \\ 
        9&+  & Altruism(A3) & Love to help others. \\ 
        10&+  & Altruism(A3) & Am concerned about others. \\ 
        11&$-$  & Altruism(A3) & Am indifferent to the feelings of others. \\ 
        12&$-$  & Altruism(A3) & Take no time for others. \\ 
        13&$-$  & Cooperation(A4) & Love a good fight. \\ 
        14&$-$  & Cooperation(A4) & Yell at people. \\ 
        15&$-$  & Cooperation(A4) & Insult people. \\ 
        16&$-$  & Cooperation(A4) & Get back at others. \\ 
        17&$-$  & Modesty(A5) & Believe that I am better than others. \\ 
        18&$-$  & Modesty(A5) & Think highly of myself. \\ 
        19&$-$  & Modesty(A5) & Have a high opinion of myself. \\ 
        20&$-$  & Modesty(A5) & Boast about my virtues. \\ 
        21&+  & Sympathy(A6) & Sympathize with the homeless. \\ 
        22&+ & Sympathy(A6) & Feel sympathy for those who are worse off than myself. \\ 
        23&$-$  & Sympathy(A6) & Am not interested in other people's problems. \\ 
        24&$-$& Sympathy(A6) & Try not to think about the needy. \\ 
        \bottomrule
    \end{tabular}
    }
\end{table}

\FloatBarrier

\section{Real Data Analysis: Additional Results}\label{append:rd compete}

In this section, we present the estimated loading matrix and correlation matrix of the three competing models. The estimated correlation matrix of the three models, denoted by $\hat{\Phi}_{\text{CFA}}$, $\hat{\Phi}_{\text{CBF}}$, $\hat{\Phi}_{\text{EBF}}$, are shown in \eqref{eq:phi CFA}, \eqref{eq:phi CBF}, and \eqref{eq:phi EBF}. The estimated loading matrix of the three models, denoted by $\hat{\Lambda}_{\text{CFA}}$, $\hat{\Lambda}_{\text{CBF}}$, $\hat{\Lambda}_{\text{EBF}}$, are shown in \eqref{eq:loading CFA}, \eqref{eq:loading CBF}, and \eqref{eq:loading EBF}. 

\begin{equation}\label{eq:phi CFA}
\resizebox{0.5\textwidth}{!}{$
\begin{aligned}
\hat{\Phi}_{\text{CFA}} = \left(\begin{array}{cccccc}
        1 & 0.33 & 0.44 & 0.43 & -0.06 & 0.37 \\ 
        0.33 & 1 & 0.42 & 0.62 & 0.25 & 0.37 \\ 
        0.44 & 0.42 & 1 & 0.39 & 0.15 & 0.80 \\ 
        0.43 & 0.62 & 0.39 & 1 & 0.11 & 0.30 \\ 
        -0.06 & 0.25 & 0.15 & 0.11 & 1 & 0.16 \\ 
        0.37 & 0.37 & 0.80 & 0.30 & 0.16 & 1 \\ 
    \end{array}\right),
\end{aligned}
$}
\end{equation}

\bigskip

\begin{equation}\label{eq:phi CBF}
\resizebox{0.5\textwidth}{!}{$
\begin{aligned}
\hat{\Phi}_{\text{CBF}} = \left(\begin{array}{ccccccc}
        1 & 0 & 0 & 0 & 0 & 0 & 0 \\
        0 & 1 & 0.01 & 0.24 & 0.03 & -0.07 & 0.25 \\ 
        0 & 0.01 & 1 & 0.12 & 0.27 & 0.34 & 0.22 \\ 
        0 & 0.24 & 0.12 & 1 & -0.08 & 0.18 & 0.74 \\ 
        0 & 0.03 & 0.27 & -0.08 & 1 & 0.25 & 0.05 \\ 
        0 & -0.07 & 0.34 & 0.18 & 0.25 & 1 & 0.17 \\ 
        0 & 0.25 & 0.22 & 0.74 & 0.05 & 0.17 & 1 \\ 
    \end{array}\right),
\end{aligned}
$}
\end{equation}

\bigskip

\begin{equation}\label{eq:phi EBF}
\resizebox{0.5\textwidth}{!}{$
\begin{aligned}
\hat{\Phi}_{\text{EBF}} = \left(\begin{array}{ccccccc}
        1 & 0 & 0 & 0 & 0 & 0 & 0 \\ 
        0 & 1 & 0.12 & 0.18 & 0.24 & 0.12 & -0.02 \\ 
        0 & 0.12 & 1 & 0.50 & 0.11 & 0.95 & 0.33 \\ 
        0 & 0.18 & 0.50 & 1 & 0.13 & 0.74 & 0.24 \\ 
        0 & 0.24 & 0.11 & 0.13 & 1 & 0.09 & -0.14 \\ 
        0 & 0.12 & 0.95 & 0.74 & 0.09 & 1 & 0.31 \\ 
        0 & -0.02 & 0.33 & 0.24 & -0.14 & 0.31 & 1 \\ 
    \end{array}\right).
\end{aligned}
$}
\end{equation}

\begin{equation}\label{eq:loading CFA}
\resizebox{0.5\textwidth}{!}{$
\begin{aligned}
\hat{\Lambda}_{\text{CFA}} = \left(\begin{array}{cccccc}
        0.85 & 0 & 0 & 0 & 0 & 0 \\ 
        0.73 & 0 & 0 & 0 & 0 & 0 \\ 
        0.76 & 0 & 0 & 0 & 0 & 0 \\ 
        0.87 & 0 & 0 & 0 & 0 & 0 \\ 
        0 & 0.89 & 0 & 0 & 0 & 0 \\ 
        0 & 0.64 & 0 & 0 & 0 & 0 \\ 
        0 & 0.92 & 0 & 0 & 0 & 0 \\ 
        0 & 0.39 & 0 & 0 & 0 & 0 \\ 
        0 & 0 & 0.51 & 0 & 0 & 0 \\ 
        0 & 0 & 0.61 & 0 & 0 & 0 \\ 
        0 & 0 & 0.67 & 0 & 0 & 0 \\ 
        0 & 0 & 0.57 & 0 & 0 & 0 \\ 
        0 & 0 & 0 & 0.55 & 0 & 0 \\ 
        0 & 0 & 0 & 0.71 & 0 & 0 \\ 
        0 & 0 & 0 & 0.81 & 0 & 0 \\ 
        0 & 0 & 0 & 0.71 & 0 & 0 \\ 
        0 & 0 & 0 & 0 & 0.71 & 0 \\ 
        0 & 0 & 0 & 0 & 0.90 & 0 \\ 
        0 & 0 & 0 & 0 & 1.12 & 0 \\ 
        0 & 0 & 0 & 0 & 0.33 & 0 \\ 
        0 & 0 & 0 & 0 & 0 & 0.70 \\ 
        0 & 0 & 0 & 0 & 0 & 0.71 \\ 
        0 & 0 & 0 & 0 & 0 & 0.65 \\ 
        0 & 0 & 0 & 0 & 0 & 0.65 \\ 
    \end{array}\right),
\end{aligned}
$}
\end{equation}

\begin{equation}\label{eq:loading CBF}
\resizebox{0.5\textwidth}{!}{$
\begin{aligned}
\hat{\Lambda}_{\text{CBF}} = \left(\begin{array}{ccccccc}
        0.42 & 0.73 & 0 & 0 & 0 & 0 & 0 \\ 
        0.35 & 0.64 & 0 & 0 & 0 & 0 & 0 \\ 
        0.30 & 0.72 & 0 & 0 & 0 & 0 & 0 \\ 
        0.53 & 0.69 & 0 & 0 & 0 & 0 & 0 \\ 
        0.46 & 0 & 0.83 & 0 & 0 & 0 & 0 \\ 
        0.49 & 0 & 0.41 & 0 & 0 & 0 & 0 \\ 
        0.57 & 0 & 0.71 & 0 & 0 & 0 & 0 \\ 
        0.47 & 0 & 0.11 & 0 & 0 & 0 & 0 \\ 
        0.25 & 0 & 0 & 0.45 & 0 & 0 & 0 \\ 
        0.24 & 0 & 0 & 0.60 & 0 & 0 & 0 \\ 
        0.43 & 0 & 0 & 0.51 & 0 & 0 & 0 \\ 
        0.41 & 0 & 0 & 0.41 & 0 & 0 & 0 \\ 
        0.28 & 0 & 0 & 0 & 0.70 & 0 & 0 \\ 
        0.54 & 0 & 0 & 0 & 0.43 & 0 & 0 \\ 
        0.68 & 0 & 0 & 0 & 0.41 & 0 & 0 \\ 
        0.66 & 0 & 0 & 0 & 0.27 & 0 & 0 \\ 
        0.30 & 0 & 0 & 0 & 0 & 0.73 & 0 \\ 
        -0.14 & 0 & 0 & 0 & 0 & 0.93 & 0 \\ 
        -0.03 & 0 & 0 & 0 & 0 & 1.09 & 0 \\ 
        0.38 & 0 & 0 & 0 & 0 & 0.35 & 0 \\ 
        0.15 & 0 & 0 & 0 & 0 & 0 & 0.73 \\ 
        0.16 & 0 & 0 & 0 & 0 & 0 & 0.75 \\ 
        0.38 & 0 & 0 & 0 & 0 & 0 & 0.52 \\ 
        0.27 & 0 & 0 & 0 & 0 & 0 & 0.58 \\ 
    \end{array}\right),
\end{aligned}
$}
\end{equation}

\begin{equation}\label{eq:loading EBF}
\resizebox{0.5\textwidth}{!}{$
\begin{aligned}
\hat{\Lambda}_{\text{EBF}} = \left(\begin{array}{ccccccc}
        0.35 & 0 & 0 & 0 & 0 & 0 & 0.77 \\ 
        0.29 & 0 & 0 & 0 & 0 & 0 & 0.67 \\ 
        0.27 & 0 & 0 & 0 & 0 & 0 & 0.73 \\ 
        0.45 & 0 & 0 & 0 & 0 & 0 & 0.74 \\ 
        0.65 & 0.65 & 0 & 0 & 0 & 0 & 0 \\ 
        0.55 & 0.32 & 0 & 0 & 0 & 0 & 0 \\ 
        0.69 & 0.61 & 0 & 0 & 0 & 0 & 0 \\ 
        0.45 & 0 & 0.10 & 0 & 0 & 0 & 0 \\ 
        0.18 & 0 & 0.52 & 0 & 0 & 0 & 0 \\ 
        0.19 & 0 & 0 & 0 & 0 & 0.57 & 0 \\ 
        0.37 & 0 & 0 & 0 & 0 & 0.55 & 0 \\ 
        0.32 & 0 & 0.52 & 0 & 0 & 0 & 0 \\ 
        0.50 & 0 & 0 & 0 & 0.18 & 0 & 0 \\ 
        0.68 & -0.27 & 0 & 0 & 0 & 0 & 0 \\ 
        0.80 & -0.21 & 0 & 0 & 0 & 0 & 0 \\ 
        0.69 & 0 & 0 & 0 & 0 & 0 & 0.10 \\ 
        0.38 & 0 & 0 & 0 & 0.67 & 0 & 0 \\ 
        0.01 & 0 & 0 & 0 & 0.93 & 0 & 0 \\ 
        0.13 & 0 & 0 & 0 & 1.09 & 0 & 0 \\ 
        0.44 & 0 & 0 & 0 & 0.29 & 0 & 0 \\ 
        0.20 & 0 & 0 & 0.74 & 0 & 0 & 0 \\ 
        0.18 & 0 & 0 & 0.75 & 0 & 0 & 0 \\ 
        0.29 & 0 & 0 & 0 & 0 & 0.62 & 0 \\ 
        0.27 & 0 & 0 & 0.59 & 0 & 0 & 0 \\ 
    \end{array}\right),
\end{aligned}
$}
\end{equation}

\newpage
\bibliographystyle{apalike}
\bibliography{ref}

\end{document}